\documentclass[pra, notitlepage, showpacs, floatfix, reprint, nobalancelastpage,citeautoscript]{revtex4-1}

\usepackage[T1]{fontenc}
\usepackage[latin9]{inputenc}

\usepackage{amsmath}
\usepackage{amssymb}
\usepackage{bbm}
\usepackage{braket}
\usepackage{xcolor}
\usepackage{pifont}
\newcommand{\cmark}{\ding{51}}%
\newcommand{\xmark}{\ding{55}}%
\usepackage[mathscr]{euscript}
\usepackage[shortlabels]{enumitem}

\allowdisplaybreaks

\usepackage{graphicx}
\usepackage[colorlinks=true]{hyperref}  
\hypersetup{
    bookmarks=true,         
    unicode=false,          
    pdftoolbar=true,        
    pdfmenubar=true,        
    pdffitwindow=false,     
    pdfstartview={FitH},    
    pdftitle={Pairing in graphene-based moire superlattices},    
    pdfauthor={Mathias Scheurer},     
    pdfsubject={},   
    pdfcreator={},   
    pdfproducer={}, 
    pdfkeywords={} {} {}, 
    pdfnewwindow=true,      
    colorlinks=true,       
    linkcolor=blue, 
    citecolor=blue,        
    filecolor=magenta,      
    urlcolor=blue           
}

\newcommand{\equref}[1]{Eq.~(\ref{#1})}
\newcommand{\equsref}[2]{Eqs.~(\ref{#1}) and (\ref{#2})}
\newcommand{\secref}[1]{Sec.~\ref{#1}}
\newcommand{\figref}[1]{Fig.~\ref{#1}}
\newcommand{\refcite}[1]{Ref.~\onlinecite{#1}} 
\newcommand{\refscite}[1]{Refs.~\onlinecite{#1}}

\newcommand{\tableref}[1]{Table~\ref{#1}}
\newcommand{\appref}[1]{Appendix~\ref{#1}}

\newcommand{\pdagger}{{\phantom{\dagger}}}
\newcommand{\diff}{\mathrm{d}}
\newcommand{\sign}{\,\text{sign}}
\renewcommand{\approx}{\simeq}

\renewcommand{\vec}[1]{\boldsymbol{#1}}

\definecolor{wrongultramarine}{rgb}{1,0.5,0}
\begin{document}
\title{Pairing in graphene-based moir\'e superlattices}
\author{Mathias S.~Scheurer}
\affiliation{Department of Physics, Harvard University, Cambridge, MA
02138, USA}

\author{Rhine Samajdar}
\affiliation{Department of Physics, Harvard University, Cambridge, MA
02138, USA}

\begin{abstract}
We present a systematic classification and analysis of possible pairing instabilities in graphene-based moir\'e superlattices. Motivated by recent experiments on twisted double-bilayer graphene showing signs of triplet superconductivity, we analyze both singlet and triplet pairing separately, and describe how these two channels behave close to the limit where the system is invariant under separate spin rotations in the two valleys, realizing an SU(2)$_+$ $\times$ SU(2)$_-$ symmetry. Further, we discuss the conditions under which singlet and triplet can mix via two nearly degenerate transitions, and how the different pairing states behave when an external magnetic field is applied. The consequences of the additional microscopic or emergent approximate symmetries relevant for superconductivity in twisted bilayer graphene and ABC trilayer graphene on hexagonal boron nitride are described in detail.
We also analyze which of the pairing states can arise in mean-field theory and study the impact of corrections coming from ferromagnetic fluctuations.
For instance, we show that, close to the parameters of mean-field theory, a nematic mixed singlet-triplet state emerges. 
Our study illustrates that graphene superlattices provide a rich platform for exotic superconducting states, and allow for the admixture of singlet and triplet pairing even in the absence of spin-orbit coupling.
\end{abstract}

\maketitle

\section{Introduction}
Experiments on twisted bilayers of graphene have recently revealed interaction-induced insulating phases and superconductivity when the relative angle between the layers is fine-tuned to yield almost flat moir\'e bands, which enhances the impact of electronic correlations \cite{2018Natur.556...80C,SupercondTBGExp,Yankowitz1059,2019arXiv190306513L}. Due to the strong-coupling nature of the problem, which is corroborated by tunneling spectroscopy measurements \cite{PasupathySTM,NadjPergeSTM,AndreiSTM,YazdaniSTM,STMReview}, the form and mechanism of the insulating and superconducting phases are still under debate, despite considerable theoretical effort \cite{2018arXiv180400627B,2018arXiv180506906W,2018PhRvB..97w5453G,KoshinoFuModel,FuModel1,AshvinModelTBG,2018arXiv180704382L,EmergentSymmetries,2018PhRvB..98g5154D,CenkeLeon,2018PhRvB..98g5109T,2018PhRvB..98h5436F,VafekModel,2018SSCom.282...38L,2018arXiv181008642K,2018arXiv181202550S,2018PhRvX...8d1041I,2018PhRvL.121u7001L,2018PhRvB..98t5151S,2018PhRvB..98s5101S,RafaelsPaperTBG,2018PhRvB..98v0504P,2019arXiv190100083L,2018PhRvB..98x1407K,2018PhRvB..98x1412C,2018PhRvL.121y7001W,2019arXiv190100500L,HUANG2019310,PhysRevLett.122.026801,2019JPCM...31f5601C,2019arXiv190301701C,2019PhRvB..99l1407R,2019arXiv190407875W,LiangFu,YiZhuangPairing,2019PhRvB..99m4515R,2019PhRvB..99o5413A}. 
Another graphene-based moir\'e system that displays both superconducting and correlated insulating behavior is ABC-stacked trilayer graphene on hexagonal boron nitride \cite{2019NatPh..15..237C,SuperconductivityInTrilayer}. In this case, the moir\'e pattern results from the difference in lattice constants, and it can be controlled by application of a vertical electric field \cite{PhysRevLett.122.016401,SenthilABCTrilayer}.

The most recent member of the family of strongly correlated graphene superlattice systems is twisted double-bilayer graphene \cite{2019arXiv190306952S,ExperimentKim,PabllosExperiment}, where two individually aligned AB-stacked graphene bilayers are twisted with respect to one another. As theoretical calculations show \cite{2019PhRvB..99g5127Z,2019arXiv190108420R,2019arXiv190300852C,2019arXiv190308685L,KoshinosPaper,FirstModel,2019arXiv190600623H}, flat electronic bands can be realized by tuning the twist angle and a vertical electric field. Similar to the abovementioned graphene moir\'e systems, both correlated insulating \cite{2019arXiv190306952S,ExperimentKim,PabllosExperiment} and superconducting \cite{2019arXiv190306952S,ExperimentKim} phases are observed in experiment. However, in stark contrast to twisted bilayer and trilayer graphene, the superconducting transition temperature is found to increase linearly with a weak in-plane magnetic field \cite{ExperimentKim}, which is a strong indication of triplet pairing \cite{2019arXiv190308685L,2019arXiv190407875W}. Furthermore, the gap of the correlated insulating phase is seen to increase with an applied magnetic field, indicating ferromagnetic order \cite{2019arXiv190306952S,ExperimentKim,PabllosExperiment}. There are also clear experimental indications of ferromagnetism in twisted bilayer \cite{Sharpe605,2019arXiv190306513L,IlaniExperiment} and ABC trilayer graphene \cite{FMTrilayer}.

\begin{figure*}[tb]
   \centering
    \includegraphics[width=\linewidth]{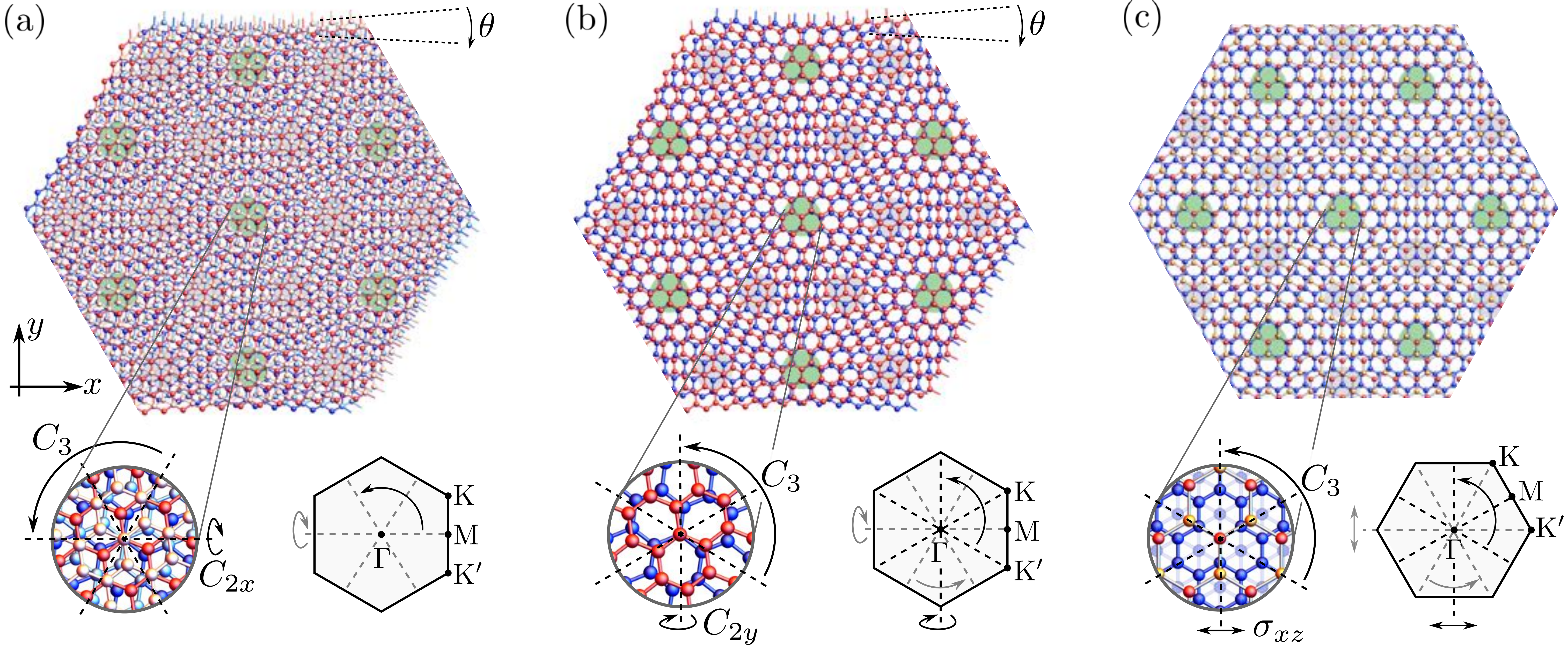}
    \caption{Geometry (top), lattice symmetries (bottom left), and Brillouin zone with symmetries (bottom right) for (a) twisted double-bilayer graphene, (b) twisted bilayer graphene, and (c) ABC-stacked trilayer graphene on hexagonal boron nitride. In all three cases, we assume a commensurate superlattice structure for simplicity. For (a) twisted double-bilayer, we only show ABAB stacking, which exhibits a $\pi$-rotation symmetry $C_{2x}$ along the $x$-axis. For ABBA stacking, we have a $C_{2y}$ symmetry instead. Our symmetry analysis of pairing applies to both stacking orders as the applied electric field breaks both of these in-plane rotation symmetries. In (b) twisted bilayer graphene, superconductivity emerges without any applied electric field and, hence, the $C_{2y}$ symmetry of the lattice has to be taken into account. In addition, the system is approximately invariant under a $C_6$ symmetry \cite{AshvinModelTBG} as indicated in gray in the Brillouin zone in (b). In (c), we show only the top boron nitride layer and one of the graphene layers for image clarity in the main panel; the other two graphene layers are indicated in the close-up view of the lattice in light blue. We assume no additional twist such that the moir\'e pattern is solely due to the lattice mismatch. This leads to the reflection symmetry $\sigma_{xz}$, which for the effective two-dimensional description of the system can be viewed as an in-plane rotation $C_{2y}$ as is done in the main text. Also, this system is believed to exhibit an approximate $C_6$ symmetry \cite{SenthilABCTrilayer} as indicated in gray in the respective Brillouin zone.}
    \label{LatticesAndSymmetries}
\end{figure*}

In this paper, we study the possible pairing states in graphene moir\'e superlattices. Motivated by the recent experimental signatures of triplet pairing, we pay special attention to the triplet channel, and possible mixed singlet and triplet phases. While the weak spin-orbit coupling in graphene seems to disfavor the latter class of phases, projections of the Coulomb interaction on the relevant moir\'e bands evince that the interaction terms that couple the spin degrees of freedom of the two valleys, $v=\pm$, of the system are much weaker than other interaction terms that do not \cite{KoshinoFuModel,SenthilABCTrilayer,2019arXiv190308685L}. Together with the nearly valley-diagonal band structure, this indicates that the system is approximately invariant under independent spin rotations in the two valleys. As has been pointed out before \cite{CenkeLeon,YiZhuangPairing}, the associated SU(2)$_+$ $\times$ SU(2)$_-$ symmetry renders the singlet and triplet pairing channels degenerate. This paper will address the questions: ({\it i\/}) under which conditions can singlet and triplet mix when the SU(2)$_+$ $\times$ SU(2)$_-$ symmetry is only weakly broken, and ({\it ii\/}) which triplet state transforms into which singlet upon reversing the sign of the symmetry-breaking interactions? In this way, we map out all possible phase diagrams close to the SU(2)$_+$ $\times$ SU(2)$_-$-invariant limit. 

In light of the narrow bandwidth and strong correlations of the graphene-based moir\'e superlattices we are interested in, our analysis will begin with a comprehensive study of exact constraints resulting from symmetry. The symmetry-based classification will then be supplemented with energetics, by studying which of the pairing states can be realized in the weak-coupling limit and what changes in the presence of additional fluctuation corrections.
As it has the smallest set of symmetries, we will begin our classification with twisted double-bilayer graphene: while the lattice is invariant under threefold  rotation, $C_3$, perpendicular to the graphene sheets, and under a twofold in-plane rotation [see \figref{LatticesAndSymmetries}(a)], the latter is broken due to the vertical electric field that is applied to tune the band structure and to induce superconductivity. 
It seems currently unclear whether the superconducting state coexists with the likely ferromagnetic correlated insulator and whether, at least in part of the phase diagram, there is a thermal transition directly from the (paramagnetic) normal metal to superconductivity without any ferromagnetic order. For this reason, we will analyze two scenarios separately: (I) there is no ferromagnetic order around the critical temperature, $T_c$, of superconductivity, and (II) there is ferromagnetic order already at $T>T_c$ that coexists microscopically with superconductivity for $T<T_c$ (or at the minimum, the associated ferromagnetic moments couple significantly to the superconducting order parameter).
We will begin with the analysis of the superconducting states transforming under the IRs of the point group $C_3$ assuming time-reversal symmetry in the high-temperature phase---this is relevant for case (I) above. In order to capture scenario (II), we will later add the coupling to the time-reversal-symmetry breaking magnetic moments and examine how it affects the superconducting transition. This allows us to determine which of the pairing states are compatible with the linear increase of the critical temperature with small magnetic field, $B$, and to describe the possible phase diagrams in the temperature-$B$ plane.

We also generalize our discussion to twisted bilayer graphene and ABC trilayer graphene on hexagonal boron nitride. Here, we have to take into account an additional twofold rotation symmetry, $C_2$, perpendicular to the plane of the system and an in-plane rotation symmetry $C_{2y}$ along the $y$-axis; these symmetries are either realized as exact microscopic symmetries of the lattice or as approximate emergent symmetries \cite{AshvinModelTBG,EmergentSymmetries,SenthilABCTrilayer} of those systems, see \figref{LatticesAndSymmetries}(b) and (c).

\subsection{Brief summary of the main results}
Due to the length of the paper, here, we provide a very concise summary of the key results of this work for the convenience of the reader:
\begin{enumerate}[(1),leftmargin=2\parindent]
    \item We analyze the consequences of the enhanced SU(2)$_+$ $\times$ SU(2)$_-$ spin symmetry, taking into account the possibility of several consecutive superconducting transitions with their difference in transition temperatures vanishing in the limit where SU(2)$_+$ $\times$ SU(2)$_-$ becomes exact. The resulting complete sets of possible phases for the relevant symmetry groups $C_3$ and $D_3$ (or, equivalently, $D_6$, see \secref{TwistedBilayerGraphene}) are summarized in Tables~\ref{TableSummaryOfPairingC3}, \ref{SummaryOfPairingComplexRepC3}, and \ref{SummaryOfPairingD3}. 
    \item As follows from these tables, all point groups and all of their irreducible representations (IRs) allow for singlet-triplet admixed phases in the \textit{absence} of any spin-orbit coupling. As opposed to the conventional mechanism of singlet-triplet admixture, which is based on a \textit{reduced} symmetry \cite{GorkovRashbaAdmixture}, here, it results from (the proximity to) an \textit{enhanced} spin symmetry.
    \item To supplement these purely symmetry-based considerations with energetics, we analyze which of those states can be realized in single-band mean-field theory, \textit{i.e.}, whether there exists a form of the effective electron-electron interaction that can stabilize the superconducting state when treated within the mean-field approximation; the result is indicated in the last column in Tables \ref{TableSummaryOfPairingC3}, \ref{SummaryOfPairingComplexRepC3}, and \ref{SummaryOfPairingD3}. This identifies the most important pairing states from a weak-coupling perspective. The presence of any of the remaining pairing phases---as might eventually be established in future experiments---must result from the strong-coupling and/or interband nature of superconductivity.
    \item We also study corrections to mean-field theory coming from ferromagnetic fluctuations, within a simplified phenomenological approach in \secref{FluctuationInducedSC} that is justified microscopically in \appref{MicroscopicDerivationFluctuations}. We analyze two limits. First, we consider the case of weak fluctuations in order to lift the residual degeneracies within mean-field theory. We find that out of the two possible phase diagrams for the IR $E$ close to mean-field theory, shown in \figref{SchematicPDForIRE}, the one in part (b) [part (a)] is favored for spin (orbital) ferromagnetic fluctuations. This reveals that a nematic mixed singlet-triplet phase is a natural candidate pairing phase in graphene moir\'e superlattices.
    Second, we analyze which pairing states are favored in the case where the fluctuation corrections dominate over the mean-field contributions (see last column in the tables mentioned above).
    \item We study the coupling of the superconducting states to the magnetic field, $B$, and examine which states can give rise to a linear increase of the critical temperature for small $B$: if SU(2)$_+$ $\times$ SU(2)$_-$ is broken significantly, triplet pairing has to dominate for $B=0$ and there are only three possible triplet states as leading instabilities for $B\neq0$. In the case where SU(2)$_+$ $\times$ SU(2)$_-$ is an approximate symmetry, even singlet pairing at $B=0$ can yield a linear increase. For instance, the possible phase diagrams in the presence of a magnetic field for pairing in the trivial IR $A$ of $C_3$ are shown in \figref{PDInMagneticField}.

\end{enumerate}

\subsection{Relation to other works}
Let us briefly comment on the relation of this article to other works in the literature. While our classification also contains the pure singlet states, which have been subject to intense scrutiny in twisted bilayer graphene, we are mainly interested in elucidating the consequences of the enhanced SU(2)$_+$ $\times$ SU(2)$_-$ spin symmetry with respect to subsequent transitions and the associated nontrivial interplay of singlet and triplet pairing.

In the context of twisted double-bilayer graphene, where, recently, signs of triplet pairing have been discovered, \refcite{XenkeFollowUp} mainly focuses on the correlated insulating phase in this system, whereas \refcite{2019arXiv190308685L} also discusses pairing. We extend the work of \refcite{2019arXiv190308685L} by allowing for momentum-dependent pairing states, contrasting weakly and significantly broken SU(2)$_+$ $\times$ SU(2)$_-$ symmetry, investigating admixed singlet and triplet phases, analyzing fluctuation corrections to mean-field theory, and mapping out the phase diagram in the presence of a magnetic field. In a follow-up work \cite{OurFollowUp}, we will complement the analysis of this paper by a microscopic energetic study specifically for twisted double-bilayer graphene. 
In \refcite{OurFollowUp}, we discuss which IR is expected to be favored, the form of the associated basis functions, and the impact of disorder on superconductivity.

\subsection{Structure of the paper}
This paper is organized as follows: as described above, we start with twisted double-bilayer graphene. In \secref{ModelAndSymmetries}, we introduce the model and the action of the relevant symmetries. We first discuss pairing in the trivial IR of the point group of the system in \secref{TrivialRepresentation} and then generalize to the complex IR $E$ in \secref{ComplexRepresentation}. Section \ref{FluctuationInducedSC} demonstrates how strong fluctuations can yield significant corrections to mean-field theory.
We extend our analysis to twisted bilayer graphene and ABC trilayer graphene in \secref{ComparisonWithOtherSystems}, and explore the consequences of the additional microscopic and emergent symmetries relevant to those systems.
A discussion of our results can be found in \secref{SummaryDiscussion}.

\section{Model and symmetries}
\label{ModelAndSymmetries}
We first focus on the (nearly flat) conduction band of twisted double-bilayer graphene which appears to host the superconducting phase observed experimentally \cite{2019arXiv190306952S,ExperimentKim}, and later discuss the modifications for the related moir\'e systems, bilayer and trilayer graphene. Owing to the presence of a gap to other bands in the relevant parameter regime \cite{2019arXiv190300852C,FirstModel,KoshinosPaper,2019arXiv190308685L}, it is reasonable to describe the superconducting instability in a single-band picture. We stress, however, that many of our conclusions are symmetry-based and thus, also apply when several bands are taken into account. Exceptions are provided by the energetic mean-field and fluctuation considerations, where we will specifically comment on the consequences of interband effects that might be present in these systems \cite{YazdaniSTM}.

Denoting the corresponding electronic creation and annihilation operators by $c^{}_{\vec{k}\sigma v}$, where $\vec{k}$ is crystal momentum, $\sigma$ spin, and $v=\pm$ represents the valleys, the general pairing term can be written as
\begin{equation}
    \mathcal{H}^{}_{\text{SC}} = \sum_{\vec{k}} c^\dagger_{\vec{k}\sigma v} \left( \Delta_{\vec{k}} i\sigma^{}_y\tau^{}_x \right)_{\sigma v,\sigma' v'} c^\dagger_{-\vec{k}\sigma' v'} + \text{H.c.}. \label{GeneralFormOfPairing}
\end{equation}
Here and in the following, $\sigma_j$ and $\tau_j$ are Pauli matrices in spin and valley space, respectively, and the $4\times 4$ matrix $\Delta_{\vec{k}}$ is the superconducting order parameter. 
In \equref{GeneralFormOfPairing}, we have already made the assumptions that only Cooper pairs with zero net momentum form and that superconductivity preserves translational symmetry. Due to the proximity of superconductivity to ferromagnetic order \cite{2019arXiv190306952S,ExperimentKim}, relaxing this assumption could be interesting, but we leave this for future work. Consequently, we need not consider IRs of the full space group but rather, can concentrate on the point group $\mathcal{G}$ of the system and time-reversal $\Theta$.

In this regard, we study two different point groups: an approximate point group,
\begin{equation}
    \mathcal{G}_1 = C_3 \times \text{SU}(2)_+ \times \text{SU}(2)_- \times U(1)_v, \label{ApproximateSymmetryGroup}
\end{equation}
where $C_3$ is the crystalline point group, SU(2)$_{\pm}$ is spin rotation in valley $v=\pm$, and $U(1)_v$ corresponds to valley charge conservation. As argued in \refcite{2019arXiv190308685L}, the intervalley ``Hund's'' coupling $J$ is much smaller than the intravalley-density interaction $V$. In combination with the fact that the noninteracting band structure only has very small valley mixing, the system is invariant under \equref{ApproximateSymmetryGroup} to a good approximation.
In the presence of a finite Hund's coupling, \equref{ApproximateSymmetryGroup} is reduced to
\begin{equation}
    \mathcal{G}_2 = C_3 \times \text{SU}(2)_s \times U(1)_v, \label{ReducedSymmetryGroup}
\end{equation}
where SU(2)$_s$ is global spin rotation. To define these symmetries more precisely, we specify their representation on the electronic field operators:
\begin{subequations}
\begin{align}
    C_3:\, &c^{}_{\vec{k}} \,\longrightarrow \,  c^{}_{C_3\vec{k}} \\
    \text{SU}(2)_s:\, &c^{}_{\vec{k}} \,\longrightarrow \, e^{i\vec{\varphi}\cdot \vec{\sigma}} c^{}_{\vec{k}} \\
    \text{SU}(2)_{\pm}:\, &c^{}_{\vec{k}} \,\longrightarrow \, \left(P_{\pm }e^{i\vec{\varphi}\cdot \vec{\sigma}} + P_{\mp} \right) c^{}_{\vec{k}}, \\
    \text{U}(1)_v:\, &c^{}_{\vec{k}} \,\longrightarrow \, e^{i\varphi \tau_z} c^{}_{\vec{k}},
\end{align}\end{subequations}
with $P_{\pm} = (\tau_0 \pm \tau_z)/2$ being the valley projection operators.
Furthermore, time-reversal is represented by the antiunitary operator $\Theta$ with 
\begin{equation}
    \Theta c^{}_{\vec{k}} \Theta^\dagger = i\sigma_y\tau_x c^{}_{-\vec{k}}. \label{TRTransformation}
\end{equation}
To classify superconductivity, we proceed as usual \cite{SigristUeda} and express $\Delta_{\vec{k}}$ in \equref{GeneralFormOfPairing} in terms of the IRs $n$ (with dimension $d_n$) of the point group as
\begin{equation}
    \Delta_{\vec{k}} = \sum_n \sum_{\mu=1}^{d_n} \eta^n_\mu \chi_\mu^n(\vec{k}), \qquad \eta^n_\mu \in \mathbb{C}, \label{FormalExpansionOfDelta}
\end{equation}
where $\chi_\mu^n(\vec{k})$, $\mu=1,\dots,d_n$, are partner functions transforming under the IR $n$. Within the minimal description of pairing in \equref{GeneralFormOfPairing}, which only involves one band per valley, $\chi_\mu^n(\vec{k}) \in \mathbb{C}^{4\times 4}$ are matrices in spin and valley space.

In our case, the point group has the form $\mathcal{G}_j = C_3 \times U(1)_v \times \mathcal{G}^s_j$ with $\mathcal{G}^s_1 = \text{SU}(2)_+ \times \text{SU}(2)_- \approx \text{SO}(4)$ and $\mathcal{G}^s_2 = \text{SU}(2)_s$. As a consequence, the IRs of $\mathcal{G}_j$ have the form $n=n_{C_3} \times n_{v} \times n_{s}$ where $n_{C_3}$, $n_{v}$, and $n_{s}$ are IRs of $C_3$, $U(1)_v$, and $\mathcal{G}_j^s$, respectively. We can thus rewrite \equref{FormalExpansionOfDelta} more explicitly as
\begin{equation}
    \Delta_{\vec{k}} = \sum_{n_{C_3},n_{v},n_{s}} \sum_{\mu_1=1}^{d_{n_{C_3}}}\sum_{\mu_2=1}^{d_{n_{v}}}\sum_{\mu_3=1}^{d_{n_{s}}} \eta^n_{\mu_1\mu_2\mu_3} \chi_{\mu_1}^{n_{C_3}}(\vec{k})\chi_{\mu_2}^{n_{v}}\chi_{\mu_3}^{n_{s}}. \label{ExplicitExpansion}
\end{equation}
In order to classify superconducting states, we need to consider the different IRs of $C_3$, $U(1)_v$, and $\mathcal{G}_j^s$. 

Let us begin our discussion of IRs with $U(1)_v$. While it has, in general, countably infinite IRs (one-dimensional and with character $e^{i m_v \varphi}$, $m_v\in \mathbb{Z}$), only three are relevant here as all representations with $|m_v|>1$ cannot be realized with only two valleys. First, there is the trivial representation, $m_v=0$, with $\chi^{m_v=0} = a \tau^{}_0 + b \tau^{}_z$ with \textit{a priori} unknown $a,b$. Recalling the extra factor of $\tau^{}_x$ in \equref{GeneralFormOfPairing}, this translates to purely intervalley pairing. Secondly, the pair of complex conjugate representations with $m_v=\pm 1$ has to be considered. Note that due to time-reversal symmetry, the complex representations cannot be discussed separately. Here, the basis functions read as $\chi^{m_v=\pm 1} = \tau^{}_x \pm i \tau^{}_y$; as such, this corresponds to purely intravalley pairing. 

We thus see that $U(1)_v$ prohibits the mixing of inter- and intravalley pairing. As time-reversal (\ref{TRTransformation}) interchanges the valleys along with sending $\vec{k}\rightarrow -\vec{k}$ and we assume zero-momentum Cooper pairs, we will restrict our discussion to intervalley pairing, \textit{i.e.}, $m_v=0$ for the rest of the paper.

As is well known \cite{dresselhaus2008group}, $C_3$ has the following IRs, both of which are one-dimensional: the trivial one, $A$, and the complex representation $E$ (and its complex conjugate partner). We analyze each of these IRs  in Secs.~\ref{TrivialRepresentation} and \ref{ComplexRepresentation}, and in both cases, discuss the differences between $\mathcal{G}^s_1$ and $\mathcal{G}^s_2$; we will also see how the states ``connect'' once $\mathcal{G}^s_1$ is weakly broken to $\mathcal{G}^s_2$ due to a small but finite value of the Hund's coupling.

\section{Trivial representation of the crystalline point group}
\label{TrivialRepresentation}
For simplicity, we begin with the trivial representation $A$ of $C_3$, which is real and one-dimensional. In fact, the following discussion will not be modified as long as the IR is real and one-dimensional and there is no crystalline symmetry relating the two valleys. Interestingly, the last assumption is violated in twisted bilayer graphene and trilayer graphene on boron nitride; see \secref{ComparisonWithOtherSystems} for a detailed discussion of the associated modifications. 

As already mentioned, we consider only intervalley pairing which corresponds to a real and one-dimensional IR as well. This means that the order parameter in \equref{ExplicitExpansion} has the form
\begin{equation}
    \left(\Delta_{\vec{k}}\right)_{\sigma v,\sigma' v'} = \delta^{}_{v,v'}\chi^{A}(\vec{k},v)\sum_{\mu=1}^{d_{n_{s}}} \eta^{n_s}_{\mu}   \left(\chi_{\mu}^{n_{s}}(v)\right)_{\sigma\sigma'},  \label{MostImpCase}
\end{equation}
where $\chi^{A}(\vec{k},v)$ is invariant under $\vec{k}\rightarrow g\vec{k}$ for all generators $g$ of the crystalline point group (here, we only have $g=C_3$).

\subsection{Limit of exact SU(2)$_+\times\,  $SU(2)$_-$ symmetry}
\label{ExactSU2SU2Symmetry}
To proceed further, we have to inspect the scenarios for both $\mathcal{G}_1^s$ and $\mathcal{G}_2^s$. We start with the former, \textit{i.e.}, we assume that the Hund's coupling is zero. Inserting \equref{MostImpCase} in the general pairing Hamiltonian (\ref{GeneralFormOfPairing}), we obtain a pairing term of the form
\begin{alignat}{1}
    \mathcal{H}^{}_{\text{SC}} &= \sum_{\vec{k},v}  c^\dagger_{\vec{k}\sigma v} \left( M^{}_{\vec{k}v} i\sigma^{}_y \right)_{\sigma,\sigma'} c^\dagger_{-\vec{k}\sigma' \bar{v}} + \text{H.c.}, \label{SimplifiedFormOfHighSymPairing} \,\\
    \nonumber M^{}_{\vec{k}v} &=\chi^{A}(\vec{k},v) \Delta^{}_{v}, 
\end{alignat}
where $\bar{v}=\mp$ for $v=\pm$, and $M_{\vec{k}v}$ as well as $\Delta_{v} = \sum_\mu \eta_\mu^{n_s} \chi_\mu^{n_s}(v)$ are matrices in spin space. Fermi-Dirac statistics implies
\begin{equation}
    M^{}_{\vec{k}v} = \sigma^{}_y M_{-\vec{k}\bar{v}}^T\,\sigma^{}_y. \label{FermiDiracStatConstr}
\end{equation}
Rewriting pairing in terms of singlet and triplet as $M^{}_{\vec{k}v} = \sigma^{}_0\Delta^s_{\vec{k}v} + \vec{\sigma}\cdot \vec{d}_{\vec{k}v}$, \equref{FermiDiracStatConstr} is equivalent to $\Delta^s_{\vec{k}v} = \Delta^s_{-\vec{k}\bar{v}}$ and $\vec{d}^{}_{\vec{k}v} = -\vec{d}^{}_{-\vec{k}\bar{v}}$, as expected.

We now study the stable superconducting phases in this channel by writing down the most general Ginzburg-Landau expansion constrained by the symmetries
\begin{subequations}\begin{align}
    \Theta: \, &M_{\vec{k}v} \, \longrightarrow \,  M^\dagger_{\vec{k}v}, \\
    \text{SU}(2)_+ \times \text{SU}(2)_-: \, &M_{\vec{k}v} \, \longrightarrow \,  e^{-i\vec{\varphi}_+\cdot \vec{\sigma}}M_{\vec{k}v} e^{i\vec{\varphi}_-\cdot \vec{\sigma}}. \label{SU2SU2Rotations}
\end{align}\label{AllSymmetriesConstr}\end{subequations}
Due to the constraint (\ref{FermiDiracStatConstr}) stemming from Fermi-Dirac statistics, we express the free energy in terms of one valley only (say $v=+$) as $\mathcal{F} = \mathcal{F}[M_{\vec{k}+}=\chi^{A}(\vec{k},+) \Delta_{+}]$, and the pairing in the other valley just follows from \equref{FermiDiracStatConstr}. 
The most general free energy to quartic order in $\Delta_{+}$, invariant under \equref{AllSymmetriesConstr} and $\Delta_{v}\rightarrow e^{i\varphi}\Delta_{v}$, reads as 
\begin{alignat}{1}
\label{ExpansionOfFreeEnergy}
    \mathcal{F} &\sim \,\frac{a(T)}{2} \text{tr}\left[ \Delta_{+}^\dagger \Delta_{+}^\pdagger  \right] + \frac{b_1}{4} \left(\text{tr}\left[ \Delta_{+}^\dagger \Delta_{+}^\pdagger  \right]\right)^2 \\ 
    \nonumber & + \frac{b_2}{2}\,\text{tr}\left[ \Delta_{+}^\dagger \Delta_{+}^\pdagger \Delta_{+}^\dagger \Delta_{+}^\pdagger  \right] +\frac{b_3}{4} \left| \text{tr}\left[ \sigma_y \Delta_{+}^\pdagger \sigma^{}_y \Delta_{+}^T  \right] \right|^2.
\end{alignat}
Note that $| \text{tr}[ \sigma_y \Delta_{+}^\pdagger \sigma_y \Delta_{+}^T]|^2/2= \text{tr}[ \Delta_{+}^\pdagger \sigma_y \Delta_{+}^T\Delta_{+}^* \sigma_y \Delta_+^\dagger]$, so the latter is not an independent term to consider. It further holds that $|\text{tr}[ \sigma_y \Delta_{+}^\pdagger \sigma_y \Delta_{+}^T  ] |^2/2 = (\text{tr}[ \Delta_{+}^\dagger \Delta_{+}^\pdagger ])^2 - \text{tr}[ \Delta_{+}^\dagger \Delta_{+}^\pdagger \Delta_{+}^\dagger \Delta_{+}^\pdagger]$, which allows us to set $b_3 = 0$ in the following without loss of generality.

Using the singular-value decomposition of $\Delta_{+}$, it is straightforward to find all symmetry-inequivalent minima of \equref{ExpansionOfFreeEnergy}. There are two different states depending on the sign of $b_2$ which we label by $A_{m_v=0}(\Delta^s;\vec{d})$, where $\Delta^s$ and $\vec{d}$ refer to the singlet and the triplet vector, respectively, $A$ indicates the trivial IR of $C_3$, and $m_v=0$ signifies intervalley pairing (IR of U(1)$_v$ with $m_v=0$). If $b_2>0$, we get $\Delta_+ \propto \sigma^{}_0$, \textit{i.e.}, $M_{\vec{k},\pm} = \lambda_{\pm \vec{k}} \sigma^{}_0$ with $\lambda^{}_{C_3 \vec{k}} = \lambda^{}_{\vec{k}}$; according to the notation introduced above, this state will be labeled as $A_{m_v=0}(1;0,0,0)$. There are (infinitely) many different equivalent representations of this state since, for instance, the transformations in \equref{SU2SU2Rotations} mix the singlet and triplet components---as described by the isomorphism $\text{SU}(2)_+ \times \text{SU}(2)_- \approx \text{SO}(4)$. However, for the sake of notational clarity, we will henceforth only show one convenient representative of each state. The $A_{m_v=0}(1;0,0,0)$ state preserves time-reversal symmetry and breaks SU(2)$_+$ $\times$ SU(2)$_-$ down to SU(2)$_s$ [rotations of the total spin, \textit{i.e.}, $\vec{\varphi}_+=\vec{\varphi}_-$ in \equref{SU2SU2Rotations}].

On the other hand, if $b_2<0$, we find $\Delta_+ \propto \sigma^{}_0 + \sigma^{}_z$, which corresponds to $A_{m_v=0}(1;1,0,0)$. For this phase, the order parameter in \equref{SimplifiedFormOfHighSymPairing} assumes the form $M_{\vec{k},\pm} = \lambda_{\pm \vec{k}} \left( \sigma^{}_0 \pm \sigma^{}_z \right)$ with $\lambda_{C_3 \vec{k}} = \lambda_{\vec{k}}$. This state preserves time-reversal symmetry too, but it breaks SU(2)$_+$ $\times$ SU(2)$_-$ down to O(2)$_s$ (with $\vec{\varphi}_+=\vec{\varphi}_- = \varphi \,\hat{\vec{e}}_z$), \textit{i.e.}, rotations of the total spin along a single axis.

\subsection{Turning on the Hund's coupling}\label{NonZeroHundsCoupl}
In reality, there is, of course, a finite Hund's coupling that reduces $\mathcal{G}^s_1 =$ SU(2)$_+$ $\times$ SU(2)$_-$ to only global spin rotations, $\mathcal{G}^s_2 =$ SU(2)$_s$, already in the high-temperature phase. In \refcite{2019arXiv190308685L}, the Hund's coupling $J$ has been estimated to be about 60 times smaller than the intravalley interaction $V$. Note, however, $J$ might be enhanced due to loop corrections. For this reason, we first classify the possible instabilities in the absence of an approximate SU(2)$_+$ $\times$ SU(2)$_-$ symmetry and then, analyze how the different states ``connect'' for small values of $J$ and whether admixtures of singlet and triplet are possible.

To introduce our notation, we will begin with the classification for the reduced symmetry group $\mathcal{G}_2$ in \equref{ReducedSymmetryGroup}; in that case, we have either singlet or triplet pairing:

    \paragraph{Singlet:} This corresponds to the $d_{n_s}$\,$=$\,$1$ one-dimensional IR of $\mathcal{G}^s_2$ with $\chi^{n_s}$\,$=$\,$\sigma_0$ in \equref{MostImpCase}. The pairing Hamiltonian simply has the form
    \begin{equation}
        \mathcal{H}^{}_{\text{SC}} = \sum_{\vec{k},v} \lambda^s_{\vec{k}v} c^\dagger_{\vec{k}\sigma v} \left( i\sigma^{}_y \right)_{\sigma,\sigma'} c^\dagger_{-\vec{k}\sigma' \bar{v}} + \text{H.c.},
    \end{equation}
    with $\lambda^s_{\vec{k}v}$\,$=$\, $\lambda^s_{-\vec{k}\bar{v}}$ and $\lambda^s_{C_3 \vec{k}v}$\,$=$\,$ \lambda^s_{\vec{k}v}$.
    All symmetries of the high-temperature phase are preserved. We refer to this state as $A^{1_s}_{m_v=0}$ with the $1_s$ referring to spin singlet, $m_v=0$ to intervalley pairing, and $A$ to the trivial representation of $C_3$.
    \paragraph{Triplet:} This pairing channel is associated with the three-dimensional IR of $\mathcal{G}^s_2$. A possible choice of basis functions is $\chi^{n_s}_\mu(v) = \sigma_\mu$, $\mu=1,2,3$, in \equref{MostImpCase}. As it is a multidimensional representation, the free energy has to be expanded beyond quadratic order. Writing $\vec{d} = (\eta_1^{n_s},\eta_2^{n_s},\eta_3^{n_s})$, we have up to quartic order
    \begin{equation}
        \mathcal{F} \sim a(T) \, \vec{d}^\dagger \vec{d} + b^t_1 \left(\vec{d}^\dagger \vec{d}\right)^2 + b^t_2 \lvert \vec{d}^* \times \vec{d} \rvert^2. \label{GLExpTriplet}
    \end{equation}
    Observe that $\lvert \vec{d}^T \vec{d} \rvert^2$ is not an independent quartic term since $\lvert \vec{d}^T \vec{d} \rvert^2= (\vec{d}^\dagger \vec{d})^2 - \lvert \vec{d}^* \times \vec{d}\rvert^2$.
   The free energy in \equref{GLExpTriplet} has two stable minima. For $b^t_2>0$, we have $\vec{d} \propto (1,0,0)^T$ and the corresponding pairing term is
    \begin{equation}
        \mathcal{H}^{}_{\text{SC}} = \sum_{\vec{k},v} \lambda^t_{\vec{k}v} c^\dagger_{\vec{k}\sigma v} \left( \sigma^{}_x i\sigma^{}_y \right)_{\sigma,\sigma'} c^\dagger_{-\vec{k}\sigma' \bar{v}} + \text{H.c.},
    \end{equation}
    with $\lambda^t_{\vec{k}v} = -\lambda^t_{-\vec{k}\bar{v}}$ and $\lambda^t_{C_3 \vec{k}v} = \lambda^t_{\vec{k}v}$. As is easily seen, this term preserves time-reversal symmetry and breaks SU(2)$_s$ down to spin rotation along a single axis. This state will be referred to as unitary triplet and denoted by the symbol $A^{3_s}_{m_v=0}(1,0,0)$, where the three components just indicate the direction of the triplet vector.
    If, instead, $b^t_2< 0$, we obtain $\vec{d}\propto(1,i,0)^T$, whence
    \begin{equation}
        \mathcal{H}^{}_{\text{SC}} = \sum_{\vec{k},v} \lambda^t_{\vec{k}v} c^\dagger_{\vec{k}\sigma v} \left( ( \sigma_x + i \sigma_y) i\sigma_y \right)_{\sigma,\sigma'} c^\dagger_{-\vec{k}\sigma' \bar{v}} + \text{H.c.},
    \end{equation}
    with $\lambda^t_{\vec{k}v}$ as above. This is a nonunitary triplet state. It breaks time-reversal symmetry and will be denoted by $A^{3_s}_{m_v=0}(1,i,0)$.

One might wonder what kind of interaction or band structure would favor $A^{3_s}_{m_v=0}(1,i,0)$ over $A^{3_s}_{m_v=0}(1,0,0)$ and vice versa. In mean-field theory, as detailed in \appref{MicrosCopGLExpansion}, it is straightforward to show by evaluation of a one-loop diagram that
\begin{equation}
    b^t_2 = T \sum_{\omega_n} \int \frac{\diff^2 \vec{k}}{(2\pi)^2} \frac{|\lambda_{\vec{k}}^t|^4}{(\omega_n^2 +\xi^2_{\vec{k}+})^2}. \label{QuarticTermMeanField}
\end{equation}
Here, $\omega_n$ are fermionic Matsubara frequencies and $\xi_{\vec{k}+}$ is the electronic band energy in valley $v=+$ of the nearly flat band hosting superconductivity. 
We observe that $b^t_2>0$ holds irrespective of microscopic details and hence, $A^{3_s}_{m_v=0}(1,0,0)$ is generically favored if we neglect corrections beyond mean-field theory (such as residual interactions or frequency dependence of pairing). Intriguingly, there have been experimental reports \cite{PhysRevB.82.174511} of intrinsically nonunitary pairing in LaNiC$_2$, \textit{i.e.}, nonunitary triplet pairing born out of a paramagnetic normal state. Thus, there is reason to believe that we cannot generically exclude this state, but we do not expect it to show up in any simple mean-field computation. 

\subsubsection{How do the states connect in the $J=0$ limit?}

Next, we establish how the three possible states, $A^{1_s}_{m_v=0}$, $A^{3_s}_{m_v=0}(1,0,0)$, and $A^{3_s}_{m_v=0}(1,i,0)$, connect to the two derived in the previous subsection with enhanced SU(2)$_+$ $\times$ SU(2)$_-$ symmetry, namely $A_{m_v=0}(1;0,0,0)$ and $A_{m_v=0}(1;1,0,0)$.
To this end, we decompose the Ginzburg-Landau expansion (\ref{ExpansionOfFreeEnergy}) into singlet and triplet by writing $\Delta_{+} = \Delta^s + \vec{\sigma}\cdot\vec{d}$. Since $\text{tr}\left[ \Delta_{+}^\dagger \Delta_{+}^\pdagger  \right] = |\Delta^s|^2 + \vec{d}^\dagger \vec{d}$, singlet and triplet are degenerate at quadratic order in $\mathcal{F}$ as a consequence of the enhanced SU(2)$_+$ $\times$ SU(2)$_-$ symmetry. For nonzero $J$, this degeneracy is lifted and we have
\begin{equation}
    \mathcal{F} \sim a(T) \left(|\Delta^s|^2 + \vec{d}^\dagger \vec{d}\right)  + \delta a(T) \left(|\Delta^s|^2 - \vec{d}^\dagger \vec{d}\right), \label{GeneralQuadraticGLExp}
\end{equation}
where $\delta a$ can be made arbitrarily small as $J\rightarrow 0$. Neglecting, for now, the ``back action'' of the superconducting order parameter that condenses first on the second one (as described by higher-order terms in the Ginzburg-Landau expansion), we conclude that there are two superconducting transitions at $T_{c,0}^\pm$\,$=$\, $T^{}_{c,0} \pm \Delta T^{}_{c}$ with $\Delta T_{c} = |\delta a\,(T_{c,0})|/\alpha$, taking $a(T) \sim \alpha(T-T_{c,0})$ near $T_{c,0}$. The extra index $0$ in $T_{c,0}^\pm$ highlights the fact that the aforementioned higher-order terms in the Ginzburg-Landau expansion can significantly affect the lower transition temperature, $T_c^- \neq T_{c,0}^-$; of course, this has no effect on the higher transition temperature, $T_c^+=T_{c,0}^+$.

Before analyzing these corrections, it is useful to estimate the temperature scale $\Delta T_c$. Using the expected result, $T_c^\pm \approx \Lambda \exp(-1/[(V\pm J)\nu])$ of mean-field theory (from the linearized gap equations)---where $\Lambda$ is the cutoff and $\nu$ the density of states at the Fermi level---leads to
\begin{equation}
    \frac{\Delta T_c}{T_{c,0}} \sim \frac{|J|}{V^2\nu^{}}. \label{EstimateOfTc}
\end{equation}
The large density of states, taken together with the estimated value of $V$---which is larger than even the bandwidth \cite{2019arXiv190308685L} of the flat bands---and the relation $J\ll V$, implies that $\Delta T^{}_c \ll T_{c,0}$ \footnote{Taking $\nu \approx 1/W$ with bandwidth $W \approx 10\,\textrm{meV}$ and $V\approx 35\,\textrm{meV}$, $J\approx 0.6\,\textrm{meV}$ \cite{2019arXiv190308685L}, we estimate $\Delta T_c/T_{c,0} \approx 0.5\%$.}; the temperature/energy scale $\Delta T_c$ is most likely too small to be visible in experiments. While the estimate above is only based on mean-field theory, it indicates at least that it is important to study the behavior of superconductivity in the limit of small $\Delta T^{}_c/T_{c,0}$ (and hence, weakly broken SU(2)$_+$ $\times$ SU(2)$_-$ symmetry), accounting for the possibility of two transitions and mixing of singlet and triplet pairing (despite the absence of spin-orbit coupling). Moreover, we will see that nearly degenerate singlet and triplet pairing also has crucial consequences for the behavior of superconductivity in the presence of a magnetic field.

While we postpone the analysis of magnetic fields to \secref{MagneticField}, here, we investigate the possibility of an admixture of singlet and triplet in the presence of time-reversal symmetry [relevant to scenario (I) defined in the introduction]. As anticipated above, this requires also considering the quartic terms of \equref{ExpansionOfFreeEnergy}. We find
\begin{alignat}{1}
    \nonumber\mathcal{F} &\sim a(T) \left(|\Delta^s|^2 + \vec{d}^\dagger \vec{d}\right)  + \delta a \left(|\Delta^s|^2 - \vec{d}^\dagger \vec{d}\right)  \\
    \nonumber & + (b_1+b_2) |\Delta^s|^4 + (b_1+b_2) \left(\vec{d}^\dagger \vec{d}\right)^2 + b_2 \left|\vec{d}^*\times \vec{d}\right|^2 \\ 
    &+ 2(b_1+2b_2) |\Delta^s|^2 \vec{d}^\dagger \vec{d} + 2b_2\text{Re}\left[ \left(\Delta^s\right)^2 \vec{d}^\dagger \vec{d}^*\right], \label{FullExpansionWithHigherSym}
\end{alignat}
neglecting corrections to the quartic terms coming from finite $J$.

\begin{figure}
   \centering
    \includegraphics[width=\linewidth]{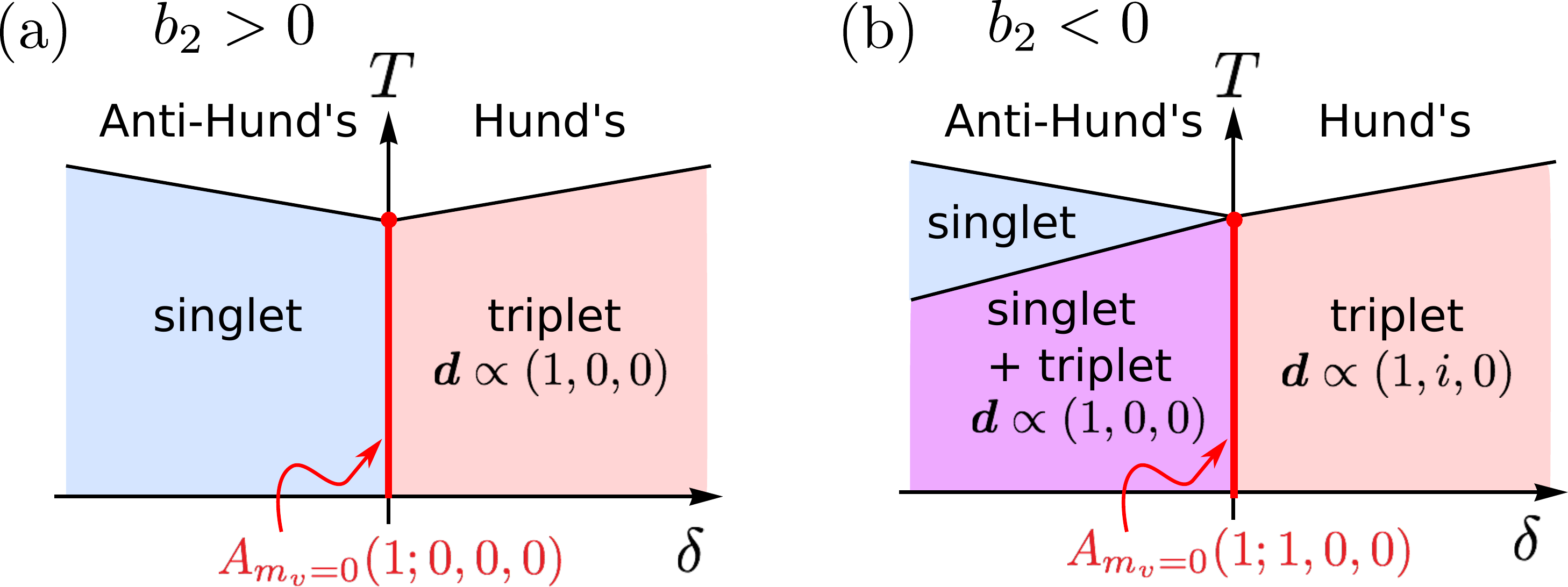}
    \caption{Schematic phase diagram as a function of temperature, $T$, and $\delta = \frac{\delta a}{\alpha T_{c,0}} \approx \frac{J}{V^2\nu} $ close to the SU(2)$_+$ $\times$ SU(2)$_-$ invariant point ($\delta =0$) obtained by minimizing the free energy in \equref{FullExpansionWithHigherSym}. Parts (a) and (b) correspond to $b_2>0$ and $b_2<0$, respectively, and are, hence, associated with the pairing states $A_{m_v=0}(1;0,0,0)$ and $A_{m_v=0}(1;1,0,0)$ at $\delta = 0$ as indicated in red. In our notation for the pairing $A_{m_v}(\Delta^s;\vec{d})$, $m_v$ is the valley quantum number, and $\Delta^s$ and $\vec{d}$ are singlet and triplet pairing amplitudes, respectively. As the pure singlet and the unitary triplet state for $b_2 > 0$ (the mixed singlet-triplet phase and the nonunitary triplet for $b_2 < 0$) transform into each other under reversing the sign of $\delta \propto J$, we will refer to them as Hund's partners.}
    \label{SchematicPD}
\end{figure}

Looking at the first transition with the higher transition temperature, we assess which of the two distinct triplet states, $A^{3_s}_{m_v=0}(1,0,0)$ and $A^{3_s}_{m_v=0}(1,i,0)$, and the singlet state  can be stabilized by starting from $A_{m_v=0}(1;0,0,0)$ or $A_{m_v=0}(1;1,0,0)$ and turning on a finite Hund's coupling $J$. For this purpose, we can neglect the coupling terms in the third line of \equref{FullExpansionWithHigherSym}. Clearly, if $\delta a<0$ (``anti-Hund's coupling''), we get a singlet state for both $A_{m_v=0}(1;0,0,0)$ and $A_{m_v=0}(1;1,0,0)$. A straightforward way of establishing which of the triplet states is realized when $\delta a > 0$ (``conventional'' Hund's coupling) proceeds by evaluating their respective free energy in \equref{FullExpansionWithHigherSym}. One finds that the state $A^{3_s}_{m_v=0}(1,0,0)$ is realized if $b_2 > 0$; otherwise, $A^{3_s}_{m_v=0}(1,i,0)$ is favored. This brings us to the conclusion that
\begin{subequations}\begin{align}
    A_{m_v=0}(1;0,0,0)  &\longrightarrow  A^{1_s}_{m_v=0} \text{ or }A^{3_s}_{m_v=0}(1,0,0), \label{PairingStatea} \\
    A_{m_v=0}(1;1,0,0) &\longrightarrow  A^{1_s}_{m_v=0} \text{ or }A^{3_s}_{m_v=0}(1,i,0), \label{PairingStateb}
\end{align}\end{subequations}
at the first transition (see the schematic phase diagram in \figref{SchematicPD}). This result is just a consequence of the fact that the form $\Delta_+ \propto \sigma_0$ for the $A_{m_v=0}(1;0,0,0)$ state we had chosen in the previous section can alternatively be written as $\Delta_+ \propto \sigma_x$ due to the SU(2)$_+$ $\times$ SU(2)$_-$ symmetry and thus, explicitly assumes the form of the unitary triplet state. Similarly, $\Delta_+ \propto \sigma_0 + \sigma_z$ used above for $A_{m_v=0}(1;1,0,0)$ can also be written as $\Delta_+ \propto \sigma_x + i \sigma_y$. This is why it transitions into the nonunitary triplet state, upon turning on a nonzero Hund's coupling.     

In order to determine whether there is a second transition, we have to include the coupling terms between singlet and triplet in the third line of \equref{FullExpansionWithHigherSym}. To illustrate that these terms can be crucial, we consider the case $\delta a>0$ and $b_2>0$, \textit{i.e.}, the triplet state $A^{3_s}_{m_v=0}(1,0,0)$ condenses first. This leads to the coupling between singlet and triplet $2c|\Delta^s|^2|\vec{d}(T)|^2$, $c=b_1+b_2$, in the free energy, where we have made use of the fact that a relative phase of $\pi/2$ between singlet and triplet is energetically most favorable. As a result of $|\vec{d}(T)|^2 = (\delta a-a(T))/(2 c)$, which is valid as long as there is no additional singlet pairing, the growing triplet component induces the extra term
\begin{equation}
    2c|\Delta^s|^2|\vec{d}(T)|^2 = (\delta a -a(T))|\Delta^s|^2,
\end{equation}
which is always larger than the ``bare'' quadratic term of singlet pairing [in the first line of \equref{FullExpansionWithHigherSym}]. Accordingly, there is no second transition (at least close to $T_{c,0}$ where our Ginzburg-Landau approach is valid) into a state that has a nonzero singlet component. We also checked that \equref{FullExpansionWithHigherSym} does not allow for a first-order transition. 

Similarly, all other cases can be scrutinized and one finds that if triplet dominates, there is no second transition. However, if singlet has a larger transition temperature ($\delta a < 0$), there is a second transition into a phase with singlet and triplet pairing when $b_2<0$. This transition happens at the temperature
\begin{equation}
    T_c^- = T_{c,0}\left(1 +  \, \frac{c- |b_2|}{|b_2|}\delta \right), \quad \delta \equiv \frac{\delta a}{\alpha T_{c,0}} \approx \frac{J}{V^2\nu}.
\end{equation}
The stability of the Ginzburg-Landau expansion only requires $c$\,$>$\,$0$ and $c$\,$>$\,$-b_2$, so both $T_c^-$\,$<$\,$T_{c,0}^-$ and $T_c^-$\,$>$\,$T_{c,0}^-$ are possible. More importantly, unless $|b_2|/c$ is fine-tuned to be of order $\delta$, generically, $T_c^- \rightarrow T_{c,0}$ as $J \rightarrow 0$ and the two transitions, if present, are likely too close to be experimentally discernible. Due to the term $2b_2\,\text{Re}[ \left(\Delta^s\right)^2 \vec{d}^\dagger \vec{d}^*]$ in the free energy, we obtain the unitary triplet vector $\vec{d} = d_0 (1,0,0)^T$ with $\Delta^s d_0^* \in \mathbb{R}$ (same phase). This is to be expected as $\Delta_+ \propto \sigma_0 + \sigma_z$ for the ``parent'' state $A_{m_v=0}(1;1,0,0)$. 

A summary of these results is provided by the schematic phase diagrams in \figref{SchematicPD}. We observe that the proximity to the enlarged symmetry in spin space, SU(2)$_+$ $\times$ SU(2)$_-$, favors the possibility of having a nonzero triplet component: for $b_2<0$, even a negative Hund's coupling (anti-Hund's) allows for $\vec{d} \neq 0$ and leads to the exotic possibility of significant ($d_0 \approx \Delta^s$ for $T_{c}^--T > \Delta T_c$) singlet-triplet mixing in spite of the absence of spin-orbit coupling. 

\newcommand{\exspace}{\hspace{0.9em}}
\begin{table*}[htb]
\begin{center}
\caption{Summary of the different intervalley pairing states transforming under the trivial representation of the point group $C_3$ in the absence of a magnetic field. For notational convenience, we neglect the extra label $m_v=0$ to indicate intervalley pairing. $\lambda_{\vec{k}}$ is a real-valued and Brillouin-zone-periodic function that is invariant under $C_3$. To lowest order, we can take $\lambda_{\vec{k}}$ to be independent of $\vec{k}$. We also indicate the minimal number of nodes, which state it transforms to when setting $J=0$ [``SO(4) parent''] and reversing the sign of $J$ (``Hund's partner''), and whether the state can be found in a single-band mean-field (MF) computation neglecting residual interactions and/or when the ferromagnetic (FM) fluctuation corrections discussed in \secref{FluctuationInducedSC} dominate. In the last line, $\eta$ describes the temperature-dependent strength of admixing of the unitary triplet state.}
\label{TableSummaryOfPairingC3}
\begin{ruledtabular}
 \begin{tabular} {cccccc} 
   \exspace  Pairing \exspace    & \exspace $M_{\vec{k}+}$ \exspace   & \exspace Nodes \exspace & \exspace SO(4) parent \exspace & \exspace Hund's partner \exspace & \exspace MF/FM \exspace  \\ \hline
$A^{1_s}$ & $\lambda^{}_{\vec{k}}\sigma^{}_0$   & none  &  $A(1;0,0,0)$      & $A^{3_s}(1,0,0)$ & \cmark/\cmark  \\
$A^{3_s}(1,0,0)$ & $\lambda^{}_{\vec{k}}\sigma^{}_x$   & none  &  $A(1;0,0,0)$      & $A^{1_s}$ & \cmark/\xmark  \\
$A^{3_s}(1,i,0)$ & $\lambda^{}_{\vec{k}}(\sigma^{}_x+i\sigma^{}_y)$   & $\downarrow$ gapless/none  &  $A(1;1,0,0)$      & $A^{1_s}+A^{3_s}(1,0,0)$ & \xmark/\cmark  \\
$A^{1_s}+A^{3_s}(1,0,0)$ & $\lambda^{}_{\vec{k}}(\sigma^{}_0 + \eta\, \sigma^{}_x )$   & none  &  $A(1;1,0,0)$      & $A^{3_s}(1,i,0)$ & \xmark/\cmark  \\
 \end{tabular}
 \end{ruledtabular}
\end{center}
\end{table*}

It is noteworthy that all the states are fully gapped (more precisely, they have no symmetry-enforced nodes) except for the nonunitary triplet $A^{3_s}_{m_v=0}(1,i,0)$, which is gapped for one spin species while the other is completely gapless. The admixture of singlet and unitary triplet has two unequal gaps for the two spin species both of which are finite as long as the magnitudes of singlet and triplet are not fine-tuned to be equal. All the states, along with their order parameters and properties, are summarized in \tableref{TableSummaryOfPairingC3}.

We finally comment on the nature of the thermal phase transition for the different superconducting states once fluctuations of the order parameter are taken into account. Neglecting stray fields, the transition into the singlet phase $A^{1_s}$ is expected to be a BKT transition with quasi-long-range order of the complex-valued order parameter $\Delta^s$ below the transition temperature. For the triplet states, it is important to keep in mind that $\vec{d}$ cannot even have quasi-long-range order as it transforms as a three-component vector under spin-rotation. For the unitary triplet state [with order parameter manifold $(S_2 \times S_1)/\mathbb{Z}_2$] a BKT transition of the composite charge-$4e$ order parameter $\vec{d}^T\vec{d}$ is possible and is associated with the (un)binding of half vortices. This is different for the nonunitary state [with order parameter manifold $S_3/\mathbb{Z}_2\approx \text{SO}(3)$] where $\vec{d}^T\vec{d}=0$ and no BKT transition into a quasi-long-range-ordered superconductor is expected. For the case of the two consecutive transitions in \figref{SchematicPD}(b) with $\delta < 0$, we first expect a BKT transition into a singlet phase followed by a crossover at which the triplet vector becomes nonzero.

However, we point out that, even in the simplest case of the singlet $A^{1_s}$, there are significant corrections to the BKT transition resulting from stray fields and mirror vortices \cite{2007PhRvB..75f4514K}, which make the observation of a pristine BKT transition in a (charged) superconductor difficult. We believe that the current status of experiments does not allow one to exclude pairing phases that will not exhibit quasi-long-range order and a BKT transition in the limit of infinite system size.

\subsubsection{Expectations within mean-field theory}\label{ExpectationsMeanField}
Lastly, we evaluate what a na\"{\i}ve mean-field computation is expected to yield. In fact, from \equref{QuarticTermMeanField}, we already know that the prefactor of the term $\left|\vec{d}^*\times \vec{d}\right|^2$ in \equref{FullExpansionWithHigherSym} must be positive within mean-field theory and therefore, it holds that $b_2>0$. For completeness, we mention that in the mean-field approximation, $b_1 = 0$, as shown in \appref{MicrosCopGLExpansion}.
Consequently, a single-band mean-field computation will generally favor \figref{SchematicPD}(a) over (b); in other words, only half of the phases proposed in this section can be found in mean-field, which we also indicate in the last column of \tableref{TableSummaryOfPairingC3}. 

However, there is no fundamental mechanism prohibiting the mixing of singlet and triplet via two transitions (see, e.g., \refcite{PhysRevB.72.144522}) and there are multiple reasons why we can effectively have $b_2 < 0$ (and $b_1 > 0$ to ensure stability): for instance, strong residual interactions and fluctuations have been shown to modify the values of the quartic terms in the free energy significantly \cite{RafaelsNematicFlucs,LiangFu}, thereby stabilizing phases that are otherwise not possible in the mean-field approximation. Given the small bandwidth and the underlying strong-coupling features of the problem \cite{PasupathySTM,NadjPergeSTM,AndreiSTM,YazdaniSTM,STMReview}, it is plausible that there are sizable corrections to mean-field theory. In addition, we recognize that there are other corrections arising from interband pairing, and that disorder can also dress the Ginzburg-Landau expansion. Moreover, it is unclear whether adding frequency dependence to the gap function could be of relevance. 

In \secref{FluctuationInducedSC}, we will analyze the impact of ferromagnetic fluctuations, which are expected to be relevant for graphene moir\'e systems \cite{2019arXiv190306952S,ExperimentKim,PabllosExperiment,Sharpe605,2019arXiv190306513L,IlaniExperiment,FMTrilayer}, and find that these generically decrease the value of $b_2$; if sufficiently strong, these fluctuations will favor the phase diagram in \figref{SchematicPD}(b).

\subsection{In the presence of a magnetic field}\label{MagneticField}
We now generalize the Ginzburg-Landau expansion to also include the coupling to a Zeeman field $\vec{M}_Z=(M_Z^x,M_Z^y,M_Z^z)$ and an (in-plane) orbital coupling $\vec{M}_O = (M_O^x,M_O^y)$. Both of these terms can either be due to an applied external magnetic field or due to the correlated insulating state. This enables us to discuss ({\it i\/}) the behavior of the superconducting critical temperature $T_c^+$ as a function of an external magnetic field in the absence of any ferromagnetic moments associated with the correlated insulating state [case (I) defined in the introduction]. At the same time, we can study ({\it ii\/}) how the transition temperature and the order parameter of superconductivity is affected by the potentially coexisting ferromagnetic order [case (II)].

\subsubsection{Leading superconducting transition}
We first turn our attention to the leading superconducting transition with the highest temperature $T_c^+$; potential subsequent superconducting transitions at lower temperatures are addressed later in \secref{QuarticTermsInMagneticField}.
For the goal of studying the first transition, we can restrict ourselves to quadratic order in the order parameter. Only keeping terms up to quadratic order in the magnetic field as well, we obtain
\begin{alignat}{1}
    \nonumber\mathcal{F}_M &\sim a(T) \left(|\Delta^s|^2 + \vec{d}^\dagger \vec{d}\right)  + \delta a \left(|\Delta^s|^2 - \vec{d}^\dagger \vec{d}\right) \\ 
    \nonumber& + 2 \delta c^{}_1 \vec{M}_Z \cdot \text{Im}\left(\vec{d}^* \Delta^s\right) + i c^{}_2 \vec{M}_Z \cdot \vec{d}^* \times \vec{d} \\
    \nonumber& + (c^{}_3 \vec{M}_Z^2 +  c^{}_4 \vec{M}_O^2) \left(|\Delta^s|^2 + \vec{d}^\dagger \vec{d}\right) \\ 
     &+ (\delta c^{}_5 \vec{M}_Z^2 + \delta c^{}_6 \vec{M}_O^2)  \left(|\Delta^s|^2 - \vec{d}^\dagger \vec{d}\right). \label{GinzburgLandauExpansion} 
\end{alignat}
While the prefactors $\delta a$, $\delta c_1$, $\delta c_5$, and $\delta c_6$ are necessarily zero in the limit $J\rightarrow 0$, where the SU(2)$_+ \times$ SU(2)$_-$ symmetry becomes exact, all remaining terms can be nonzero (and different in their values) at $J=0$. Notice that the third term has not been considered in \refcite{2019arXiv190308685L}; this term arises only when both singlet and triplet are allowed for and leads to the admixture of a unitary triplet state with a singlet superconductor. The vanishing of $\delta a$ and $\delta c_6$ at $J=0$ is an obvious consequence of the enhanced SU(2)$_+ \times$ SU(2)$_-$ symmetry. To see that $\delta c_1$ also has to vanish as $J\rightarrow 0$, let us take $\vec{M}_Z$ along the $z$ direction; this breaks SU(2)$_+ \times$ SU(2)$_-$ down to $O(2)_+\times O(2)_-$, \textit{i.e.}, the system is only invariant under $c_{\vec{k}v} \rightarrow e^{i\varphi_v \sigma_z}c_{\vec{k} v}$. Performing this transformation with $\varphi_+=0$ and $\varphi_- = \pi/2$, we get $(\Delta^s,d_z) \rightarrow (id_z,i\Delta^s)$ and hence, $\delta c_1 \rightarrow -\delta c_1$. With the same argument, it can be proven that $\delta c_5$ has to go to zero as $J\rightarrow 0$. In \appref{MicrosCopGLExpansion}, we show that $\delta c_1 = 0$ in mean-field theory within the single-band description, even when SU(2)$_+ \times$ SU(2)$_-$ is broken; this results from an emergent valley-exchange symmetry within the single-band mean-field approximation.

In discussing the highest critical temperature and the corresponding order parameter for $\vec{M}_Z, \vec{M}_O \neq 0$, it is instructive to first look at the linear-in-field terms in \equref{GinzburgLandauExpansion}. We find two different cases. If $|c_2 M_Z| + \delta a > \sqrt{(\delta c_1 M_Z)^2 + \delta a^2}$, one obtains a pure triplet state of the type $A^{3_s}_{m_v=0}(1,i,0)$. Choosing $\vec{M}_Z= M_Z \vec{e}_x$ with $M_Z>0$, the triplet vector is given by $\vec{d} = \left(0,1,\sign(c_2) i\right)^T$ and the critical temperature is
\begin{equation}
    T_c = T_{c,0} + \left(\delta a+|c_2 M_Z|\right)/\alpha. \label{Tc1}
\end{equation}
Else, if $|c_2 M_Z| + \delta a < \sqrt{(\delta c_1 M_Z)^2 + \delta a^2}$, one finds an admixture of singlet and triplet with order parameter
\begin{equation}
        \Delta^s = \Delta_0,\quad \vec{d} = i\vec{e}_x \Delta_0 \frac{\delta c_1 M_Z}{\sqrt{(\delta c_1M_Z)^2 + \delta a^2} - \delta a}. \label{MixtureInMagneticField}
\end{equation}
The transition temperature in this case is
\begin{equation}
    T_c = T_{c,0} + \sqrt{\delta a^2 +(\delta c_1 M_Z)^2}/\alpha. \label{Tc2}
\end{equation}
We see from \equref{MixtureInMagneticField} that there is an approximately equal mixing of singlet and triplet for $|\delta c_1| M_Z \gg |\delta a|$ while in the opposite limit, $|\delta c_1| M_Z \ll |\delta a|$, either singlet or triplet dominates depending on whether $\delta a <0$ or $\delta a >0$. The relative phase of $\pi/2$ between singlet and triplet makes the pairing state break time-reversal symmetry as is required in order to couple linearly to magnetic moments.

To understand how the approximate SU(2)$_+$ $\times$ SU(2)$_-$ symmetry can naturally explain the linear-in-magnetic-field behavior, we first consider case (II), \textit{i.e.}, there is already microscopically coexisting ferromagnetic order (or there is at least a significant coupling between superconductivity and the ferromagnetic moments) at $T_c^+$. Then, $\vec{M}_Z$ and $\vec{M}_O$ should be thought of as the combination of the applied external magnetic field and the ferromagnetic order parameter. In this scenario, it is apt to assume $|\delta a| \ll \text{max}(\delta c_1 M_Z,c_2 M_Z)$ and we generically obtain a linear increase of the critical temperature with magnetic field [see \equsref{Tc1}{Tc2}]. If $c_2 > \delta c_1$, we obtain the nonunitary triplet state with $\vec{d} \propto (1,i,0)^T$, which we expect close to the $J=0$ line, while $\delta c_1 > c_2$ leads to the admixture of singlet and triplet with $\vec{d} \propto (1,0,0)^T$. As $\delta c_1$ vanishes for $J=0$ and in single-band mean-field theory (even when $J\neq 0$), we expect the former scenario to be more likely, which will favor the nonunitary triplet state as the leading instability.

In the case of scenario (I), we should view $\vec{M}_Z$ and $\vec{M}_O$ in \equref{GinzburgLandauExpansion} as resulting entirely from the Zeeman and orbital coupling of the external magnetic field alone. 
For large magnetic fields where $|\delta a| \ll \text{max}\,(\delta c_1 M_Z,c_2 M_Z)$, the same conclusions as above will apply and $T_c$ will generically vary linearly with the field. 
However, for sufficiently small magnetic fields, we have $|\delta a| \gg \text{max}(\delta c_1 M_Z,c_2 M_Z)$. In this limit, only $\delta a > 0$ favoring the nonunitary triplet pairing $A^{3_s}_{m_v=0}(1,i,0)$ is consistent with the transition temperature changing linearly with magnetic field. Alternatively, the system could ultimately be in a singlet state at $M_Z=0$ (\textit{i.e.}, $\delta a < 0$) but the magnitude of $\delta a$ is sufficiently small such that the ``rounding off'' of $T^+_c(M_Z)$ at low $M_Z$ cannot be seen in experiment.

\begin{figure}[tb]
   \centering
    \includegraphics[width=\linewidth]{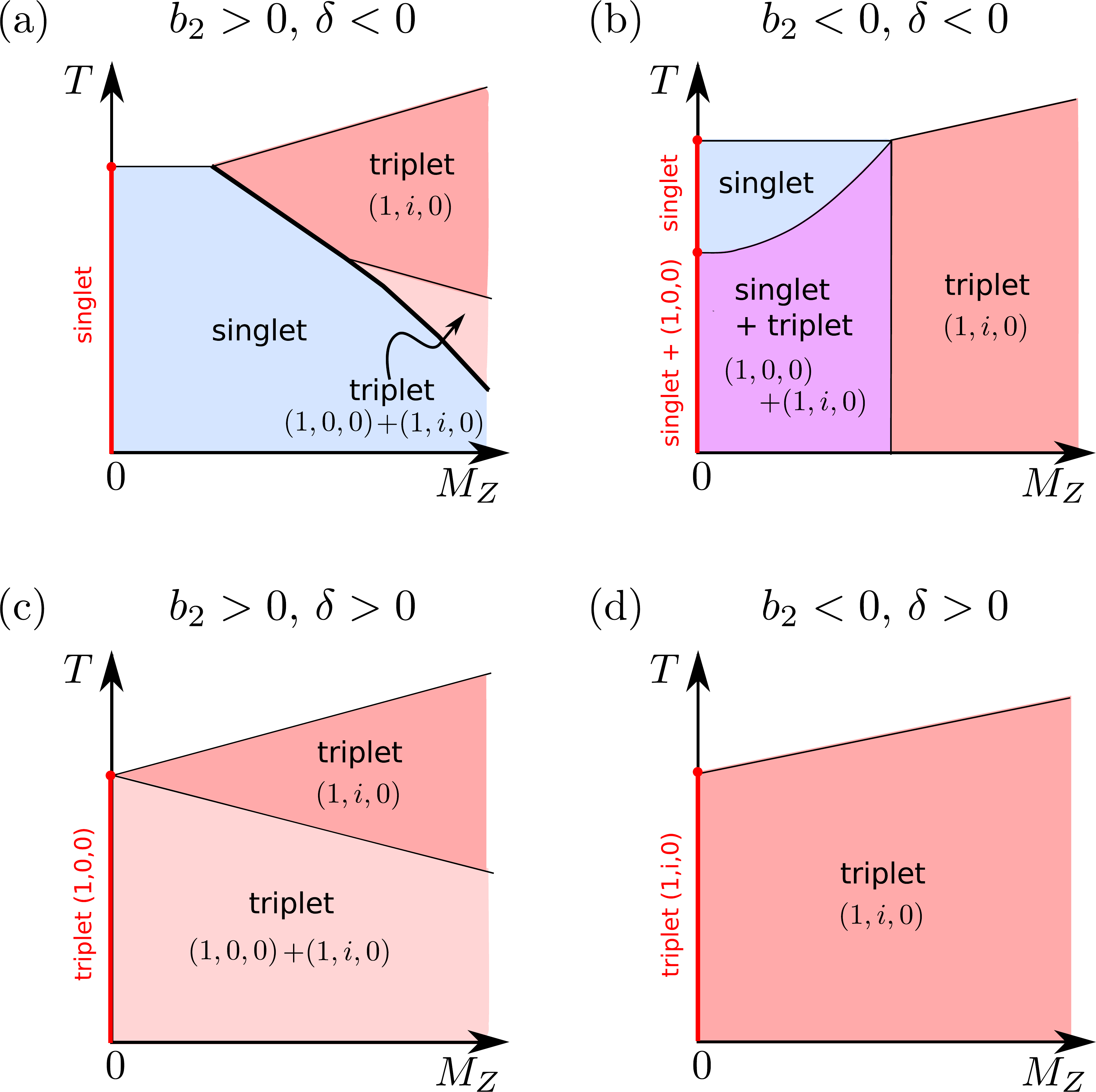}
    \caption{Phase diagram as a function of temperature $T$ and Zeeman field $\vec{M}_Z =  M_Z \vec{e}_x$ when (a,b) singlet dominates at low fields and (c,d) triplet dominates, which we determine by minimizing \equref{IncludingQuartic}.  Thin (thick) black lines correspond to second (first) order transitions. The phases for $M_Z=0$ are indicated in red and we recover the four different possible temperature dependences of \figref{SchematicPD}. Recall from \secref{ExpectationsMeanField} that $b_2 > 0$ is expected in mean-field theory. However, as we will see in \secref{FluctuationInducedSC}, strong ferromagnetic fluctuations will favor $b_2 < 0$. As symmetry requires $\delta c_1$ to be proportional to the Hund's coupling $J$, we have set $\delta c_1 = 0$ here. For nonzero $\delta c_1$, the singlet superconducting phases will contain an admixture of unitary triplet as described by \equref{MixtureInMagneticField} and a first-order transition into a singlet state (with unitary triplet admixture) will be possible at lower temperatures and nonzero Zeeman field in part (c). Note that the transition temperature from the normal state into the singlet superconductor is constant as we neglect here the nonlinear coupling to the magnetic field. A discussion of the latter can be found in \secref{NonLinearCouplingOfMagnField}.}
    \label{PDInMagneticField}
\end{figure}

\subsubsection{Quartic terms and sub-leading transitions}
\label{QuarticTermsInMagneticField}
Having examined the first superconducting transition that takes place upon cooling the system down starting from the normal state, we now assess whether and what type of subsequent superconducting transitions can occur. In this context, we need to include terms quartic in the superconducting order parameter and extend \equref{GinzburgLandauExpansion} to 
\begin{alignat}{1}
    \mathcal{F}_M &\sim (a(T) + \delta a) |\Delta^s|^2   + (a(T) - \delta a) \sum_{s=\pm,0} |d_s|^2 \label{IncludingQuartic} \\ 
    \nonumber& + 2 \delta c_1 M_Z \text{Im}\left(d^*_0 \Delta^s\right) + c_2 M_Z \left(|d_-|^2 - |d_+|^2 \right) \\
    \nonumber& + (b_1+b_2) \left(|\Delta^s|^4 + |d_0|^4\right) + (b_1+2b_2) \left(|d_+|^4 + |d_-|^4\right) \\
    \nonumber& + 2(b_1+2b_2) |\Delta^s|^2\sum_{s=\pm,0} |d_s|^2 -4b_2\text{Re}\left[ d_0^2 d^*_+d^*_-\right] \\
    \nonumber& + 2b_2 \text{Re}\left[ (\Delta^s)^2 ((d_0^*)^2 + 2 d_+^*d_-^*)\right] \\
    \nonumber& + 2b_1 |d_+|^2|d_-|^2 + 2(b_1 + 2b_2) |d_0|^2(|d_+|^2 + |d_-|^2),
\end{alignat}
where we kept only the terms linear in magnetic field, took $\vec{M}_Z$ along the $z$-axis, and re-expressed the triplet in the form $\vec{d}$\,$=$\,$d_+ (1,i,0)/\sqrt{2} + d_-(1,-i,0)/\sqrt{2}+d_0 (0,0,1)$. This parametrization is more convenient in the presence of a magnetic field than that used in \equref{FullExpansionWithHigherSym}. Additionally, we have neglected the impact of the magnetic field on the quartic terms.

Taking $\delta c_1$\,$=$\,$0$ (as it has to vanish for $J$\,$=$\,$0$), the different possible phase diagrams are summarized in \figref{PDInMagneticField}.
The possibility illustrated in part (c) of \figref{PDInMagneticField} corresponds to the picture put forward by \citet{PhysRevLett.30.81} for He$^3$ in the presence of a magnetic field, which might very well also apply to twisted double-bilayer graphene \cite{ExperimentKim,2019arXiv190308685L}. The difference with \refcite{PhysRevLett.30.81} is that we do not get a third transition since we work with a one-dimensional IR of the spatial point group.

However, there are three other options, depicted in \figref{PDInMagneticField}(a), (b), and (d), that we cannot easily exclude given the experimental data: owing to the strong-coupling properties of the problem at hand, a nonunitary triplet state might be dominant at $M_Z=0$, as seems to be the case in LaNiC$_2$ \cite{PhysRevB.82.174511} and is favored by our fluctuation approach of \secref{FluctuationInducedSC}; under this condition, only one transition is expected even when $M_Z \neq 0$ [see \figref{PDInMagneticField}(d)]. It could also be that singlet dominates without a magnetic field instead. We can see in \figref{PDInMagneticField}(a) and (b) that, in these two cases, triplet shows up and $T_c$ increases linearly when $M_Z > 2|\delta a|/c_2$. The small value of $\Delta T_c/T_{c,0}$ estimated in \equref{EstimateOfTc} suggests that resolving this initial region, where $T_c$ is constant as a function of $M_Z$, is experimentally challenging.

\subsubsection{Nonlinear couplings in a magnetic field}\label{NonLinearCouplingOfMagnField}
We finally come back to the quadratic couplings to the magnetic field, associated with the terms with prefactors $c_{3,4}$ and $\delta c_{5,6}$ in \equref{GinzburgLandauExpansion}. We first notice that this will lead to an additional quadratic suppression of the leading transition temperatures in \figref{PDInMagneticField}; in particular, the transition temperature into the singlet state in part (a) and (b) will not be field-independent any more. More interestingly, the suppression of singlet and triplet is enforced to be nearly identical for small $J$ due to the SU(2)$_+$ $\times$ SU(2)$_-$ symmetry, $\delta c_{5,6} \ll c_{3,4}$. Resultantly, if the effective $J$ relevant for superconductivity is indeed small, the nonlinear terms $\propto \vec{M}_Z^2, \vec{M}_O^2$ are not expected to affect the competition between singlet and triplet significantly and the qualitative form of the phase diagrams in \figref{PDInMagneticField} is not modified.

\section{Complex representation of $C_3$}
\label{ComplexRepresentation}
In this section, we extend our previous analysis to the complex IR $E$ of the spatial point group $C_3$. Time-reversal symmetry necessitates treating the representation and its complex-conjugate partner on an equal footing. Alternatively, one can think of a two-dimensional (reducible) representation with partner functions transforming as $x$ and $y$ under $C_3$. 

Akin to our discussion earlier, we first study the case of nonzero Hund's coupling, $J\neq 0$, with point group $\mathcal{G}_2$ in \equref{ReducedSymmetryGroup}, which enables us to distinguish between singlet and triplet pairing. After discussing all symmetry-allowed singlet and triplet states separately, we will derive the phase diagrams analogous to \figref{SchematicPD}: we will examine how these states ``connect'' when adiabatically changing the Hund's coupling from negative to positive values, and whether singlet and triplet can mix when $J$ is small and the SU(2)$_+$ $\times$ SU(2)$_-$ symmetry is only weakly broken.

\subsection{Nonzero Hund's coupling}\label{SingeltTripletSepForComplexRep}
To proceed with singlet pairing, we parametrize $M_{\vec{k}v}$ in \equref{SimplifiedFormOfHighSymPairing} according to
\begin{equation}
    M_{\vec{k}+} = \sum_{\mu=\pm}\eta^{}_\mu \left( X^{}_{\vec{k}} + i\mu\, Y^{}_{\vec{k}} \right)\sigma^{}_0, \label{SingletPairingMatrix}
\end{equation}
while $M_{\vec{k}-}$ is determined by the Fermi-Dirac constraint (\ref{FermiDiracStatConstr}); $X_{\vec{k}}$ and $Y_{\vec{k}}$ are real-valued functions that are continuous on the Brillouin zone and transform as $k_x$ and $k_y$ under $C_3$. A one-parameter family of possible choices for the lowest-order functions (\textit{i.e.}, with minimal number of sign changes in the Brillouin zone) is given by
\begin{subequations}
\begin{equation}
    (X_{\vec{k}},Y_{\vec{k}})^T = R_{\phi} \left(X^{(1)}_{\vec{k}},Y^{(1)}_{\vec{k}}\right)^T \label{parametrizationOfBasisFuncs}
\end{equation}
with arbitrary $\phi \in [0,2\pi)$, where $R_{\phi}$ is a $2\times 2$ matrix describing rotations by angle $\phi$, $R_{\phi} = e^{i\phi\sigma_y}$, and
\begin{align}
    X^{(1)}_{\vec{k}}&=\frac{2}{\sqrt{3}}\sin (\sqrt{3}k_x/2) \cos (k_y/2), \\
    Y^{(1)}_{\vec{k}}&=\frac{2}{3}\left(\sin k_y + \cos (\sqrt{3}k_x/2) \sin (k_y/2)\right).
\end{align}\label{ExampleBasisFuncs}\end{subequations}
Both $X_{\vec{k}}$ and $Y_{\vec{k}}$ have to vanish at $\Gamma$, $K$ and $K'$ as these momenta are invariant under $C_3$. Further, both $X_{\vec{k}}$ and $Y_{\vec{k}}$ must have lines of zeros going through these high symmetry points; the orientation of these lines is, however, not fixed due to the absence of additional reflection or in-plane rotation symmetries---this is different from the situation for twisted bilayer and trilayer graphene in \secref{ComparisonWithOtherSystems}. For \equref{ExampleBasisFuncs}, the orientation of these zeros changes with $\phi$.

With the parametrization defined in \equref{SingletPairingMatrix}, the relevant symmetries act as follows
\begin{subequations}
\begin{align}
    C_3:\quad (\eta_+,\eta_-) \, &\longrightarrow \, (\omega \eta_+, \omega^* \eta_- ), \quad \omega = e^{i\frac{2\pi}{3}}, \\
    \Theta:\quad (\eta_+,\eta_-) \, &\longrightarrow \, (\eta_-^*, \eta_+^* ).
\end{align}\label{TrafoComplexSinglet}\end{subequations}
It readily follows from \equref{TrafoComplexSinglet} that the most general free energy up to quartic order reads as
\begin{align}
    \mathcal{F} \sim a(|\eta_+|^2 + |\eta_-|^2) + b^s_1(|\eta_+|^2 + |\eta_-|^2)^2 + b^s_2 |\eta_+|^2|\eta_-|^2.  \label{SingletTwoComponentExpansion}
\end{align}
The sign of $b^s_2$ therefore distinguishes between two different singlet phases: if $b^s_2 > 0$, we have $(\eta_+,\eta_-) = (1,0)$, which corresponds to
\begin{equation}
            M_{\vec{k}+} =  \left( X_{\vec{k}} + i\, Y_{\vec{k}} \right)\sigma^{}_0. \label{ChiralSinglet}
\end{equation}
Exactly as in \secref{TrivialRepresentation}, we always show only one out of the many symmetry-equivalent representations of the order parameter---instead of using a general parametrization of a phase---to make the notation and the discussion of properties of the superconducting state more easily accessible.
The state in \equref{ChiralSinglet} breaks time-reversal symmetry but preserves $C_3$ (and spin-rotation symmetry). We refer to this state as a chiral singlet superconductor and denote it by $E^{1s}(1,i)$ in the following. It is fully gapped (unless the Fermi surfaces go through the $\Gamma$, $K$, or $K'$ point) and has been investigated extensively in the recent literature on pairing in twisted bilayer graphene \cite{YiZhuangPairing,RafaelsPaperTBG,2019arXiv190407875W,2018PhRvB..98x1407K,2018PhRvB..97w5453G,HUANG2019310,2019arXiv190301701C,2019arXiv190100500L,2018PhRvL.121u7001L,2018PhRvB..98h5436F}. 

Conversely, if $b^s_2 < 0$, we find that $|\eta_+| = |\eta_-|$ at the minimum of \equref{SingletTwoComponentExpansion}. As the relative phase $\varphi$ between $\eta_+$ and $\eta_-=\eta_+ e^{i\varphi}$ is not fixed by \equref{SingletTwoComponentExpansion}, one might naively conclude that higher order terms have to be considered. In fact, in sixth order, there is indeed the contribution
\begin{equation}
        c_1 \text{Re}\left[ \eta_+^3 (\eta_-^*)^3 \right] + c_2 \, \text{Im}\left[ \eta_+^3 (\eta_-^*)^3 \right], \quad c_{1,2} \in \mathbb{R},\label{SexticTerm}
\end{equation}
and the relative phase $\varphi$ will depend on $c_1/c_2$. However, upon reinserting $\eta_-=\eta_+ e^{i\varphi}$ into \equref{SingletPairingMatrix}, we notice that $\varphi \neq 0$ simply corresponds to rotating the basis functions $X_{\vec{k}}$ and $Y_{\vec{k}}$ into each other, which does not change their transformation behavior under $C_3$ [$\varphi$ is directly related to $\phi$ in \equref{parametrizationOfBasisFuncs}]. Consequently, we can set $\varphi=0$ without loss of generality, which implies
\begin{equation}
        M^{}_{\vec{k}v} = \Delta^{}_s X^{}_{\vec{k}} \sigma^{}_0.
\end{equation}
This state, which we call $E^{1s}(1,0)$, breaks $C_3$ but preserves time-reversal symmetry; this is the nematic singlet phase.

Within a single-band mean-field description (see \appref{MicrosCopGLExpansion}), we find $b^s_1=b^s_2/2 > 0$. As such, mean-field theory generically favors the chiral singlet superconductor over the nematic state $E^{1s}(1,0)$; this has been noted before in the context of twisted bilayer graphene \cite{YiZhuangPairing} and \refcite{LiangFu} discusses how strong fluctuations can stabilize the nematic phase. 

Turning to triplet pairing, we now modify the parametrization \eqref{SingletPairingMatrix} to
\begin{align}
        M^{}_{\vec{k}+} = \sum_{\mu=\pm}\sum_{\nu=1}^3\eta^{}_{\mu\nu} \left( X^{}_{\vec{k}} + i\mu\, Y^{}_{\vec{k}} \right)\sigma^{}_\nu, \label{parametrizationOfTripletComplexIR}
\end{align}
where $X_{\vec{k}}$ and $Y_{\vec{k}}$ are defined exactly as before. For simplicity, we
introduce the complex-vector notation, $\vec{d}_\mu = (\eta_{\mu,1},\eta_{\mu,2},\eta_{\mu,3})^T$, $\mu=\pm$. The representations of the symmetries now read as
\begin{subequations}\begin{align}
    C_3:\quad (\vec{d}_+,\vec{d}_-) \quad &\longrightarrow \quad (\omega \vec{d}_+, \omega^* \vec{d}_- ),  \\
    \Theta:\quad (\vec{d}_+,\vec{d}_-) \quad &\longrightarrow \quad (\vec{d}_-^*, \vec{d}_+^* ), \\
    \text{SU(2)}_{s}:\quad (\vec{d}_+,\vec{d}_-) \quad &\longrightarrow \quad (\mathcal{R}\vec{d}_+, \mathcal{R}\vec{d}_-),
\end{align}\label{TrafoOf2DTriplet}\end{subequations}
with $\mathcal{R} \in \text{SO}(3)$ and $\omega = e^{i\frac{2\pi}{3}}$.
The most general free-energy expansion is given by 
\begin{alignat}{1}
    \nonumber\mathcal{F} &\sim a\sum_{\mu=\pm} \vec{d}_\mu^\dagger\vec{d}^\pdagger_\mu + b^t_1\left(\sum_{\mu=\pm} \vec{d}_\mu^\dagger\vec{d}^\pdagger_\mu\right)^2 + b^t_2 (\vec{d}^\dagger_+\vec{d}^\pdagger_+)(\vec{d}^\dagger_-\vec{d}^\pdagger_-)\\
    &\quad + b^t_3 |\vec{d}_+^\dagger \vec{d}^\pdagger_-|^2 + b^t_4 |\vec{d}_+^T \vec{d}^\pdagger_-|^2 + b^t_5 \sum_{\mu=\pm} |\vec{d}_\mu^T \vec{d}^\pdagger_\mu|^2 \label{GLExpansionTripletComplexIR}
\end{alignat}
up to quartic order, where $b^t_j$\,$\in$\,$\mathbb{R}$; the different symmetry-allowed phases follow from the stable minima of the free energy. When minimizing \equref{GLExpansionTripletComplexIR}, we take into account that the relative phase between $\vec{d}_+$ and $\vec{d}_-$ can always be absorbed into a redefinition of the basis functions $X_{\vec{k}}$ and $Y_{\vec{k}}$, as for the singlet above. In total, we find eight distinct triplet states which we label by $E^{3_s}(a)$ through $E^{3_s}(h)$. Phase diagrams describing which of these phases is realized for a given configuration of the quartic couplings $b^t_{j}$ can be found in \appref{DetailsOfComplexRepresentation}; here, we list all the phases, describe their properties, and refer to \figref{fig:BdG} for an illustration of their respective spectra and densities of states:

\begin{figure}[tb]
    \centering
    \includegraphics[width=\linewidth]{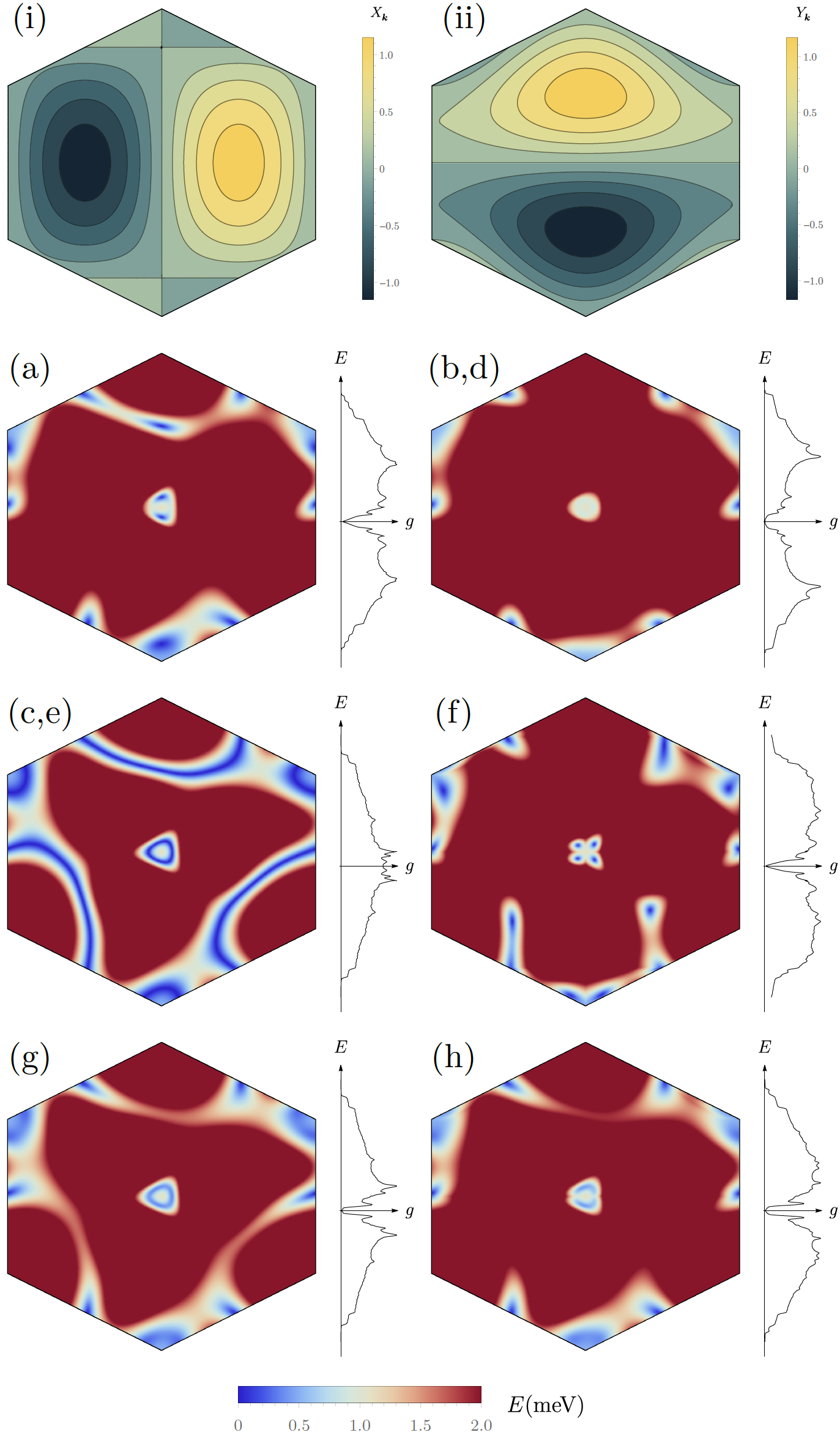}
    \caption{(i, ii) The lowest lattice harmonics of the basis functions [Eq.~\eqref{ExampleBasisFuncs}]. (a--h) The momentum dependence of the gap and the density of states, $g(E)$, for the pairing phases, $E^{3_s}(a)$ through $E^{3_s}(h)$. The Bogoliubov-de Gennes excitation spectrum is calculated using the band structure of the system predicted by the continuum model \cite{KoshinosPaper}, assuming a pairing term of the form of Eq.~\eqref{SimplifiedFormOfHighSymPairing} with an overall scale of $\Delta_0 = 4$\,meV. The nodal points/lines are demarcated in dark blue; note that the states ($b$), ($d$), ($g$), and ($h$), taking $\alpha = \pi/4$, are fully gapped, ($a$) and ($f$) have nodal points, and ($c$), ($e$) have nodal lines, as is also visible in $g(E)$.}
    \label{fig:BdG}
\end{figure}{}

\begin{enumerate}[(a),leftmargin=2\parindent]
        \item This state, labeled as $E^{3_s}(a)$, can be represented by $\vec{d}_+$\,$=$\,$\vec{d}_-$\,$=$\,$\left(1,0,0 \right)^T$ with the associated order parameter $M_{\vec{k}+}$\,$=$\,$X_{\vec{k}}\, \sigma_x$. More physically, it corresponds to a \textit{nematic unitary triplet phase}. It preserves time-reversal symmetry, but breaks both SU(2)$_s$ spin-rotation symmetry [down to O(2)] and $C_3$ rotational symmetry. This state has two symmetry-enforced nodal points at each Fermi surface around the $K$, $K'$, or $\Gamma$ point. Owing to the lack of any reflection symmetry (cf. the discussion of $D_3$ in \secref{ComparisonWithOtherSystems} below), the positions of these nodal points are not pinned to any specific direction.
        \item One representative configuration of this phase is given by $\vec{d}_+$\,$=$\,$(1,-i,0)^T/2$ and $ \vec{d}_-$\,$=$\,$(1,i,0)^T/2$; it can thus be seen as a \textit{helical triplet}, consisting of two time-reversed copies of states with opposite chirality. The order parameter can be more explicitly written as $M_{\vec{k}+}$\,$=$\,$ X_{\vec{k}} \sigma_x + Y_{\vec{k}} \sigma_y$, which can alternatively be thought of as a 2D analogue of the Balian-Werthamer state of the B-phase of superfluid $^3$He \cite{vollhardt2013superfluid}. This state, denoted by $E^{3_s}(b)$ in the following, only has point nodes at $\Gamma$, $K$, and $K'$, \textit{i.e.}, it is expected to exhibit a full gap for generic Fermi surfaces not going through these high-symmetry points. It preserves time-reversal symmetry. While this state breaks spin-rotation symmetry as well as $C_3$, the product of $C_3$ and a rotation in spin space along $\sigma_z$ with angle $2\pi/3$ is preserved; this can be viewed as the spontaneous formation of spin-orbit coupling. 
        \item Here, we can write $\vec{d}_+$\,$=$\,$\vec{d}_-$\,$=$\,$ (1,i,0)^T/2$; hence, $M_{\vec{k}+}$\,$=$\,$ X_{\vec{k}}(\sigma_x + i \sigma_y)$. This is a \textit{nematic nonunitary triplet} state which breaks time-reversal symmetry and $C_3$. One spin-species will be gapless while the other will have nodal lines (\textit{i.e.}, point nodes on the Fermi surface).
        \item The triplet vectors in this phase can be written as $\vec{d}_+$\,$=$\,$ (1,0,0)^T$, $\vec{d}_-$\,$=$\,$0$ leading to $M_{\vec{k}+}$\,$=$\,$\left( X_{\vec{k}} + i\, Y_{\vec{k}} \right)\sigma^{}_x$. As one of the two chiralities is preferred over the other ($|\vec{d}_+| \neq |\vec{d}_-|$), this state can be referred to as \textit{chiral unitary triplet}. It is a 2D analogue of the A-phase of $^3$He \cite{vollhardt2013superfluid}. It breaks SU(2)$_s$ spin-rotation symmetry [down to O(2)] and time-reversal, but preserves $C_3$. Except for $\Gamma$, $K$, and $K'$, this state has no symmetry-imposed nodal points. In fact, its spectrum is identical to that of the helical triplet $E^{3_s}(b)$, which is why we group these two states together in \figref{fig:BdG}.
        \item For this state, we have $\vec{d}_+$\,$=$\,$(1,i,0)^T$, $\vec{d}_-$\,$=$\,$0$, \textit{i.e.}, $M_{\vec{k}+}$\,$=$\,$\left( X_{\vec{k}} + i\, Y_{\vec{k}} \right)\left(\sigma^{}_x + i\sigma^{}_y \right)$. It consists of only one of the two time-reversed copies with opposite chirality of the $E^{3_s}(b)$ state discussed above and, thus, is a \textit{chiral nonunitary triplet state.} This state can be seen as an analogue of the $A_1$-phase of $^3$He \cite{vollhardt2013superfluid}. It preserves $C_3$, but breaks SU(2)$_s$ spin-rotation symmetry [down to O(2)] and time-reversal. Here, one of the spin components will be gapless while the other is fully gapped (as before, except for the high symmetry points $\Gamma$, $K$, and $K'$ which are generically not on the Fermi surface). Note that although the spectrum of this state is not strictly identical to that of the nematic nonunitary triplet $E_s^{3_s}(c)$, we grouped them together in \figref{fig:BdG} as their respective plots are practically indistinguishable; this is related to the fact that, in both cases, the low-energy spectrum is dominated by the Fermi surface of one of the spin species.
        \item In this phase, $\vec{d}_+$\,$=$\,$(1,0,0)^T$, $\vec{d}_-$\,$=$\,$(0,1,0)^T$, implying $M_{\vec{k}+}^{}$\,$=$\,$\left( X_{\vec{k}} + i\, Y_{\vec{k}} \right) \sigma^{}_x$\,$+$\,$\left( X_{\vec{k}} - i\, Y_{\vec{k}} \right) \sigma^{}_y$. The state can, thus, be thought of as a superposition of two chiral unitary triplets with orthogonal spin polarizations or, when inserted into \equref{SimplifiedFormOfHighSymPairing}, as Cooper pairs of electrons with spin polarization $\downarrow\downarrow$ ($\uparrow\uparrow$) and orbital basis function $X_{\vec{k}}+Y_{\vec{k}}$ ($X_{\vec{k}}-Y_{\vec{k}}$). Time-reversal, $C_3$, and spin-rotation symmetry are all broken. The excitation spectrum  is given by $E_{\pm}(\vec{k}) =\sqrt{\xi_{\vec{k}+}^2 +2(X_{\vec{k}} \pm Y_{\vec{k}})^2}$, so it is characterized by ``two gaps'', given by $|X_{\vec{k}} \pm Y_{\vec{k}}|$, both of which are forced to vanish at two points for each Fermi surface enclosing $K$, $K'$, and $\Gamma$. While the number of nodes of this state and of $E^{3_s}(a)$ are the same, the spin degrees of freedom on the Fermi surface have nodes at the same two momenta for $E^{3_s}(a)$. For $E^{3_s}(f)$, however, the two spin species have nodal points at different momenta.
        \item Denoted by $E^{3_s}(g)$, this phase has $\vec{d}_+$\,$=$\,$\cos \,(\alpha)$ $(1,i,0)^T/\sqrt{2}$,\, $\vec{d}_- = \sin (\alpha)\, (0,0,1)^T$, where the parameter $\alpha$ varies continuously with $b^t_j$ in the part of the phase diagram where this state is realized. The corresponding order parameter can be written as 
        $
            M_{\vec{k}+} = \cos (\alpha)\, \left(X_{\vec{k}} + i\, Y_{\vec{k}} \right) (\sigma^{}_x+i\sigma^{}_y)/\sqrt{2} 
            + \sin (\alpha)\, \left(X_{\vec{k}} - i\, Y_{\vec{k}} \right) \sigma^{}_z,
        $
        $0<\alpha<\pi$, and can be viewed as a superposition of a chiral nonunitary triplet state and a unitary state with opposite chirality. This state breaks time-reversal symmetry, spin-rotation invariance, and $C_3$ but preserves the product of $C_3$ and spin rotation by angle $2\pi/3$ along $\sigma_z$. So, similar to the state $E^{3_s}(b)$ above, this state spontaneously entangles rotations in spin and real space and its spectrum, see \figref{fig:BdG}(g), is $C_3$ invariant. It is fully gapped (again, as long as the Fermi surfaces do not go through $\Gamma$, $K$, and $K'$), with two different gaps $[(1\pm g_\alpha)\left(X^2_{\vec{k}} + Y^2_{\vec{k}}\right)]^{1/2}$, where $g_\alpha = \cos\alpha \sqrt{1+\sin^2\alpha}$.
        \item Finally, for the triplet phase $E^{3_s}(h)$, one has $\vec{d}_+$ $=$ $(\cos \alpha,0, i \sin \alpha)^T$, $\vec{d}_- = (0,\cos \alpha,-i\sin \alpha)^T$,  which yields $M_{\vec{k}+} = \cos (\alpha)[(X_{\vec{k}} + i\, Y_{\vec{k}}) \sigma_x + (X_{\vec{k}} - i\, Y_{\vec{k}}) \sigma_y] - 2\sin(\alpha)  Y_{\vec{k}}  \sigma_z$. It can be seen as a superposition of the states $E^{3_s}(a)$ and $E^{3_s}(f)$ to which it reduces for $\alpha=\pi/2$ and $\alpha=0$; it will have two nodal points for $\alpha$ close to these limiting cases, but can be fully gapped for other values of $\alpha$. For $\alpha\neq \pi/2$, this state breaks time-reversal, $C_3$, and spin-rotation symmetry.
\end{enumerate}
In \appref{MicrosCopGLExpansion}, we show that $b^t_1=b^t_3/2=-b^t_4/2=-2b^t_5 > 0$ and $b^t_2 = 0$ within a single-band mean-field description. Minimizing \equref{GLExpansionTripletComplexIR} yields that the phases $E^{3_s}(b)$ and $E^{3_s}(d)$ have the lowest energy and are exactly degenerate for this configuration of quartic couplings. This degeneracy within mean-field theory, which was noted before in \refcite{CenkeLeon}, will be lifted by corrections resulting, e.g., from residual interactions. In \secref{FluctuationInducedSC}, we will find that $E^{3_s}(b)$ ($E^{3_s}(d)$) is favored in the presence of ferromagnetic spin (orbital) fluctuations. We will also see that significant fluctuations can stabilize phases other than the two, $E^{3_s}(b)$ and $E^{3_s}(d)$, favored in mean-field theory.

\subsection{Approximate SU(2)$_+$ $\times$ SU(2)$_-$}
After having classified singlet and triplet separately, we now focus on small Hund's coupling for which SU(2)$_+$ $ \times$ SU(2)$_-$ is an approximate symmetry, and singlet and triplet are nearly degenerate at the quadratic level of the free energy. This requires studying them on an equal footing and generalizing the parametrization in \equsref{SingletPairingMatrix}{parametrizationOfTripletComplexIR} to include both singlet and triplet, \textit{i.e.}, extending the summation over $\nu$ in \equref{parametrizationOfTripletComplexIR} to $\nu=0,1,2,3$. In analogy with \secref{ExactSU2SU2Symmetry}, we use $2 \times 2$ matrices and write 
\begin{equation}
    M^{}_{\vec{k}v} = \sum_{\mu=\pm} \left( X^{}_{\vec{k}} + i\mu\, Y^{}_{\vec{k}} \right) \Delta^{}_\mu, \quad \Delta_\mu = \sum_{\nu=0}^3 \eta^{}_{\mu\nu} \sigma^{}_\nu.
\end{equation}
It is easy to see that the symmetries act according to
\begin{subequations}
\begin{align}
        C_3&:\, (\Delta_+,\Delta_-) \, \longrightarrow \, (\omega \Delta_+, \omega^* \Delta_- ), \\
    \Theta&:\, (\Delta_+,\Delta_-) \, \longrightarrow \, (\Delta_-^\dagger,\Delta_+^\dagger), \\
    \mathcal{G}^s_1&:\, \Delta_\mu \, \longrightarrow \,e^{-i\vec{\varphi}_+\cdot \vec{\sigma}}\Delta_\mu e^{i\vec{\varphi}_-\cdot \vec{\sigma}},
\end{align}\end{subequations}
where, recall, $\mathcal{G}^s_1 \equiv \text{SU}(2)_+ \times \text{SU}(2)_-$. Imposing $\text{SU}(2)_+ \times \text{SU}(2)_-$ as an exact symmetry, the most general free energy up to quartic order reads as
\begin{alignat}{1}
    \nonumber\mathcal{F} &\sim a \sum_{\mu=\pm} \text{tr}[\Delta_\mu^\dagger \Delta_\mu^\pdagger] + \frac{b_1}{4} \left(\sum_{\mu=\pm} \text{tr}[\Delta_\mu^\dagger \Delta_\mu^\pdagger]\right)^2 \\
    \nonumber&+ \frac{b_2}{2} \sum_{\mu=\pm} \text{tr}[\Delta_\mu^\dagger \Delta_\mu^\pdagger\Delta_\mu^\dagger \Delta_\mu^\pdagger] \\
    \nonumber&+ \frac{b_3}{4} \text{tr}[\Delta_+^\dagger \Delta_+^\pdagger]\,\text{tr}[\Delta_-^\dagger \Delta_-^\pdagger] + \frac{b_4}{4} \left|\text{tr}[\Delta_+^\dagger \Delta_-^\pdagger] \right|^2\\
    &+ \frac{b_5}{2} \left(\text{tr}[\Delta_+^\dagger \Delta_+^\pdagger\Delta_-^\dagger \Delta_-^\pdagger] + \text{tr}[\Delta_-^\pdagger \Delta_-^\dagger\Delta_+^\pdagger \Delta_+^\dagger] \right). \label{MatrixExpansionOfFreeEnergyComplRep}
\end{alignat}
At first glance, one might think that there are additional terms with extra factors of $\sigma_y$, similar to the last term in \equref{ExpansionOfFreeEnergy}. However, as before, all of them can be related to combinations of the terms already present in \equref{MatrixExpansionOfFreeEnergyComplRep} as outlined in \appref{DetailsOfComplexRepresentation}. 

Following the procedure applied in \secref{TrivialRepresentation} to the one-dimensional IR $A$, we now add a small quadratic term, $\delta a\sum_{\mu} ( |\Delta^s_\mu|^2 - \vec{d}_\mu^\dagger\vec{d}^\pdagger_\mu)$, where $\Delta_\mu^s$ and $\vec{d}_\mu$ are the singlet and triplet component of $\Delta_\mu$ in \equref{MatrixExpansionOfFreeEnergyComplRep}, \textit{i.e.}, $\Delta_\mu = \sigma_0 \Delta^s_\mu + \vec{\sigma}\cdot \vec{d}_\mu$. This term breaks $\text{SU}(2)_+ \times \text{SU}(2)_-$ and hence, makes singlet and triplet inequivalent. It allows us to study which of the different singlet and triplet states defined above can mix, and to identify ``Hund's partners'', \textit{i.e.}, which states transform into each other when changing the sign of the Hund's coupling $J$ and accordingly, of $\delta a$. This generalizes the phase diagrams in \figref{SchematicPD} and \tableref{TableSummaryOfPairingC3} to the complex representation.

\renewcommand{\exspace}{\hspace{0.9em}}
\begin{table*}[tb]
\begin{center}
\caption{Summary of possible pairing states transforming under the complex representation $E$ of $C_3$. The labeling of the pairing states and their symmetry properties can be found in the main text. The states are ordered by pure singlet, triplet, and admixtures of singlet and triplet. The latter are only expected generically when the SU(2)$_- \times$ SU(2)$_+$ symmetry is weakly broken. We use $X_{\vec{k}}$ and $Y_{\vec{k}}$ to denote real-valued continuous functions on the Brillouin zone that transform as $k_x$ and $k_y$ under $C_3$ [see, e.g., \equref{ExampleBasisFuncs}]. The temperature-dependent coefficient $\eta$ describes the admixture of a triplet/singlet pairing at a second transition to a purely singlet/triplet one. Furthermore, $a,b\in\mathbb{R}$ vary continuously with system parameters. The minimal number of nodes on any Fermi surface enclosing the $\Gamma$, $K$, or $K'$ point is indicated in the column ``Nodes''. As before, two states are referred to as Hund's partners if they transform into each other under reversing the sign of the Hund's coupling, see, e.g., \figref{SchematicPDForIRE}. As singlet and triplet mix for both $\delta > 0$ and $\delta < 0$, there are no Hund's partners for $E^{3_s}(g)$ and $E^{3_s}(h)$; the corresponding mixed phases, contained in the last two lines of the table, are their own Hund's partners.}
\label{SummaryOfPairingComplexRepC3}
\begin{ruledtabular}
 \begin{tabular} {ccccc} 
   \exspace  Pairing \exspace    & \exspace $M_{\vec{k}+}$ \exspace   & \exspace Nodes \exspace & \exspace Hund's partner \exspace & \exspace \hspace{-1.2em} MF/FM \exspace  \\ \hline
$E^{1_s}(1,0)$ & $X_{\vec{k}}\sigma_0$   & 2 points       & $E^{3_s}(a)$ & \xmark/\cmark  \\
$E^{1_s}(1,i)$ & $(X_{\vec{k}}+ i\, Y_{\vec{k}})\sigma_0$   & 0       & $E^{3_s}(d)$ & \cmark/\cmark   \\ \hline 
$E^{3_s}(a)$ & $X_{\vec{k}}\sigma_x$   & 2 points & $E^{1_s}(1,0)$ & \xmark/\xmark   \\
$E^{3_s}(b)$ & $X_{\vec{k}}\sigma_x+Y_{\vec{k}}\sigma_y$   & 0 & $E^{1_s}(0,i) + E^{3_s}(a)$ & \cmark/\xmark   \\
$E^{3_s}(c)$ & $X_{\vec{k}}(\sigma_x + i\sigma_y)$   & $\downarrow$ gapless/2 points & $E^{1_s}(1,0) + E^{3_s}(a)$ & \xmark/\cmark   \\
$E^{3_s}(d)$ & $(X_{\vec{k}}+i \,Y_{\vec{k}})\sigma_x$   & 0 & $E^{1_s}(1,i)$ & \cmark/\cmark   \\
$E^{3_s}(e)$ & $(X_{\vec{k}}+i \,Y_{\vec{k}})(\sigma_x + i\sigma_y)$   & $\downarrow$ gapless/0 & $E^{1_s}(1,i) + E^{3_s}(d)$ & \xmark/\cmark   \\
$E^{3_s}(f)$ & $(X_{\vec{k}}+i \,Y_{\vec{k}})\sigma_x + (X_{\vec{k}}-i \,Y_{\vec{k}})\sigma_y$   & 2 points & $E^{1_s}(1,-i) + E^{3_s}(d)$ & \xmark/\xmark   \\ 
$E^{3_s}(g)$ & $ a(X_{\vec{k}} + i\, Y_{\vec{k}} ) (\sigma_x+i\sigma_y) + b (X_{\vec{k}} - i\, Y_{\vec{k}} ) \sigma_z$   & 0 & --- & \xmark/\xmark   \\ 
$E^{3_s}(h)$ & $ a[(X_{\vec{k}} + i\, Y_{\vec{k}}) \sigma_x + (X_{\vec{k}} - i\, Y_{\vec{k}}) \sigma_y] + b Y_{\vec{k}}  \sigma_z$   & 0 & --- & \xmark/\xmark   \\ \hline 
$E^{1_s}(0,i) + E^{3_s}(a)$ & $i\, Y_{\vec{k}} \sigma_0 + \eta X_{\vec{k}}\sigma_x$   & 0 & $E^{3_s}(b)$ & \cmark/\xmark   \\
$E^{1_s}(1,0) + E^{3_s}(a)$ & $X_{\vec{k}}( \sigma_0 + \eta \sigma_x)$   & 2 points  & $E^{3_s}(c)$ & \xmark/\cmark   \\
$E^{1_s}(1,i) + E^{3_s}(d)$ & $(X_{\vec{k}}+i \,Y_{\vec{k}})( \sigma_0 + \eta \sigma_x)$   & 0  & $E^{3_s}(e)$ & \xmark/\cmark  \\
 $E^{1_s}(1,-i) + E^{3_s}(d)$ & $(X_{\vec{k}}+i \,Y_{\vec{k}})\sigma_0 + \eta  (X_{\vec{k}}-i \,Y_{\vec{k}})\sigma_x$ & 2 points  & $E^{3_s}(f)$ & \xmark/\xmark \\
 $E^{3_s}(g) + E^{1_s}(1,-i)$ & $ a(X_{\vec{k}} + i\, Y_{\vec{k}} ) (\sigma_x+i\sigma_y) +   (X_{\vec{k}} - i\, Y_{\vec{k}} ) (b \sigma_z + \eta  \sigma_0)$ & 0  & $ E^{1_s}(1,-i) + E^{3_s}(g)$ & \xmark/\xmark  \\
 $E^{3_s}(h) + E^{1_s}(1,0)$ & $a[(X_{\vec{k}} + i\, Y_{\vec{k}}) \sigma_x + (X_{\vec{k}} - i\, Y_{\vec{k}}) \sigma_y] + b Y_{\vec{k}}  \sigma_z + \eta X_{\vec{k}}\sigma_0$ & 0  & $ E^{1_s}(1,0) + E^{3_s}(h)$ & \xmark/\xmark
 \end{tabular}
\end{ruledtabular}
\end{center}
\end{table*}

We find that, out of the eight different triplet states $E^{3_s}(a)$ to $E^{3_s}(h)$, only two---$E^{3_s}(a)$ and $E^{3_s}(d)$---do not allow for a singlet-triplet admixture when reversing the sign of $J$ (or $\delta a$) so that singlet has the higher transition temperature. The reason for the absence of an admixture is the same as sketched by way of example in \secref{NonZeroHundsCoupl}: besides pure singlet and pure triplet terms, the quartic terms in \equref{MatrixExpansionOfFreeEnergyComplRep} also contain couplings between singlet and triplet, as is readily seen by inserting the parametrization $\Delta_\mu = \sigma_0 \Delta^s_\mu + \vec{\sigma}\cdot \vec{d}_\mu$, $\mu=\pm$ (the full expansion can be found in \appref{DetailsOfComplexRepresentation}). At the first transition, one of either singlet or triplet becomes nonzero and hence, ``renormalizes'' the quadratic term of the other channel. In some cases, this renormalization can prohibit the presence of a second transition. In the case of phases $E^{3_s}(a)$ and $E^{3_s}(d)$, we just obtain the pure singlets $E^{1_s}(1,0)$ and $E^{1_s}(1,i)$, respectively, without a second transition. The easiest way to interpret why we do not have an admixture in these cases is to look at the associated SO(4) parent states:  the two triplets correspond to $(\eta_+;\vec{d}_+)=(\eta_-;\vec{d}_-)=(0;1,0,0)$ and $(\eta_+;\vec{d}_+)=(0;1,0,0)$, $(\eta_-;\vec{d}_-)=0$, respectively. Both of these configurations can be ``rotated'' into the pure singlets $(\eta_+;\vec{d}_+)=(\eta_-;\vec{d}_-)=(1;0,0,0)$ and $(\eta_+;\vec{d}_+)=(1;0,0,0)$, $(\eta_-;\vec{d}_-)=0$ via a $\text{SU}(2)_+ \times \text{SU}(2)_-$ transformation. 

For all other triplets, the Hund's partner is an admixed phase. Specifically, as regards $E^{3_s}(b)$ and $E^{3_s}(c)$, the Hund's partner is an admixture of a nematic singlet state and a nematic unitary triplet $E^{3_s}(a)$, with different relative phases and spatial orientations: for the former, the order parameter can be written as $i\, Y_{\vec{k}} \sigma_0 + \eta X_{\vec{k}}\sigma_x$, where $\eta$ describes the temperature-dependent strength of mixing, while it is $X_{\vec{k}}( \sigma_0 + \eta \sigma_x)$ for the latter. On any Fermi surface around one of the high-symmetry points $\Gamma$, $K$, or $K'$,  these two states have zero and two nodal points, respectively. Again, the form of the admixed state can be understood from the representation of the triplet state in terms of $(\eta_\mu;\vec{d}_\mu)$. For instance, we have $(\eta_+;\vec{d}_+)=(0;1,-i,0)$, $(\eta_-;\vec{d}_-)=(0;1,i,0)$ for $E^{3_s}(b)$, which is equivalent to $(\eta_+;\vec{d}_+)=(1;1,0,0)$, $(\eta_-;\vec{d}_-)=(-1;1,0,0)$ after applying an appropriate $\text{SU}(2)_+ \times \text{SU}(2)_-$ transformation.

Likewise, the Hund's partners of $E^{3_s}(e)$ and $E^{3_s}(f)$ are admixtures of a chiral singlet and a unitary triplet state with the same and opposite chirality, respectively. The associated order parameters can be written as $(X_{\vec{k}}+i \,Y_{\vec{k}})( \sigma_0 + \eta\, \sigma_x)$ and $(X_{\vec{k}}+i \,Y_{\vec{k}})\sigma_0 + \eta\,  (X_{\vec{k}}-i \,Y_{\vec{k}})\sigma_x$. While the first of the two states has two fully established gaps, given by $(1\pm \eta)\sqrt{X^2_{\vec{k}}+Y^2_{\vec{k}}}$ (with $\pm$ referring to the spin species), the other has two gaps, $|X_{\vec{k}}|$ and $|Y_{\vec{k}}|$, with distinct momentum dependencies; it, thus, exhibits two point nodes per Fermi surface which occur at different positions for the two spin species, similar to the associated triplet phase $E^{3_s}(f)$.

In general, admixing a singlet component at a second transition to a triplet state is less likely to occur as a singlet state has less options to ``adapt'' (the order parameter comprises two complex numbers for $E$) than a triplet state (for which, the order parameter comprises six complex numbers). While this is not possible for the one-dimensional representation $A$ (see \figref{SchematicPD}), the IR $E$ does allow for this scenario but only for the triplet states $E^{3_s}(g)$ and $E^{3_s}(h)$: for small $\delta a<0$, we find a second transition where an additional chiral (nematic) singlet component is admixed to $E^{3_s}(g)$ ($E^{3_s}(h)$). As both pure triplet states can be fully gapped, the same holds for the admixed phases. The admixture of the extra singlet component does not change the symmetries of $E^{3_s}(g)$ and $E^{3_s}(h)$ listed in \secref{SingeltTripletSepForComplexRep} above. Reversing the sign of $\delta a$ to small positive values, we obtain the same admixed phase. The only difference is that the first transition is a singlet transition into a chiral (nematic) phase and the secondary triplet $E^{3_s}(g)$ [$E^{3_s}(h)$] becomes nonzero at a lower transition temperature.

\begin{figure}[tb]
   \centering
    \includegraphics[width=\linewidth]{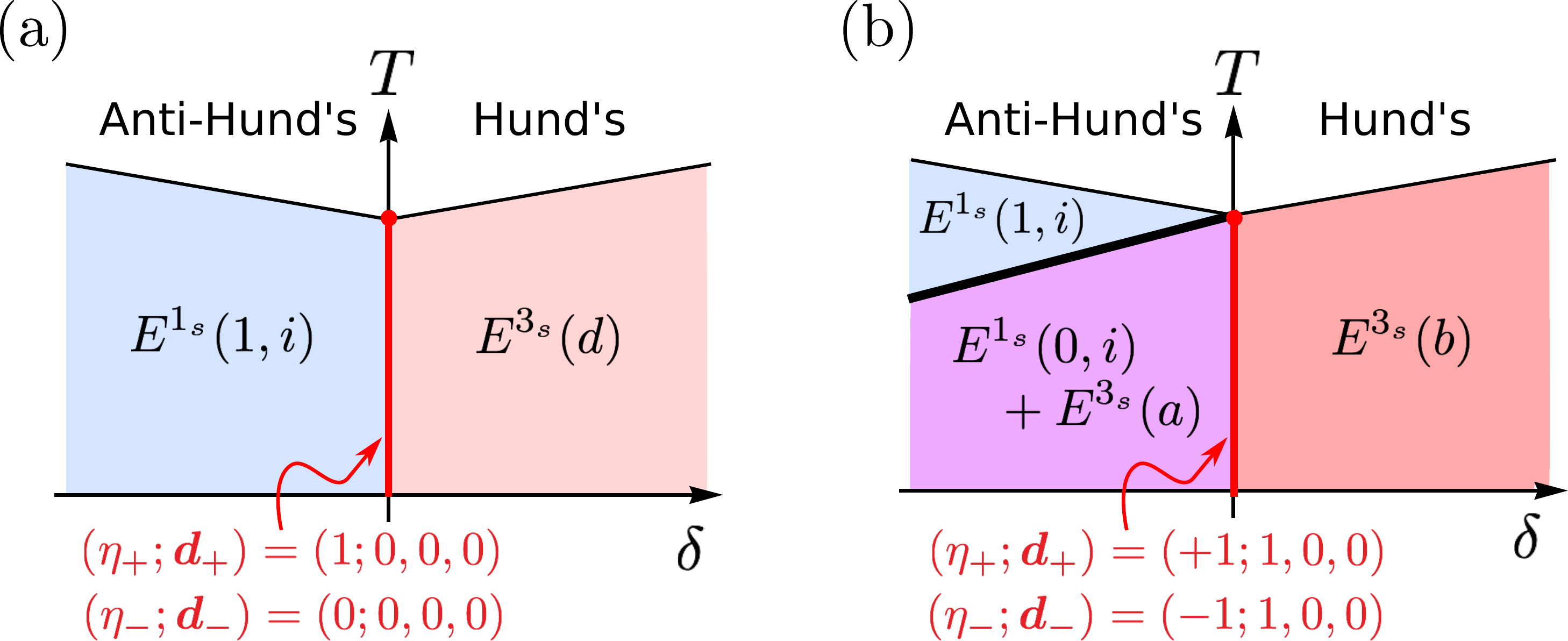}
    \caption{The two possible phase diagrams of the complex representation close to mean-field theory, using the labeling of states defined in the main text and \tableref{SummaryOfPairingComplexRepC3}. All transitions are second order, except for the one indicated by the thick line, which is first order. In \secref{FluctuationInducedSC}, we show that part (b) [part (a)] is favored when taking into account corrections to mean-field theory coming from ferromagnetic spin [orbital] fluctuations.}
    \label{SchematicPDForIRE}
\end{figure}

The key results of this section, the pure triplet/singlet states and the possible admixed phases for small $J$ along with their order parameters and properties, are summarized in \tableref{SummaryOfPairingComplexRepC3}. As already discussed above, several states are degenerate within single-band mean-field theory. Depending on the form of the corrections to mean-field theory lifting this degeneracy, there are two possible phase diagrams, shown in \figref{SchematicPDForIRE}. Interestingly, we observe that the chiral singlet, $E^{1_s}(1,i)$, is not the only possible phase close to mean-field theory for anti-Hund's coupling: as can be seen in \figref{SchematicPDForIRE}(b), a secondary phase transition into the nematic mixed singlet-triplet state $E^{1_s}(0,i) + E^{3_s}(a)$ is predicted. It is a fully gapped state with an anisotropic gap, $\sqrt{\eta ^2 X_{\vec{k}}^2 + Y_{\vec{k}}^2}$, breaking rotational symmetry. Note that this route to a nematic superconducting state, indications of which are provided by recent experiments \cite{2020arXiv200404148C}, is distinct from that of other works \cite{LiangFu,2019arXiv191007379C}. 
Of course, sufficiently large corrections to mean-field theory can in principle yield any of the phases listed in \tableref{SummaryOfPairingComplexRepC3}; we will come back to these corrections in \secref{FluctuationInducedSC} below.

Let us finally discuss the impact of   fluctuations of the order parameter on the thermal phase transitions. As readily follows from the respective order parameter manifolds, the singlet phases in \tableref{SummaryOfPairingComplexRepC3} exhibit a conventional BKT transition, the triplets (a), (b), (d), (f), (g), and (h) will be charge-$4e$ superconductors where only spin-rotation invariant combinations of the triplet vector assume quasi-long-range order at finite temperature, and the triplets (c) and (e) will only display a crossover. However, as pointed out above, none of these three classes of transitions can currently be excluded based on the experimental data.

\subsection{Behavior in a magnetic field}\label{ComplexRepInMagnField}
Finally, we turn our attention to the behavior of the pairing states of the complex representation in the presence of a Zeeman field, $\vec{M}_Z$, and in-plane orbital coupling $\vec{M}_O$, along the same lines as \secref{MagneticField}. 
From \equsref{TrafoComplexSinglet}{TrafoOf2DTriplet}, it follows that there are three possible coupling terms linear in the field and quadratic in the superconducting order parameter given by 
\begin{alignat}{1}
    \nonumber\Delta\mathcal{F}_{M}^{E} &\sim  \vec{M}_Z \cdot\sum_\mu \left[\delta c_1^E\, \text{Im}\left(\vec{d}_\mu^*\eta^\pdagger_\mu\right) +   c_2^E \mu\,  \text{Re}\left(\vec{d}_\mu^*\eta^\pdagger_\mu\right)\right] \\ 
    & + ic_3^E \vec{M}_Z \cdot\sum_\mu  \vec{d}_\mu^*\times \vec{d}^\pdagger_\mu.
\label{CouplingInMagnFieldCompl}\end{alignat}
Notice that, exactly as for the IR $A$, there is no linear coupling to the in-plane orbital field, which is prohibited by time-reversal and $C_3$ rotation symmetry. While the first term in \equref{CouplingInMagnFieldCompl} is again forced to vanish for $J\rightarrow 0$ [for the same reason as $\delta c_1$ in \equref{GinzburgLandauExpansion}], the second singlet-triplet-mixing coupling, $c_2^E$, is not constrained to be zero for $J=0$. However, the emergent symmetry in the single-band mean-field description of \appref{MicrosCopGLExpansion}, leads to $c_2^E=0$, so it is natural to expect $c_2^E \ll c_3^E$ such that the last term in \equref{CouplingInMagnFieldCompl} describes the dominant linear coupling to the magnetic field---even when $J$ is small. As expounded in \appref{MicrosCopGLExpansion}, the expression for $c_3^E$ is identical in form to that for $c_2$ in \equref{GinzburgLandauExpansion}. As such, the linear increase of the (first) superconducting transition temperature with small magnetic fields seen in experiment does not permit one to distinguish between the IRs $A$ and $E$.

There is one difference between the pairing states of the two IRs worth mentioning here: while the form of the leading triplet vector in a magnetic field is completely fixed to be $\vec{d} \propto (1,i,0)^T$ for the one-dimensional IR $A$, the complex IR allows for either the nematic nonunitary $E^{3_s}(c)$ or the chiral nonunitary $E^{3_s}(e)$ pairing for nonzero $\vec{M}_Z$. Which of the two is realized, depends on the value of the quartic terms in \equref{GLExpansionTripletComplexIR}: if $b_2^t + b_3^t > 0$, the state $E^{3_s}(e)$ will be preferred while the opposite sign corresponds to $E^{3_s}(c)$. Within single-band mean-field theory, we find $b_2^t=0$ and $b_3^t>0$, which leads to phase $E^{3_s}(e)$. In the next section, we will see that additional ferromagnetic fluctuations will further enlarge the positive value of $b_2^t + b_3^t$ and consequently, not affect the mean-field prediction that $E^{3_s}(e)$ is the leading triplet state with the highest transition temperature in the presence of a magnetic field.

\section{Fluctuation-induced superconductivity}\label{FluctuationInducedSC}
Among the plethora of possible superconducting phases outlined in this paper, only a few can be realized in single-band mean-field theory (see Tables~\ref{TableSummaryOfPairingC3}, \ref{SummaryOfPairingComplexRepC3}, and \ref{SummaryOfPairingD3}). This originates from the fact that, within single-band mean-field theory, the ratio of the quartic terms is fixed and only one state or two degenerate states can occur for each IR. However, the presence of sizable correlations in the nearly flat bands of graphene moir\'e systems is expected to give rise to significant corrections to mean-field theory. This has recently been demonstrated for the case of charge-density-wave fluctuations in twisted bilayer graphene \cite{LiangFu}, and in the context of nematic fluctuations in the iron-based superconductors \cite{RafaelsNematicFlucs}. In this section, we study how corrections associated with ferromagnetic fluctuations will split the mean-field degeneracies and, if sufficiently strong, realize phases distinct from mean-field theory.

To this end, we will first focus on spin fluctuations. This is prompted by experiments \cite{2019arXiv190306952S,ExperimentKim,PabllosExperiment}, which indicate a spin-polarized correlated insulating state in twisted double-bilayer graphene, and by the fact that the superconducting phase emerges when doping out of this polarized state. Likewise, we also expect ferromagnetic fluctuations to play an important role in twisted bilayer \cite{Sharpe605,2019arXiv190306513L,IlaniExperiment} and trilayer graphene \cite{FMTrilayer}. In particular, in the latter two systems, however, these fluctuations will likely not only be of spin but also of orbital origin. This is why we will also discuss orbital fluctuations.

As it is known to capture the essential physics \cite{LiangFu,RafaelsNematicFlucs}, we focus in the main text on a phenomenological Ginzburg-Landau-like approach (that does not explicitly take into account fluctuations with nonzero momentum and frequencies), but provide a systematic microscopic derivation in \appref{MicroscopicDerivationFluctuations}. 
Representing the ferromagnetic spin moment in valley $v=\pm$ by $\vec{m}_v$, we parametrize its contribution to the free energy as
\begin{equation}
    \mathcal{F}_{m} = \frac{1}{2} \sum_{v,v'} \left(\hat{\chi}^{-1}\right)_{vv'} \vec{m}^{}_v\cdot \vec{m}^{}_{v'}, \quad \hat{\chi} = \begin{pmatrix} \chi & \delta \chi \\ \delta \chi & \chi  \end{pmatrix}. \label{parametrizationOfFluctsMagn}
\end{equation}
In this expression, $\hat{\chi}$ plays the role of the spin susceptibility (with $|\delta \chi| < \chi$ to ensure stability) and we expect $\delta \chi > 0$ close to a phase where the spin moments in the two valleys are aligned. The ratio $\delta \chi /\chi$ controls how strongly the SU(2)$_+ \times$ SU(2)$_-$ symmetry is broken down to SU(2)$_s$.

\subsection{Trivial representation}
Focusing first on the one-dimensional IR $A$ of $C_3$, the magnetic moments couple to the superconducting order parameter in \secref{TrivialRepresentation} according to
\begin{equation}
    \mathcal{F}^A_{m\Delta} = c^{}_2 \sum_{v=\pm} \vec{m}^{}_v \cdot \left[i\,\vec{d}^*\times \vec{d} -2 v\, \text{Re}(\vec{d}^*\Delta^s) \right], \label{CouplingToFluctuations}
\end{equation}
where we have retained only the couplings invariant under SU(2)$_+ \times$ SU(2)$_-$ and assumed that $\delta \chi \neq 0$ in \equref{parametrizationOfFluctsMagn} is the main symmetry-breaking perturbation. Upon making the association $\vec{M}_Z = \sum_v \vec{m}_v$, we notice that $c_2$ is the same prefactor as in \equref{GinzburgLandauExpansion}. 
In the same vein as \refcite{LiangFu}, we integrate out the massive fluctuations of $\vec{m}_v$. As a consequence of the coupling (\ref{CouplingToFluctuations}), this yields corrections to the terms quartic in the superconducting order parameters in \equref{FullExpansionWithHigherSym}, which can be conveniently split into two categories. First, there are corrections that preserve the SU(2)$_+ \times$ SU(2)$_-$ symmetry; these can be restated as renormalizations of the coefficients $b_1$ and $b_2$ in \equref{FullExpansionWithHigherSym}. Corrections of the second type break this symmetry, violating the form of the free-energy expansion (\ref{FullExpansionWithHigherSym}). More explicitly, the renormalization of the free energy $\mathcal{F}$ in \equref{FullExpansionWithHigherSym} due to the presence of ferromagnetic spin fluctuations can be compactly stated as
\begin{equation}
    \mathcal{F}\, \rightarrow \, \left.\mathcal{F}\right|_{b_j\rightarrow b_j + \delta_j} - \delta_3 |\vec{d}^*\times \vec{d}|^2, \label{RenormalizationDescription}
\end{equation}
where $\delta_{1}$\,$=$\,$-\delta_2$\,$=$\,$2c_2^2 (\chi - \delta \chi)$\,$>$\,$0$ and $\delta_{3}$\,$=$\,$2  c_2^2 \delta\chi$. As required by symmetry, the contribution $\delta_3$ of the second category breaking the SU(2)$_+ \times$ SU(2)$_-$ symmetry is proportional to $\delta \chi$. 

We start with the limit $|\delta \chi| \ll \chi$, where the structure of \equref{FullExpansionWithHigherSym} is asymptotically preserved and the form of the two possible phase diagrams in \figref{SchematicPD} is unchanged. Since $\delta_2 < 0$, strong ferromagnetic fluctuations will  change the sign of $b_2$ from its positive mean-field value to negative and, as opposed to mean-field theory, favor the phase diagram in part (b) of \figref{SchematicPD} over part (a). We point out that naively taking \equref{RenormalizationDescription} alone would render the quartic free-energy expansion unstable for large enough $\chi$. However, denoting the mean-field value of $b_2$ by $b_2^0$, there exists a regime, $b_2^0/2 < c_s^2 \chi < b^0_2$, for which $b_2 < 0$ due to fluctuation corrections and the free energy in \equref{FullExpansionWithHigherSym} is stable. For larger values of $\chi$, we can imagine adding the sextic term $c(\text{tr}[ \Delta_{+}^\dagger \Delta_{+}^\pdagger])^3$ to the free energy to restore stability. 

When $\delta \chi$ is of order $\chi$, the ferromagnetic fluctuations described by \equref{parametrizationOfFluctsMagn} induce considerable SU(2)$_+ \times$ SU(2)$_-$-symmetry-breaking interactions. The presumed sign $\delta \chi > 0$ brings about a further enhancement of the term $-|\vec{d}^*\times \vec{d}|^2$ [as is obvious from \equref{RenormalizationDescription}], which favors nonunitary triplet pairing relative to the SU(2)$_+ \times$ SU(2)$_-$-invariant form of the free energy in \equref{FullExpansionWithHigherSym}. Given that $\delta \chi < \chi$, strong ferromagnetic fluctuations are still expected to change the sign of $b_2$ relative to mean-field theory. The additional effect of $\delta \chi$ lies in effecting an additional first-order transition to a nonunitary triplet state in a third transition at lower temperatures for anti-Hund's coupling in \figref{SchematicPD}(b).   

We have thus shown that significant ferromagnetic fluctuations can reverse the predictions of mean-field theory, and favor the nonunitary triplet state $A^{3_s}(1,i,0)$ and the admixed singlet-triplet phase $A^{1_s}+A^{3_s}(1,0,0)$ in \tableref{TableSummaryOfPairingC3}.

\subsection{Complex representation}
The same analysis can be performed for the complex IR $E$ of \secref{ComplexRepresentation}. In this case, the most general SU(2)$_+ \times$ SU(2)$_-$-invariant coupling between the superconducting order parameter and the spin fluctuations allows for two independent coupling constants, $c_\pm$\,$\in$\,$\mathbb{R}$, and has the form
\begin{equation}
    \mathcal{F}^E_{m\Delta} = \sum_{\mu=\pm} \sum_{v=\pm} c^{}_{v\cdot \mu}\, \vec{m}^{}_v \cdot \left[i\,\vec{d}_\mu^*\times \vec{d}^{}_\mu -2 v\, \text{Re}(\vec{d}_\mu^*\eta^{}_\mu) \right].
\label{CouplingmvForE}\end{equation}
Integrating out $\vec{m}_v$, we again obtain corrections to the free energy which are quartic in the superconducting order parameter. In the limit of SU(2)$_+ \times$ SU(2)$_-$ invariance, $\delta \chi = 0$, these corrections can be represented by renormalizations of the couplings, $b_j \rightarrow b_j + \delta b_j$, in \equref{MatrixExpansionOfFreeEnergyComplRep} with
\begin{align}\begin{split}
    \delta b_1 &= -\delta b_2 = \chi\, (c_+^2 + c_-^2)/2 >0, \\
    \delta b_3 &= -\chi\, (c_+-c_-)^2 < 0, \\
    \delta b_4 &= 0, \\
    \delta b_5 &= -\chi\, c_+ c_-.
\end{split}\label{RenormalizationForIRE}\end{align}
To study the ramifications of this result, we first consider the limit of weak fluctuations, for which $\delta b_j$ in \equref{RenormalizationForIRE} are much smaller in magnitude than the mean-field value of $b_2$. Albeit small, the corrections $\delta b_j$ are crucial here due to the exact degeneracy of the states $E^{3_s}(b)$ and $E^{3_s}(d)$ in mean-field theory observed earlier. From \equref{MatrixExpansionOfFreeEnergyComplRep} with the replacement $b_j \rightarrow b_j + \delta b_j$, we find the free-energy difference of these two states to be
\begin{equation}
    \mathcal{F}_{E^{3_s}(b)} - \mathcal{F}_{E^{3_s}(d)} = -\frac{1}{4}\chi (c_+ - c_-)^2 \left(\vec{d}^\dagger_\mu \vec{d}^\pdagger_\mu\right)^2 \leq 0, \label{SplittingDueToFluc}
\end{equation}
thereby generically favoring $E^{3_s}(b)$ along with its Hund's partner $E^{1_s}(0,i) + E^{3_s}(a)$, defined in \tableref{SummaryOfPairingComplexRepC3}; in other words, the phase diagram in \figref{SchematicPDForIRE}(b) is favored over that in part (a). In the one-band description of \appref{EnhancedSymmetryFluctuations}, it always holds that $c_+ = c_-$, which is, in turn, a consequence of an emergent valley-exchange symmetry. However, multiband effects  are expected to be present \cite{YazdaniSTM} and to lead to nonzero $|c_+ - c_-| \ll |c_+|$, which is enough to lift the degeneracy according to \equref{SplittingDueToFluc}. 

Next, we turn to the limit of strong ferromagnetic fluctuations, where the mean-field values of $b_j$ have to be treated as perturbations to the large $\delta b_j$ in \equref{RenormalizationForIRE}. As $\chi\rightarrow \infty$, we find that, out of the triplet states in \tableref{SummaryOfPairingComplexRepC3}, $E^{3_s}(e)$ has the lowest energy unless $c_+ = c_-$ or $c_+ = -c_-$. We know that $c_+ \approx c_-$ and hence, can safely neglect the latter. For the former option, $E^{3_s}(e)$ is found to be degenerate with $E^{3_s}(c)$; however, for large but finite $\chi$, the additional contribution to $b_j$ from mean-field theory lifts this degeneracy, always selecting $E^{3_s}(e)$. Out of the multitude of possible pairing states in \tableref{SummaryOfPairingComplexRepC3}, strong ferromagnetic fluctuations thus favor the chiral nonunitary triplet state $E^{3_s}(e)$ and the mixed singlet-triplet phase $E^{1_s}(1,i) + E^{3_s}(d)$. Which of these two states is realized, depends on whether singlet or triplet has the higher transition temperature (the sign of $\delta a$). 

\subsection{Orbital fluctuations}
Anticipating its relevance for twisted bilayer and trilayer graphene, here, we extend the previous analysis to the case of orbital ferromagnetic fluctuations. Due to the two-dimensional nature of the system, the in-plane orbital moments, $\vec{M}_O=(M_O^x,M_O^y)$, and the out-of-plane moment $M_O^z$ behave quite differently. Beginning with the complex representation, we already know from \secref{ComplexRepInMagnField} that there is no linear coupling to $\vec{M}_O$; however, the superconductor can couple to $M_O^z$ as
\begin{equation}
    \mathcal{F}^E_{M\Delta} = c_O^E \sum_{\mu=\pm} \mu \, M_O^z \left( |\eta_\mu|^2 + \vec{d}^\dagger_\mu \vec{d}^\pdagger_\mu  \right). \label{OrbitalCouplingFluc}
\end{equation}
For concreteness, one might think of $M_O^z$ as valley fluctuations, associated with $\sum_{\vec{k}}c_{\vec{k}}^\dagger \tau_{z} c_{\vec{k}}^\pdagger$, but our analysis is more general. Taking an energetic contribution quadratic in $M_O^z$ similar to \equref{parametrizationOfFluctsMagn} and integrating over $M_O^z$, we obtain a correction to the free-energy that can be conveniently expressed as
\begin{equation}
    b_1\rightarrow b_1 -\delta b,\quad  b_3 \rightarrow b_3 + 4\delta b,\quad \delta b > 0, \label{OrbitalFluctuationCorr}
\end{equation}
in \equref{MatrixExpansionOfFreeEnergyComplRep}. It is easily seen that taking this as a small correction to mean-field theory will now favor the phase diagram in \figref{SchematicPDForIRE}(a) over that in part (b). On the other hand, in the limit of strong orbital fluctuations, the chiral unitary, $E^{3_s}(d)$, and the chiral nonunitary triplet, $E^{3_s}(e)$, (along with their Hund's partners) will be favored. This degeneracy will be lifted by the subleading ferromagnetic spin fluctuations, which favor the latter state, $E^{3_s}(e)$ (and its Hund's partner), as readily follows from \equref{RenormalizationForIRE}. 

In the trivial representation, orbital fluctuations have no impact on which of the two possible phase diagrams in \figref{SchematicPD} is realized. This results from the fact that neither $\vec{M}_O$ (see \secref{MagneticField}) nor $M_O^z$ can couple linearly to the superconducting states and their rotational invariant quadratic forms $(M_O^z)^2$, $\vec{M}_O^2$ can only couple to $|\Delta^s|^2 + \vec{d}^\dagger\vec{d}$. Consequently, the energetic correction obtained by integrating out the orbital fluctuations will also only depend via $|\Delta^s|^2 + \vec{d}^\dagger\vec{d}$ on the superconducting states and, as such, not affect the value of $b_2$ in \equref{FullExpansionWithHigherSym} and \figref{SchematicPD}.

\subsection{In a magnetic field}
Finally, we come back to the impact of fluctuation corrections on the leading triplet phase in the presence of a magnetic field. As we have seen in \secref{ComplexRepInMagnField}, the superconducting state with the highest transition temperature in the presence of a sufficiently strong magnetic field will be a triplet phase due to the linear coupling in the second line of \equref{CouplingInMagnFieldCompl}. At the mean-field level, $b_2^t + b_3^t > 0$, which prefers $E^{3_s}(e)$ over $E^{3_s}(c)$ as the order parameter of this phase. Using the relations in \equref{DifferentCouplingsBetaB}, it is straightforward to rephrase the fluctuation corrections (\ref{RenormalizationForIRE}) and (\ref{OrbitalFluctuationCorr}) of $b_j$ in terms of $b_j^t \rightarrow b_j^t + \delta b_j^t $ in \equref{GLExpansionTripletComplexIR}. This yields $\delta b^t_1 + \delta b^t_2 = \chi (c_+ - c_-)^2 >0$ and $\delta b^t_1 + \delta b^t_2 = 4\delta b >0$ for spin and orbital fluctuations, respectively. We conclude that, as expected, ferromagnetic fluctuations do not change the mean-field prediction in this case and $E^{3_s}(e)$ is the dominant triplet order parameter in the presence of a magnetic field, for both strong and weak ferromagnetic fluctuations, and in their absence.

\section{Adding further symmetries}\label{ComparisonWithOtherSystems}
In this section, we will analyze how the results presented above are modified once the additional symmetries, twofold rotation, $C_2$, perpendicular to the plane of the system, and in-plane rotation symmetry, $C_{2y}$, are added. As shown in \figref{LatticesAndSymmetries}(b) and (c), these symmetries are relevant as either exact microscopic or approximate emergent symmetries of twisted bilayer graphene and ABC trilayer graphene on hexagonal boron nitride, both of which exhibit superconductivity \cite{SupercondTBGExp,SuperconductivityInTrilayer}. 
 
\subsection{Consequences of a $C_2$ rotation symmetry}
\label{TwistedBilayerGraphene}
One crucial difference in twisted bilayer compared to twisted double-bilayer graphene is that the former has an approximate $C_2$ symmetry \cite{AshvinModelTBG} that mixes the two valleys, \textit{i.e.}, the system is (approximately) invariant under 
\begin{equation}
    C_2:\quad c^{}_{\vec{k}} \quad\longrightarrow \quad  \tau^{}_x c^{}_{-\vec{k}}. \label{C2Symmetry}
\end{equation}
To relate to our notation used above, we assume that it is sufficient to focus on a single band for describing superconductivity in twisted bilayer graphene as well. This is quite a natural assumption and, unless stated otherwise, we expect our conclusions to hold when additional bands are taken into consideration.

This (approximate) symmetry has attracted a lot of attention in the recent theory literature \cite{VafekModel,FuModel1,KoshinoFuModel,EmergentSymmetries} of the system since it, combined with time-reversal and $C_3$, leads to a $C_6\Theta$ symmetry, which is responsible for not only the presence of (nearly gapless) Dirac cones at $K$ and $K'$ but also the (approximate) vanishing of Berry curvature in twisted bilayer graphene. If the twist axis goes through the center of a hexagon, the system has $C_6$ rotation even as a microscopic symmetry. We note in passing that the (nearly) flat bands obtained in \refscite{2019arXiv190308685L,KoshinosPaper} for double-bilayer graphene do not feature any Dirac cones but have well-separated conduction and valence bands that are characterized by nonzero Chern numbers (at least in some parameter regime); this strongly indicates that $C_2$ is not an approximate symmetry in twisted double-bilayer graphene since $C_2\Theta$ would enforce zero Berry curvature.

In a similar fashion, \refcite{SenthilABCTrilayer} has argued that the twofold symmetry (\ref{C2Symmetry}) is also an approximate symmetry for ABC trilayer graphene on hexagonal boron nitride, although it is clearly not a microscopic symmetry of the system, as can be seen in \figref{LatticesAndSymmetries}(c).

All things considered, it is currently not known whether an approximate $C_2$ symmetry is relevant for superconductivity in twisted bilayer and ABC trilayer graphene. Therefore, we will now discuss what changes for the possible superconducting instabilities once we assume that the Hamiltonian is also invariant under the transformation in \equref{C2Symmetry}. 

The $C_2$ transformation plays a special role in two dimensions as it is equivalent to $\vec{k} \rightarrow -\vec{k}$ and can, thus, significantly affect superconducting instabilities \cite{scheurer2017selection}. In graphene moir\'e superlattices, it also relates the two valleys and  ``interferes'' with the Fermi-Dirac constraint (\ref{FermiDiracStatConstr}): decomposing the pairing into singlet and triplet,
\begin{equation}
    M^{}_{\vec{k}v} =  \lambda^s_{\vec{k}v} \sigma^{}_0 \Delta^s + \lambda^t_{\vec{k}v} \vec{\sigma} \cdot \vec{d}, \label{IntervalleyPairingOrderParameter}
\end{equation}
\equref{FermiDiracStatConstr} implies that $\lambda^s_{\vec{k}v}$\,$=$\,$\lambda^s_{-\vec{k}\bar{v}}$ and $\lambda^t_{\vec{k}v}$\,$=$\,$ -\lambda^t_{-\vec{k}\bar{v}}$. Consequently, it holds (as long as the pairing matrix elements between different bands can be neglected) that
\begin{equation}
    C_2: \quad (\Delta^s,\vec{d}) \quad \longrightarrow \quad (\Delta^s,-\vec{d}), \label{C2Representation}
\end{equation}
\textit{i.e.}, all representations even (odd) in $C_2$ must be pure singlet (triplet) states and vice versa. This has a few implications worth mentioning. First, even if $C_2$ is not a good symmetry (say, it is significantly broken by interactions), SU(2)$_s$ spin-rotation invariance requires that the first transition must be into a pure singlet or triplet state and hence, the pairing must be either even or odd under $C_2$. In this sense, we can still distinguish between $p$-wave and $d$-wave pairing despite the presence of $C_2$-symmetry-breaking interactions. We emphasize that mixing will only be possible via multiple superconducting transitions (associated with admixtures of singlet and triplet) or interband pairing. The latter is expected to be quite weak given that the typical splitting between the bands at half-filling (at least a few $\textrm{meV}$ \cite{PasupathySTM}) is about or more than an order of magnitude larger than the superconducting critical temperature ($\approx$ $0.15\,\textrm{meV}$ according to \refcite{SupercondTBGExp}).  

Secondly, if we do have an enhanced $\text{SU}(2)_+ \times \text{SU}(2)_-$ symmetry (or are close to it), singlet and triplet are (nearly) degenerate. This forces the corresponding IRs of the spatial point group $D_6$ of the system, which behave identically under the subgroup $D_3$ but are even and odd under $C_2$, to be (nearly) degenerate at the quadratic level of the Ginzburg-Landau expansion. For instance, $A_1$ and $B_1$ of $D_6$ have to be degenerate, as summarized in \tableref{CharacterTableD3}. Without a Zeeman field, an extra $C_2$ symmetry with action in \equref{C2Representation} also has no consequences for the higher-order terms in the free energy since spin-rotation invariance necessitates that all of these terms are even in the triplet vector. 
The only difference arises in the presence of a Zeeman field or magnetic fluctuations: with $C_2$ symmetry, it must hold that $\delta c_1 = 0$ in \equref{GinzburgLandauExpansion} and $\delta c_1^E=0$ in \equref{CouplingInMagnFieldCompl} even when the SU(2)$_+ \times$ SU(2)$_-$ symmetry is broken. Furthermore, a $C_2$ symmetry implies $c_+ = c_-$ in \equref{CouplingmvForE}. For this reason, weak ferromagnetic spin fluctuations do not lift the degeneracy of mean-field theory if we impose an \textit{exact} $C_2$ symmetry and other types of fluctuations have to be considered. Recall, however, in both trilayer and twisted bilayer graphene, $C_2$ should only be considered as an \textit{approximate} symmetry and terms breaking this symmetry will lead to $c_+\neq c_-$, thus lifting the degeneracy of mean-field theory by, e.g., favoring \figref{SchematicPDForIRE}(b) over (a). Note that the $C_2$ symmetry also forces $c_O^E$ in \equref{OrbitalCouplingFluc} to vanish for $M_O^z$ corresponding to valley fluctuations. As such, the approximate $C_2$ symmetry does not specify whether spin or valley fluctuations are expected to be the dominant source of lifting the mean-field degeneracy. It only indicates that strong ferromagnetic fluctuations are most likely dominated by spin fluctuations. 

\begin{table}[tb]
\begin{center}
 \caption{\label{CharacterTableD3}Character table of the point group $D_{3}$ together with the corresponding basis functions and IRs of $D_6$ for singlet/triplet pairing.}
\begin{ruledtabular}
 \begin{tabular} {cccccc} 
       & $E$    & $2C_3$  & $3C_{2y}$ & Basis functions & IRs of $D_6$\\ \hline
$A_{1}$ & $1$   & $1$  &  $1$      & $x^2+y^2$/$y(3x^2-y^2)$ & $A_1$/$B_1$  \\
$A_{2}$ & $1$   & $1$  &  $-1$     & $z$/$x(x^2-3y^2)$ & $A_2$/$B_2$ \\ \hline
$E$     & $2$   & $-1$ &  $0$      & $(2xy,x^2-y^2)$/$(x,y)$ & $E_2$/$E_1$ \\ 
 \end{tabular}
 \end{ruledtabular}
\end{center}
\end{table}

In summary, when classifying superconducting states in twisted bilayer graphene or ABC trilayer graphene on hexagonal boron nitride in the absence of a Zeeman field, it is unimportant whether an approximate $C_2$ symmetry is relevant or not: singlet and triplet will always be even and odd under it. We can thus work with $D_3$ (instead of $D_6$) without loss of generality in the following. The only difference with twisted double-bilayer graphene (with finite displacement field) is an extra twofold rotation symmetry, $C_{2y}$, along the $y$-axes, see \figref{LatticesAndSymmetries}(b) and (c). Its action on the electronic operators reads as
\begin{equation}
    C^{}_{2y}:\quad c^{}_{\vec{k}} \quad\longrightarrow \quad  \tau^{}_x\, c^{}_{C_{2y}\vec{k}}, \label{C2ySymmetry}
\end{equation}
where $C_{2y}\vec{k}=(-k_x,k_y)$.
The upshot of this additional symmetry for the possible superconducting instabilities is clarified in the next subsection.

\subsection{$D_3$ versus $C_3$}
Due to the additional $C_{2y}$ symmetry, $D_3$ is a non-Abelian group and has three IRs---two one-dimensional and one two-dimensional representation (refer to the character table in \tableref{CharacterTableD3}).
It is convenient to begin with the one-dimensional IRs $A_1$ and $A_2$ and take $J\neq 0$. Since $C_{2y}$ interchanges the valleys, its action on the intervalley pairing order parameter (\ref{IntervalleyPairingOrderParameter}) can be written as 
\begin{equation}
    C^{}_{2y}: \quad \left(\lambda^s_{\vec{k}v},\lambda^t_{\vec{k}v}\right) \, \longrightarrow \, \left(\lambda^s_{-C_{2y}\vec{k}v},-\lambda^t_{-C_{2y}\vec{k}v}\right). \label{TrafoUnderC2y}
\end{equation}
So, we see that a singlet (triplet) state transforming under $A_1$ ($A_2$) has no nodes while a singlet (triplet) in the $A_2$ ($A_1$) channel has symmetry-imposed nodes on the line $k_y=0$ and along the directions rotated by $\pm \pi/3$. This creates six nodal points on any surface enclosing the $\Gamma$ point.

\begin{table*}[tb]
\begin{center}
\caption{Summary of the different intervalley pairing states classified by the IRs of the point group $D_3$. The notation closely parallels that of  \tableref{SummaryOfPairingComplexRepC3}. Here, we use $\lambda^1_{\vec{k}}$ and $\lambda^2_{\vec{k}}$ to denote continuous functions on the Brillouin zone that are even and odd under $(k_x,k_y) \rightarrow (k_x,-k_y)$, respectively, and are both invariant under $C_3$, $\lambda^j_{\vec{k}}=\lambda^j_{C_3\vec{k}}$. Furthermore, $X^\varphi_{\vec{k}}$ and $Y^\varphi_{\vec{k}}$ are rotated basis functions defined in \equref{TheRotatedBasisFuncs}; for instance, a possible choice for twisted bilayer graphene with Brillouin zone in \figref{LatticesAndSymmetries}(b) is given by $(X_{\vec{k}},Y_{\vec{k}})^T = R_{(\pi+\varphi)/2} (X^{(1)}_{\vec{k}},Y^{(1)}_{\vec{k}})^T$ with $X^{(1)}_{\vec{k}}$, $Y^{(1)}_{\vec{k}}$ in \equref{parametrizationOfBasisFuncs}. To keep the notation short, each line with reference to $\varphi_1$ or $\varphi_2$ corresponds to two distinct states with $\varphi_1=0,\pi/3$ and $\varphi_2=0,\pi/2$. The indicated number of nodal points refers to a Fermi surface enclosing the $\Gamma$ point.}
\label{SummaryOfPairingD3}
\begin{ruledtabular}
 \begin{tabular} {cccccc} 
     Pairing     &  $M_{\vec{k}+}$   &  Nodes around $\Gamma$    &  Hund's partner  &  MF/FM   \\ \hline
$A_1^{1_s}$ & $\lambda_{\vec{k}}^1\sigma_0$   & none       & $A^{3_s}_2(1,0,0)$ & \cmark/\cmark  \\
$A^{1_s}_2$ & $\lambda_{\vec{k}}^2\sigma_0$   & 6 points        & $A^{3_s}_1(1,0,0)$ & \cmark/\cmark  \\ \hline
$A^{3_s}_1(1,0,0)$ & $\lambda_{\vec{k}}^2\sigma_x$   & 6 points       & $A_2^{1_s}$ & \cmark/\xmark  \\
$A^{3_s}_2(1,0,0)$ & $\lambda_{\vec{k}}^1\sigma_x$   & none        & $A_1^{1_s}$ & \cmark/\xmark \\ $A^{3_s}_1(1,i,0)$ & $\lambda_{\vec{k}}^2(\sigma_x+i\sigma_y)$   & \hspace{-2em}$\downarrow$ gapless/6 points  &        $A^{1_s}_2 + A^{3_s}_1(1,0,0)$ & \xmark/\cmark \\
$A^{3_s}_2(1,i,0)$ & $\lambda_{\vec{k}}^1(\sigma_x+i\sigma_y)$   & \hspace{-2em}$\downarrow$ gapless/none        & $A^{1_s}_1 + A^{3_s}_2(1,0,0)$ & \xmark/\cmark \\ \hline
$A^{1_s}_1 + A^{3_s}_2(1,0,0)$ & $\lambda_{\vec{k}}^1(\sigma_0 + \eta(\sigma_x+i\sigma_y))$   & none     & $A^{3_s}_2(1,i,0)$ & \xmark/\cmark  \\
$A^{1_s}_2 + A^{3_s}_1(1,0,0)$ &  $\lambda_{\vec{k}}^2(\sigma_0 + \eta(\sigma_x+i\sigma_y))$  & 6 points  & $A^{3_s}_1(1,i,0)$ & \xmark/\cmark  \\ \hline
$E^{1_s}(1,0)_{\varphi_1}$ & $X^{\varphi_1}_{\vec{k}}\sigma_0$   & 2 points       & $E^{3_s}(a)_{\varphi_1}$ & \xmark/\cmark  \\
$E^{1_s}(1,i)$ & $(X^{0}_{\vec{k}}+i\,Y^{0}_{\vec{k}})\sigma_0$   & 0       & $E^{3_s}(d)$ & \cmark/\cmark  \\ \hline
$E^{3_s}(a)_{\varphi_1}$ & $X^{\varphi_1}_{\vec{k}}\sigma_x$   & 2 points & $E^{1_s}(1,0)_{\varphi_1}$ & \xmark/\xmark    \\
$E^{3_s}(b)$ & $X^0_{\vec{k}}\sigma_x+Y^0_{\vec{k}}\sigma_y$   & 0 & $(E^{1_s}(0,i)+ E^{3_s}(a))_{\varphi_1}$ \hspace{-1em} & \cmark/\xmark   \\
$E^{3_s}(c)_{\varphi_1}$ & $X^{\varphi_1}_{\vec{k}}(\sigma_x + i\sigma_y)$   & \hspace{-2em} $\downarrow$ gapless/2 points & $(E^{1_s}(1,0) + E^{3_s}(a))_{\varphi_1}$ \hspace{-1em} & \xmark/\cmark   \\
$E^{3_s}(d)$ & $(X^0_{\vec{k}}+i \,Y^0_{\vec{k}})\sigma_x$   & 0 & $E^{1_s}(1,i)$ & \cmark/\xmark   \\
$E^{3_s}(e)$ & $(X^0_{\vec{k}}+i \,Y^0_{\vec{k}})(\sigma_x + i\sigma_y)$   & $\downarrow$ gapless/0 & $E^{1_s}(1,i) + E^{3_s}(d)$ & \xmark/\cmark   \\
$E^{3_s}(f)_{\varphi_2}$ & $(X^{\varphi_2}_{\vec{k}}+i \,Y^{\varphi_2}_{\vec{k}})\sigma_x + (X^{\varphi_2}_{\vec{k}}-i \,Y^{\varphi_2}_{\vec{k}})\sigma_y$   & 2 points & $(E^{1_s}(1,-i) + E^{3_s}(d))_{\varphi_2}$ \hspace{-1em} & \xmark/\xmark   \\
$E^{3_s}(g)$ & $ a(X^0_{\vec{k}} + i\, Y^0_{\vec{k}} ) (\sigma_x+i\sigma_y) + b (X^0_{\vec{k}} - i\, Y^0_{\vec{k}} ) \sigma_z$   & 0 & --- & \xmark/\xmark   \\ 
$E^{3_s}(h)_{\varphi_1}$ & $ a[(X^{\varphi_1}_{\vec{k}} + i\, Y^{\varphi_1}_{\vec{k}}) \sigma_x + (X^{\varphi_1}_{\vec{k}} - i\, Y^{\varphi_1}_{\vec{k}}) \sigma_y] + b Y^{\varphi_1}_{\vec{k}}  \sigma_z$ \hspace{-4em}  & 0 & --- & \xmark/\xmark   \\ \hline
$(E^{1_s}(0,i)+ E^{3_s}(a))_{\varphi_1}$ & $i\, Y^{\varphi_1}_{\vec{k}} \sigma_0 + \eta X^{\varphi_1}_{\vec{k}}\sigma_x$   & 0 & $E^{3_s}(b)$ & \cmark/\xmark   \\
$(E^{1_s}(1,0) + E^{3_s}(a))_{\varphi_1}$ & $X^{\varphi_1}_{\vec{k}}( \sigma_0 + \eta \sigma_x)$   & 2 points  & $E^{3_s}(c)_{\varphi_1}$ & \xmark/\cmark   \\
$E^{1_s}(1,i) + E^{3_s}(d)$ & $(X^0_{\vec{k}}+i \,Y^0_{\vec{k}})( \sigma_0 + \eta \sigma_x)$   & 0  & $E^{3_s}(e)$ & \xmark/\cmark  \\
 $(E^{1_s}(1,-i) + E^{3_s}(d))_{\varphi_2}$ & $(X^{\varphi_2}_{\vec{k}}+i \,Y^{\varphi_2}_{\vec{k}})\sigma_0 + \eta  (X^{\varphi_2}_{\vec{k}}-i \,Y^{\varphi_2}_{\vec{k}})\sigma_x$ & 2 points  & $E^{3_s}(f)_{\varphi_2}$ & \xmark/\xmark \\
  $E^{3_s}(g) + E^{1_s}(1,-i)$ & $ a(X^{0}_{\vec{k}} + i\, Y^{0}_{\vec{k}} ) (\sigma_x+i\sigma_y) +   (X^{0}_{\vec{k}} - i\, Y^{0}_{\vec{k}} ) (b \sigma_z + \eta  \sigma_0)$ & 0  & $E^{1_s}(1,-i) + E^{3_s}(g)$ & \xmark/\xmark  \\
   $(E^{1_s}(1,0) + E^{3_s}(h))_{\varphi_1}$ & $a[(X^{\varphi_1}_{\vec{k}} \hspace{-0.4em}+ i Y^{\varphi_1}_{\vec{k}}) \sigma_x + (X^{\varphi_1}_{\vec{k}}\hspace{-0.4em} - i Y^{\varphi_1}_{\vec{k}}) \sigma_y]+ b Y^{\varphi_1}_{\vec{k}}  \sigma_z + \eta X^{\varphi_1}_{\vec{k}}\sigma_0$ \hspace{-3em} & 0  & $(E^{3_s}(h)+E^{1_s}(1,0) )_{\varphi_1}$ \hspace{-1em} & \xmark/\xmark
 \end{tabular}
 \end{ruledtabular}
\end{center}
\end{table*}

We can also readily understand from \equref{TrafoUnderC2y} how the one-dimensional representations ``connect'' at the SU(2)$_+$ $\times$ SU(2)$_-$ point: at the high-symmetry point, $\lambda^s_{\vec{k}v} = \lambda^t_{\vec{k}v}$, ergo $A_1^{1_s}$ and $A_2^{3_s}$ or $A_2^{1_s}$ and $A_1^{3_s}$ must meet at the $J=0$ line in \figref{SchematicPD}. We summarize these observations in \tableref{SummaryOfPairingD3}.

In addition, for the case of the two-dimensional representation $E$ of $D_3$, the $C_{2y}$ symmetry has nontrivial consequences. Once again, we take $J\neq 0$ which permits us to study singlet and triplet independently. As singlet pairing has already been analyzed in detail for twisted bilayer graphene (see, e.g., \refcite{RafaelsPaperTBG}), we are chiefly concerned with the triplet states here. We parametrize the triplet pairing as in \secref{SingeltTripletSepForComplexRep} with the sole distinction being that the basis functions $X_{\vec{k}}$ and $Y_{\vec{k}}$ are now constrained by the symmetries of $D_3$; we choose them to obey $X_{-C_{2y}\vec{k}}=-X_{\vec{k}}$ and $Y_{-C_{2y}\vec{k}}=Y_{\vec{k}}$, while transforming as $k_x$ and $k_y$ under $C_3$.
A possible choice is given by \equref{ExampleBasisFuncs} with $\phi=\pi/2$ for the Brillouin zone of twisted bilayer graphene in \figref{LatticesAndSymmetries}(b). With these conventions, the triplet vector transforms according to $(\vec{d}_+,\vec{d}_-) \rightarrow (\vec{d}_-,\vec{d}_+)$ under $C_{2y}$. This does not further constrain the quartic terms in the free energy (\ref{GLExpansionTripletComplexIR}), wherefore we can use the analysis of \secref{SingeltTripletSepForComplexRep} for the point group $C_3$, bearing in mind the caveat that the relative phase, $\varphi$, between $\vec{d}_+$ and $\vec{d}_-$ cannot be absorbed in a redefinition of the basis functions $X_{\vec{k}}$ and $Y_{\vec{k}}$ any more due to the extra $C_{2y}$ symmetry. While $\varphi$ has no consequences for $E^{3_s}(d)$ or $E^{3_s}(e)$ and can be absorbed by performing a spin rotation for the phases $E^{3_s}(b)$ and $E^{3_s}(g)$, it describes different phases for all other stable minima of \equref{GLExpansionTripletComplexIR}, and we have to go to higher order in the free-energy expansion to determine its value. 

Consider $E^{3_s}(a)$ for instance. Writing $\vec{d}_+ = (1,0,0)^T$ and $\vec{d}_- = e^{i\varphi}(1,0,0)^T$, it is easy to verify that the most general, $\varphi$-dependent sextic term to the free energy must have the form $c_1\cos(3\varphi)$ with $c_1\in \mathbb{R}$. This derives from \equref{SexticTerm} where the $C_{2y}$ symmetry forces $c_2$ to vanish. We thus find $\varphi = 2\pi n /3 $, $n\in\mathbb{Z}$, for $c_1<0$ and $\varphi = \pi/3 + 2\pi n/3$ when $c_1>0$. These two minima correspond to two different states, which can be compactly represented by defining the ``rotated'' basis functions
\begin{equation}
    (X^{\varphi}_{\vec{k}},Y^{\varphi}_{\vec{k}})^T = R_{\varphi/2} (X_{\vec{k}},Y_{\vec{k}})^T,\quad  R_{\phi} = e^{i\phi\sigma_y}, \label{TheRotatedBasisFuncs}
\end{equation}
with $X_{\vec{k}}$ and $Y_{\vec{k}}$ as introduced above. The order parameters are $M_{\vec{k}+}$\,$=$\,$X^0_{\vec{k}}\sigma_x$ and $M_{\vec{k}+} $\,$=$\,$X^{\frac{\pi}{3}}_{\vec{k}}\sigma_x$\,$\equiv$\,$(\sqrt{3}X_{\vec{k}}+Y_{\vec{k}})\sigma_x/2$ for $c_1 < 0$ and $c_1 > 0$, respectively. We denote these two states by $E^{3_s}(a)_{0}$ and $E^{3_s}(a)_{\frac{\pi}{3}}$, respectively. The first state, $E^{3_s}(a)_{0}$, preserves $C_{2y}$, but breaks $C_3$ rotation symmetry, and has a nodal line which is, as opposed to the states in \secref{SingeltTripletSepForComplexRep}, pinned to $k_y=0$. The other state, $E^{3_s}(a)_{\frac{\pi}{3}}$, however, breaks $C_{2y}$ and the nodal line is not pinned to the $k_x$ axis.

The remaining triplet states, $E^{3_s}(c)$, $E^{3_s}(f)$, and $E^{3_s}(h)$ of \secref{SingeltTripletSepForComplexRep} can be analyzed in the same way. In all cases, we find two states corresponding to two different discrete values of the relative phase $\varphi$ between $\vec{d}_+$ and $\vec{d}_-$: for $E^{3_s}(c)$ and $E^{3_s}(h)$, we find $\varphi = 0$ or $\varphi = \pi/3$ as before, whereas $E^{3_s}(f)$ requires even higher-order terms in the free energy expansion, yielding $\varphi = 0$ or $\varphi = \pi/2$. In analogy to $E^{3_s}(a)_{\varphi}$, we label the states by $E^{3_s}(c)_\varphi$, $E^{3_s}(f)_\varphi$, $E^{3_s}(h)_\varphi$; their order parameters are the same as those of the corresponding states in \secref{SingeltTripletSepForComplexRep} but with the rotated basis functions in \equref{TheRotatedBasisFuncs} using the respective value of $\varphi$. Taken together, we obtain twelve triplet states for $D_3$, which are summarized in \tableref{SummaryOfPairingD3}, instead of only eight for the point group $C_3$.

Finally, we can also ask how the different states behave for small $J$, \textit{i.e.}, whether singlet and triplet can mix and which phases are Hund's partners. Exactly as illustrated above for the pure triplet phases, we have to consider higher-order terms that determine the relative phase between the chiral, $\mu=+$, and antichiral, $\mu=-$, basis functions. As this analysis closely parallels our previous discussions, we just present the result in \tableref{SummaryOfPairingD3}. In total, there are ten symmetry-inequivalent mixed singlet and triplet phases. Seven of them are only possible if $\delta a<0$ (singlet dominates); the remaining three can be realized for either sign of $\delta a$.

\section{Discussion and conclusion}\label{SummaryDiscussion}
In this work, we have presented a systematic classification and analysis of superconducting instabilities in graphene moir\'e systems. To this end, we have focused on zero-momentum Cooper pairs formed out of electrons in different valleys. Intervalley pairing is expected to be the dominant pairing channel as time-reversal relates the two valleys. We have first analyzed singlet and triplet pairing separately since spin-orbit coupling is expected to be very weak in graphene. However, theoretical estimates of the interaction terms of twisted bilayer \cite{KoshinoFuModel}, double-bilayer \cite{2019arXiv190308685L}, and trilayer \cite{SenthilABCTrilayer} graphene indicate that these systems are approximately invariant under independent spin rotations in the two valleys, leading to an (approximate) SU(2)$_+$ $\times$ SU(2)$_-$ symmetry and the (near) degeneracy of singlet and triplet pairing. For this reason, we have also classified the pairing instabilities close to this high-symmetry point, analyzing which triplet state transforms into which singlet phase upon changing the sign of the interactions breaking the SU(2)$_+$ $\times$ SU(2)$_-$ symmetry. We have further derived the conditions under which singlet and triplet can mix despite the absence of spin-orbit coupling.

As it has the fewest symmetries, we first considered twisted double-bilayer graphene, for which there are also clear experimental indications of triplet pairing \cite{2019arXiv190306952S,ExperimentKim}. Here, a displacement field, which is required to stabilize the superconducting state, reduces the point group to $C_3$. The pairing states and their properties associated with the real representation $A$ and the complex representation $E$ of $C_3$ are summarized in Tables~\ref{TableSummaryOfPairingC3} and \ref{SummaryOfPairingComplexRepC3}, respectively. 

Being one-dimensional and real, $A$ only allows for one singlet, a unitary and a nonunitary triplet phase, and one mixed phase. The latter is expected to be relevant only if SU(2)$_+$ $\times$ SU(2)$_-$ is weakly broken and the two consecutive transitions in the schematic phase diagram of \figref{SchematicPD}(b) are very close. Using the values of the coupling constants in \refcite{2019arXiv190308685L}, we estimate the splitting to be about two orders of magnitude smaller than the critical temperature and hence, hard to see experimentally \cite{Note1}. Whether renormalization-group corrections could enhance the impact of these weak symmetry-breaking perturbations at energies of order of the transition temperature is an open question, which we leave for future work. The gap structure of the four phases transforming under $A$ is quite different: while the nonunitary triplet is gapless for one of the spin species, the singlet and unitary triplet have a single, fully established gap, and the mixed phase has two finite but distinct gaps for the two spin species. We have further shown that single-band mean-field theory will generically favor the phase diagram in \figref{SchematicPD}(a) over \figref{SchematicPD}(b). However, the small bandwidth and strong-coupling nature inherent in the problem makes the applicability of mean-field theory questionable and can lead to significant corrections which might eventually select other phases. We have illustrated these corrections for ferromagnetic fluctuations, expected to be relevant for twisted double-bilayer graphene \cite{2019arXiv190306952S,ExperimentKim,PabllosExperiment}, twisted bilayer \cite{Sharpe605,2019arXiv190306513L,IlaniExperiment}, and ABC trilayer graphene \cite{FMTrilayer}. We find that the resulting corrections will, as opposed to mean field, generally favor the phase diagram in part (b) of \figref{SchematicPD} over that in part (a).

The complex representation allows for many more states: two pure singlets, eight triplets, and, if SU(2)$_+$ $\times$ SU(2)$_-$ is only weakly broken, six distinct mixed phases. As compiled in \tableref{SummaryOfPairingComplexRepC3}, all of these three classes of states allow for nodal points and fully gapped phases. However, only the triplets can have nodal lines (residual ungapped Fermi surfaces of one spin species). Only one out of the two different triplet states of the IR $A$ allow for an admixture of singlet and triplet for weak anti-Hund's coupling but, in contrast, six out of the eight triplets transforming under $E$ do so. 

Out of the possible pairing states in \tableref{SummaryOfPairingComplexRepC3}, single-band mean-field theory favors the two triplet states $E^{3_s}(b)$ and $E^{3_s}(d)$ along with their respective Hund's partners---the nematic mixed phase $E^{1_s}(0,i)+E^{3_s}(a)$ and the chiral singlet $E^{1_s}(1,i)$. We show the associated phase diagrams in the vicinity of mean-field theory in \figref{SchematicPDForIRE}(a) and (b). We have discussed how additional weak ferromagnetic spin (orbital) fluctuations can lift the exact degeneracy of $E^{3_s}(b)$ and $E^{3_s}(d)$, generically favoring the former (latter) and, hence, the phase diagram in \figref{SchematicPDForIRE}(b) [\figref{SchematicPDForIRE}(a)]. In the limit of strong ferromagnetic fluctuations, we obtain the chiral nonunitary triplet $E^{3_s}(e)$ or, for weak anti-Hund's coupling, the mixed singlet-triplet state $E^{1_s}(1,i)+E^{3_s}(d)$ as the dominant instability.

Motivated by the experimentally observed \cite{ExperimentKim} linear increase of the transition temperature with an in-plane magnetic field in twisted double-bilayer graphene and signs of magnetism in bilayer and trilayer graphene, we have also mapped out the possible phase diagrams in the presence of a magnetic field. As expected, if the SU(2)$_+$ $\times$ SU(2)$_-$ symmetry is significantly broken, the linear increase is only consistent with triplet pairing. For pairing in the $A$ channel, there are two possible phase diagrams, shown in \figref{PDInMagneticField}(c) and (d), depending on which triplet state is realized in the absence of a magnetic field. The magnetic field fully determines the form of the leading triplet state to be $A^{3_s}(1,i,0)$ in the $A$ channel. For order parameters transforming under $E$, there are two possibilities for the leading triplet state, $E^{3_s}(c)$ or $E^{3_s}(e)$, in a magnetic field; which of the two is realized depends on the value of the quartic couplings in the free energy. Both mean-field theory and ferromagnetic fluctuations favor the $E^{3_s}(e)$ state. If, however, SU(2)$_+$ $\times$ SU(2)$_-$ is only very weakly broken, singlet pairing as the dominant instability of the system is also  consistent with the linear increase of the critical temperature; the two possible phase diagrams for the case of pairing in the IR $A$ are illustrated in \figref{PDInMagneticField}(a) and (b).

We have also derived (within mean-field theory) the key couplings, $c_2$ in \equref{GinzburgLandauExpansion} and $c_3^E$ in \equref{CouplingInMagnFieldCompl}, between the superconducting order parameter and the magnetic field $B$ that determine the slope of the increase, $\Delta T_c$, of the critical temperature with magnetic field. We found that they have the exact same mathematical form; as such, the behavior $\Delta T_c \approx 2\mu_B B$, with Bohr magneton $\mu_B$, seen in experiment \cite{ExperimentKim}, is equally surprising for both pairing channels and does not favor one channel over the other. In both cases, this might either be accidental or due to quantum critical scaling \cite{2019arXiv190308685L}.

We have also studied, in \secref{ComparisonWithOtherSystems}, the changes in the classification when there is an extra in-plane rotation symmetry, $C_{2y}$, and a twofold rotation, $C_2$, perpendicular to the plane. These two symmetries are relevant (either as exact or emergent symmetries) to twisted bilayer graphene and ABC trilayer graphene. We find that while the $C_2$ symmetry has no consequences for the classification, $C_{2y}$ not only pins the nodes of certain pairing states along high-symmetry lines but also leads to more pairing states as summarized in \tableref{SummaryOfPairingD3}. 

This work further illustrates that graphene moir\' e systems provide a very rich playground for novel strongly correlated superconducting phases. We hope that our systematic analysis of pairing in the absence and presence of magnetic fields will help future theoretical and experimental studies to pinpoint the microscopic form of the superconducting state.

\section*{Acknowledgments}

This research was supported by the National Science Foundation under Grant No.~DMR-1664842. MS also acknowledges support from the German National Academy of Sciences Leopoldina through grant LPDS 2016-12. 
We thank Andrey Chubukov, Eslam Khalaf, Alex Kruchkov, Subir Sachdev, Harley Scammell, and Ashvin Vishwanath for helpful discussions.


\appendix

\section{Microscopic Ginzburg-Landau expansion}\label{MicrosCopGLExpansion}
In this appendix, we derive the prefactors of the various free-energy expansions in the main text within mean-field theory. Unless stated otherwise, we use a single-band description.

\subsection{Without a magnetic field}
We imagine performing a mean-field decomposition in the Cooper channel and keeping only the singlet and triplet pairing of the dominant IR. The ensuing mean-field Hamiltonian for the one-band model has the form
\begin{alignat}{1}
    \mathcal{H}^{}_{\textsc{mf}} &= \sum_{\vec{k}}\xi^{}_{\vec{k}v} c^\dagger_{\vec{k}\sigma v}c^\pdagger_{\vec{k}\sigma v} \label{GeneralMeanFieldHamtilonian} \\
    \nonumber &+ \sum_{\vec{k}} c^\dagger_{\vec{k}\sigma +} \left[( \Delta^s_{\vec{k}} +  \vec{\sigma}\cdot \vec{d}_{\vec{k}})i\sigma^{}_y\right]_{\sigma,\sigma'}c^\dagger_{-\vec{k}\sigma' -}, 
\end{alignat}
where $\xi_{\vec{k}+} = \xi_{-\vec{k}-}$ due to time-reversal symmetry. In \equref{GeneralMeanFieldHamtilonian}, we have omitted a constant term, which is quadratic in the superconducting order parameter and does not affect the quartic terms we derive below. Upon integrating out the fermions in \equref{GeneralMeanFieldHamtilonian} and expanding the resulting free energy in the superconducting order parameter, the Ginzburg-Landau expansion coefficients can be obtained order by order. 

Starting with the one-dimensional real IR $A$ of $C_3$, we write $\Delta^s_{\vec{k}} = \lambda_{\vec{k}}^s\Delta^s$, $\vec{d}^s_{\vec{k}} = \lambda_{\vec{k}}^t\vec{d}$, where $\lambda_{\vec{k}}^s$ and $\lambda_{\vec{k}}^t$ are momentum-dependent basis functions that are invariant under $C_3$.
Using the generalization of \equref{FullExpansionWithHigherSym} to parametrize the free energy,
\begin{alignat}{1}
    \nonumber\mathcal{F} &\sim a(T) \left(|\Delta^s|^2 + \vec{d}^\dagger \vec{d}\right)  + \delta a \left(|\Delta^s|^2 - \vec{d}^\dagger \vec{d}\right)  + \gamma_1 |\Delta^s|^4\\ \nonumber&+ \gamma_2 \left(\vec{d}^\dagger \vec{d}\right)^2 
    + \gamma_3 \left|\vec{d}^*\times \vec{d}\right|^2 + \gamma_4 |\Delta^s|^2 \vec{d}^\dagger \vec{d}\\
    &+ \gamma_5\text{Re}\left[ \left(\Delta^s\right)^2 \vec{d}^\dagger \vec{d}^*\right], 
\end{alignat}
which allows us to account for a nonzero $J$ making singlet and triplet nonequivalent, we find
\begin{alignat}{2}
    \gamma_1 &= \mathscr{F}[|\lambda_{\vec{k}}^s|^4], \quad  &&\gamma_2 = \gamma_3 = \mathscr{F}[|\lambda_{\vec{k}}^t|^4], \\
    \gamma_4 &= 4\,\mathscr{F}[|\lambda_{\vec{k}}^t|^2|\lambda_{\vec{k}}^s|^2], \quad &&\gamma_5 = 2\,\mathscr{F}[(\lambda_{\vec{k}}^t)^2(\lambda_{\vec{k}}^{s*})^{2}]. 
\end{alignat}
To keep the expressions compact, we have defined the functional
\begin{equation}
    \mathscr{F}[f^{}_{\vec{k}}] : = T \sum_{\omega_n} \int \frac{\diff^2 \vec{k}}{(2\pi)^2} \frac{f^{}_{\vec{k}}}{(\omega_n^2 +\xi^2_{\vec{k}+})^2}.
\end{equation}
When $J=0$, we have $\lambda_{\vec{k}}^s = \lambda_{\vec{k}}^t$ and hence, obtain
\begin{equation}
    \gamma_1 = \gamma_2 = \gamma_3 = \gamma_4/4 = \gamma_5/2 > 0, \label{CoeffsIApp}
\end{equation}
which is compatible with the prefactors in \equref{FullExpansionWithHigherSym} as required from the SU(2)$_+$ $\times$ SU(2)$_-$ symmetry. On top, $\gamma_1 = \gamma_3$ is an additional constraint arising from the mean-field approximation (and not related to an exact symmetry). In terms of the prefactors in \equref{FullExpansionWithHigherSym}, it sets $b_1 = 0$, as stated in the main text. The positive sign of the coefficients in \equref{CoeffsIApp} implies that mean-field theory always favors part (a) in the phase diagram in \figref{SchematicPD}.

Similarly, we can study the complex representation of $C_3$ introduced in \secref{ComplexRepresentation} of the main text. Using the representation in \equref{SingletPairingMatrix} for the singlet pairing, $\Delta^s_{\vec{k}} = \sum_{\mu}\eta_\mu \left( X_{\vec{k}} + i\mu\, Y_{\vec{k}} \right)$, it is straightforward to show that 
\begin{equation}
    b^s_1 = b^s_2/2 = \mathscr{F}[(X^2_{\vec{k}} + Y^2_{\vec{k}})^2] > 0
\end{equation}
for the coefficients $b^s_{1,2}$ in \equref{SingletTwoComponentExpansion}. Being positive, these coefficients favor the chiral superconductor $E^{1_s}(1,i)$ as was observed earlier as well \cite{YiZhuangPairing,LiangFu}.

Finally, repeating this procedure for the triplet state with parametrization (\ref{parametrizationOfTripletComplexIR}), $\vec{d}_{\vec{k}}=\sum_{\mu} \vec{d}_\mu \left( X_{\vec{k}} + i\mu\, Y_{\vec{k}} \right)$, the coefficients in \equref{GLExpansionTripletComplexIR} evaluate to
\begin{alignat}{1}
    \nonumber b^t_1&=b^t_3/2=-b^t_4/2=-2b^t_5 = 2\,\mathscr{F}[(X^2_{\vec{k}} + Y^2_{\vec{k}})^2] > 0, \\ b^t_2&=0.
\label{MeanFieldValuesTriplet}\end{alignat}
The triplet states $E^{3_s}(b)$ and $E^{3_s}(d)$ will have the lowest energy for this configuration of quartic coefficients as argued in the main text. The degeneracy between these two states is lifted by corrections beyond the mean-field approximation, such as the ferromagnetic fluctuations of  \secref{FluctuationInducedSC}. In the presence of a magnetic field, \equref{MeanFieldValuesTriplet} uniquely determines the chiral nonunitary triplet $E^{3_s}(e)$ as the leading instability (see \secref{ComplexRepInMagnField}).

\subsection{Coupling to a magnetic field}
In this subsection, we will analyze several important coupling terms between the superconductor and the magnetic field from a weak-coupling perspective.
The microscopic form of the coupling to the Zeeman, $\vec{M}_Z$, and in-plane orbital field, $\vec{M}_O$, reads as
\begin{equation}
    \mathcal{H}_{B} = \sum_{\vec{k}} c^\dagger_{\vec{k}\sigma v} \vec{\sigma}^{}_{\sigma\sigma'} c^\pdagger_{\vec{k}\sigma' v} \cdot \vec{M}_Z + \sum_{\vec{k}} \vec{g}^{}_v(\vec{k}) c^\dagger_{\vec{k}\sigma v}c^\pdagger_{\vec{k}\sigma v}  \cdot\vec{M}_O,\label{CouplingMagnField}
\end{equation}
where we have absorbed the $g$-factor of the Zeeman coupling into the definition of $\vec{M}_Z$. This is not possible for the orbital coupling, as its $g$-factor $\vec{g}_v(\vec{k})$ depends significantly on momentum. The form of $\vec{g}_v(\vec{k})$ is determined by microscopic details such as the Bloch states. All we need here is that $\vec{g}_v(\vec{k}) = -\vec{g}_{\bar{v}}(-\vec{k})$, as follows from time-reversal symmetry (\ref{TRTransformation}), and we refer to \refcite{2019arXiv190308685L} for a microscopic derivation of its momentum dependence.

Let us first note that even when the actual interacting multiband system is not invariant under $C_2$, the single-band mean-field Hamiltonian, $\mathcal{H}_{\text{MF}} + \mathcal{H}_{B}$, is left invariant under the action of $C_2$ in \equref{C2Symmetry} if we further set $\vec{d}_{\vec{k}} \rightarrow -\vec{d}_{\vec{k}}$ in \equref{GeneralMeanFieldHamtilonian} and $\vec{M}_O \rightarrow -\vec{M}_O$. This emergent symmetry is a consequence of the special role of $C_2$ in two dimensions as it acts on $\vec{k}$ in the same manner as time-reversal and, as such, can have crucial consequences for superconducting pairing \cite{scheurer2017selection}.

In the present case, this symmetry implies that the coupling terms $\delta c_1$ in \equref{GinzburgLandauExpansion} and $\delta c_1^E$, $c_2^E$ in \equref{CouplingInMagnFieldCompl} will vanish within single-band mean-field theory as is also readily confirmed by explicit calculation; we emphasize, however, that this is not an exact statement and we have checked that a multiband mean-field description allows for nonzero values. Nonetheless, we view the vanishing of these coupling in the weak-coupling single-band limit as an indication that they are likely small in the system.

Finally, the couplings of the Zeeman term to the triplet vector in \equsref{GinzburgLandauExpansion}{CouplingInMagnFieldCompl} are also not constrained by the emergent $C_2$ symmetry. We find these to be nonzero and given by
\begin{subequations}\begin{align}
    c^{}_2 &= -4 \, \mathscr{F}\left[\xi^{}_{\vec{k}+}|\lambda_{\vec{k}}^t|^2\right], \label{FirstOfTheTwoMagnCoupls}  \\
    c_3^E &= -4 \,\mathscr{F}\left[\xi^{}_{\vec{k}+}(X_{\vec{k}}^2+Y_{\vec{k}}^2)\right], 
\end{align}\label{ZeemanCouplMicrExpr}\end{subequations}
respectively. Our main observation here is that the forms of $c_2$ and $c_3^E$ are identical: the nonuniversal part is a momentum integral which, in both cases, is weighted by a function that is invariant under $C_3$ and has no symmetry-imposed nodes on the Fermi surface. Accordingly, it is not possible to distinguish between the IRs $A$ and $E$ based on the slope of the increase of $T^+_c$ in small magnetic fields.

\section{Fluctuation corrections to mean-field}
\label{FluctuationCorrectionsDetails}
In this appendix, we provide further details on \secref{FluctuationInducedSC}.

\subsection{Microscopic derivation}\label{MicroscopicDerivationFluctuations} 
In this first part, we will derive, from a microscopic description of the system, that the prediction of the phenomenological approach of the main text provides the leading correction to the free energy of the superconductor in the limit where the mass of the fluctuations approaches zero. 

To this end, we will use the field-theoretical formalism and describe the system by the action
\begin{equation}
    \mathcal{S} = \mathcal{S}_c + \mathcal{S}_{c\Delta} + \mathcal{S}_{\phi} + \mathcal{S}_{c\phi},
\end{equation}
which consists of several contributions: first,
\begin{equation}
    \mathcal{S}_c = \int_k c^\dagger_{k\sigma v} \left(-i\omega_n + \xi_{\vec{k}v}\right) c^\pdagger_{k\sigma v}
\end{equation}
is the free-electron contribution (with Grassmann fields $c$ and $c^\dagger$, in analogy to the operators in the main text), where $\int_k \dots \equiv T\sum_{\omega_n} \sum_{\vec{k}} \dots$ with fermionic Matsubara frequencies $\omega_n = \pi T (2n+1)$, and $k=(\vec{k},\omega_n)$ comprising momentum and frequency. The second term, 
\begin{equation}
    \mathcal{S}_{c\Delta} = \int_k c^\dagger_{k\sigma +} \left[( \Delta^s_{\vec{k}} +  \vec{\sigma}\cdot \vec{d}_{\vec{k}})i\sigma^{}_y\right]_{\sigma,\sigma'}c^\dagger_{-k\sigma' -},
\end{equation}
describes pairing, similar to the second line of \equref{GeneralMeanFieldHamtilonian}, where we omitted the term proportional to the order parameter squared, since it is irrelevant for the free-energy contribution at quartic order in the superconducting state. The ferromagnetic fluctuations in valley $v$ with associated bosonic fields $\vec{\phi}_{qv} = (\phi^x_{qv},\phi^y_{qv},\phi^z_{qv})^T$ are described by the action
\begin{equation}
    \mathcal{S}_{\phi} = \frac{1}{2}\int_q \vec{\phi}_{qv} \cdot \vec{\phi}_{-qv'} \left[\hat{\chi}^{-1}(q) \right]_{vv'},
\end{equation}
where $q\equiv (\vec{q},\Omega_n)$ is the bosonic analogue of $k$, \textit{i.e.}, $\Omega_n$\,$=$\,$2\pi T n$ are bosonic Matsubara frequencies. Physically, $\hat{\chi}(\vec{q},i\Omega_n)$ plays the role of the (analytic continuation of the) dynamical spin susceptibility [as compared to the static one in \equref{parametrizationOfFluctsMagn} of the main text]. We will focus here on the $\text{SU}(2)_+ \times \text{SU}(2)_-$-symmetric limit, where $[\hat{\chi}(q)]_{vv'} = \delta_{v,v'} \chi(q)$, and take the conventional Ornstein-Zernike form 
\begin{equation}
    \chi(\vec{q},i\Omega_n) = \frac{\chi^{}_0}{\Omega_n^2 + (v \vec{q})^2 + m^2}, \label{BosonicPropagator}
\end{equation}
with velocity $v$ and mass $m$ (related to the correlation length, $\xi$, as $\xi = v/m$). Finally, the spin fluctuations couple to the electrons as described by the last contribution,
\begin{equation}
    \mathcal{S}_{c\phi} = g \int_{k}\int_q c^\dagger_{k\sigma v} \vec{\sigma}^{}_{\sigma\sigma'} c^\pdagger_{k+q\sigma' v} \cdot \vec{\phi}_{qv}, \label{FermionBosonCoupling}
\end{equation}
with a coupling constant $g$ that, of course, can be absorbed into $\chi_0$ (or vice versa), but we will keep $g$ explicit here. 

We follow \refcite{LiangFu} and focus on the leading-order correction of the coupling $g$ (second order, $\propto g^2$) to the free energy, $\mathcal{F}[\Delta_{\vec{k}},\vec{d}_{\vec{k}}]$, but emphasize that our expressions will differ from those of \refcite{LiangFu} as we consider a different type of fluctuations (centered around zero rather than finite momenta). These corrections are derived systematically by first integrating out the fermions, expanding the action to quadratic order in the bosonic fields $\vec{\phi}_{vq}$, which is sufficient to quadratic order in $g$, and integrating out the massive bosons. 

\begin{figure}
   \centering
    \includegraphics[width=\linewidth]{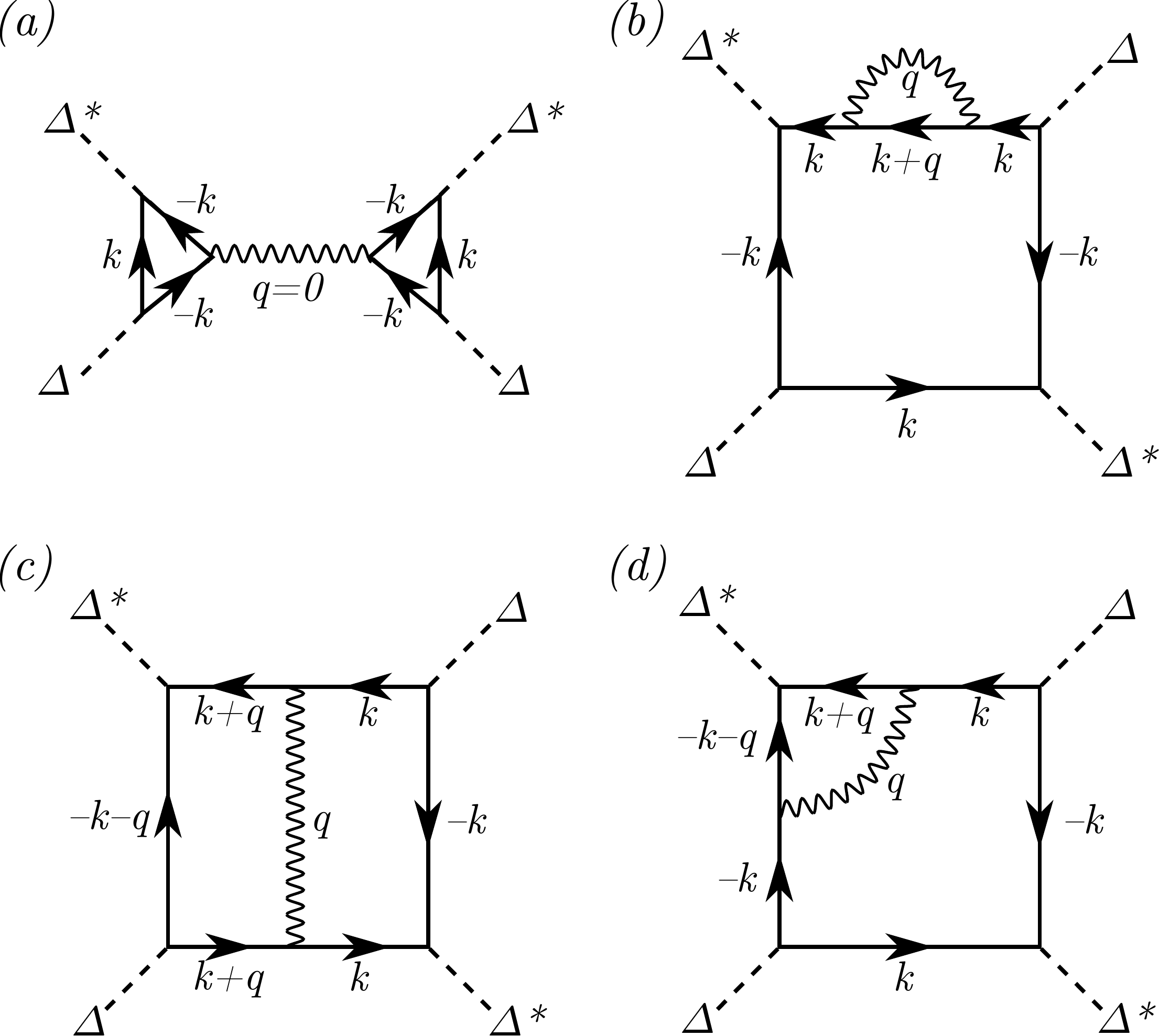}
    \caption{Diagrammatic representation of the four different types of fluctuation corrections quartic in the superconducting order parameter (schematically represented by dashed lines and $\Delta$, $\Delta^*$) to leading order in $g$ in \equref{FermionBosonCoupling}. The solid black lines with arrows are the bare electronic Green's functions, associated with $\mathcal{S}_c$, and the wavy lines denote the bosonic propagator defined in \equref{BosonicPropagator}. The diagram in (a) reproduces the contribution of the ``phenomenological approach'' of the main text, whereas the remaining diagrams in (b--d) are subleading in the limit $m\rightarrow 0$.}
    \label{Diagrams}
\end{figure}

We are interested in terms quartic in the superconducting order parameter, leaving us with the four distinct types of contributions in leading order in $g$ which are represented diagrammatically in \figref{Diagrams}. As indicated, the first diagram, in \figref{Diagrams}(a), only involves the zero-momentum and zero-frequency, $q=0$, fluctuations and it exactly captures the contributions of the simplified approach discussed in \secref{FluctuationInducedSC} of the main text.

The remaining diagrams---the self-energy in \figref{Diagrams}(b), the ladder in (c), and the vertex correction in (d) to the mean-field box-diagram---are fundamentally different: the loop-integrals involve integration over finite frequency and momentum of the bosonic fluctuations. As such, it is intuitively clear that they are less singular in the limit $m\rightarrow 0$ than the first diagram in \figref{Diagrams}(a), which is proportional to $\chi(\vec{q}=0,i\Omega_n=0) = m^{-2}$. In fact, these additional contributions can be shown to diverge with $\log (m)$ for small $m$. To illustrate this, let us consider the diagram in \figref{Diagrams}(b), which is proportional to
\begin{equation}
    D_b=\int_k \int_q \frac{\chi_0}{\Omega_n^2 + (v\vec{q})^2 + m^2} \frac{f_b(k)}{i(\omega_n+\Omega_n) - \xi_{\vec{k}+\vec{q},+}},
\end{equation}
where we introduced the function
\begin{equation}
    f_b(k) = \frac{i\omega_n+\xi_{\vec{k}+}}{(\omega^2_n+\xi_{\vec{k}+}^2)^3} (\lambda^{1}_{\vec{k}})^*(\lambda^{2}_{\vec{k}})^*\lambda^{3}_{\vec{k}}\lambda^{4}_{\vec{k}}
\end{equation}
that only depends on the fermionic momenta and frequencies. Here, $\lambda_{\vec{k}}^j$ represent the basis functions of the involved superconducting order parameters. In the following, we will cut off the $\vec{q}$ integral by $\Lambda/v$ and expand $\xi_{\vec{k}+\vec{q}} \sim \xi_{\vec{k}} + \vec{v}_{\vec{k}} \cdot \vec{q}$, allowing us to write
\begin{align}\begin{split}
    D_b&=\int_k T\sum_{\Omega_n}\int\diff \varphi \int_0^{\Lambda} E\diff E \frac{\chi_0/v^2}{\Omega_n^2 + E^2 + m^2} \\ &\qquad\times \frac{f_b(k)}{i(\omega_n+\Omega_n) - \xi_{\vec{k}+} + \hat{v}_{\vec{k}}E  \cos\varphi}, \label{SimplifiedExpr}
\end{split}\end{align}
with the dimensionless velocity ratio $\hat{v}_{\vec{k}} := |\vec{v}_{\vec{k}}|/v$. Since $|\omega_n| \geq \pi T$ and we work at finite temperature (given by the critical temperature of superconductivity), the term in the second line of \equref{SimplifiedExpr} is finite in the limit $E\rightarrow 0$. The integral, thus, diverges as $\log(m)$ at small $E$ (infrared), as stated above. The other two diagrams in \figref{Diagrams}(c) and (d) can be analyzed in the same way and are also found to be subdominant as $m\rightarrow 0$ compared to the one in \figref{Diagrams}(a). This justifies the approach of \secref{FluctuationInducedSC} microscopically.

\subsection{Enhanced symmetry in the one-band description}
\label{EnhancedSymmetryFluctuations}
In the last part of this appendix, we discuss why $c_+ \approx c_-$ in \equref{CouplingmvForE} is expected. 
From the previous subsection of this appendix, we know that the results of the main text on fluctuation-induced superconductivity are captured by zero-momentum and zero-frequency fluctuations. Writing $\vec{m}_v := \vec{\phi}_{q=0v}$ in \equref{FermionBosonCoupling} and generalizing to a momentum- and valley-dependent coupling constant, we here consider 
\begin{equation}
    \mathcal{H}_m =   \sum_{\vec{k},v} g^m_v(\vec{k}) c^\dagger_{\vec{k}\sigma v} \vec{\sigma}^{}_{\sigma\sigma'}c^\pdagger_{\vec{k}\sigma v}\cdot \vec{m}_v,
\end{equation}
where $g^m_{v}(\vec{k}) = g^m_{\bar{v}}(-\vec{k})$ as a consequence of time-reversal symmetry and we have, as before, assumed that we can focus on a single isolated electronic band. It is easy to see that $\mathcal{H}_{\text{MF}} + \mathcal{H}_m$, with $\mathcal{H}_{\text{MF}}$ in \equref{GeneralMeanFieldHamtilonian}, is again invariant under the $C_2$ symmetry in \equref{C2Symmetry} if we further replace
\begin{equation}
    \vec{d}^{}_{\vec{k}} \,\rightarrow\, -\vec{d}^{}_{\vec{k}}, \quad \vec{m}^{}_v \,\rightarrow \, \vec{m}^{}_{\bar{v}}.
\label{ActionOfEmergentC2}\end{equation}
While \equref{CouplingToFluctuations} is automatically invariant under \equref{ActionOfEmergentC2}, the coupling for the two-dimensional representation in \equref{CouplingmvForE} is invariant only if $c_+$\,$=$\,$c_-$. Consequently, multiband effects are required for nonzero $c_+ - c_-$, wherefore we expect its value to be much smaller than $c_+ + c_-$, as stated in the main text. We also checked by explicit calculation that $c_+ \neq c_-$ is possible in a multiband description.

\begin{figure*}[tb]
    \centering
    \includegraphics[width=0.25\linewidth]{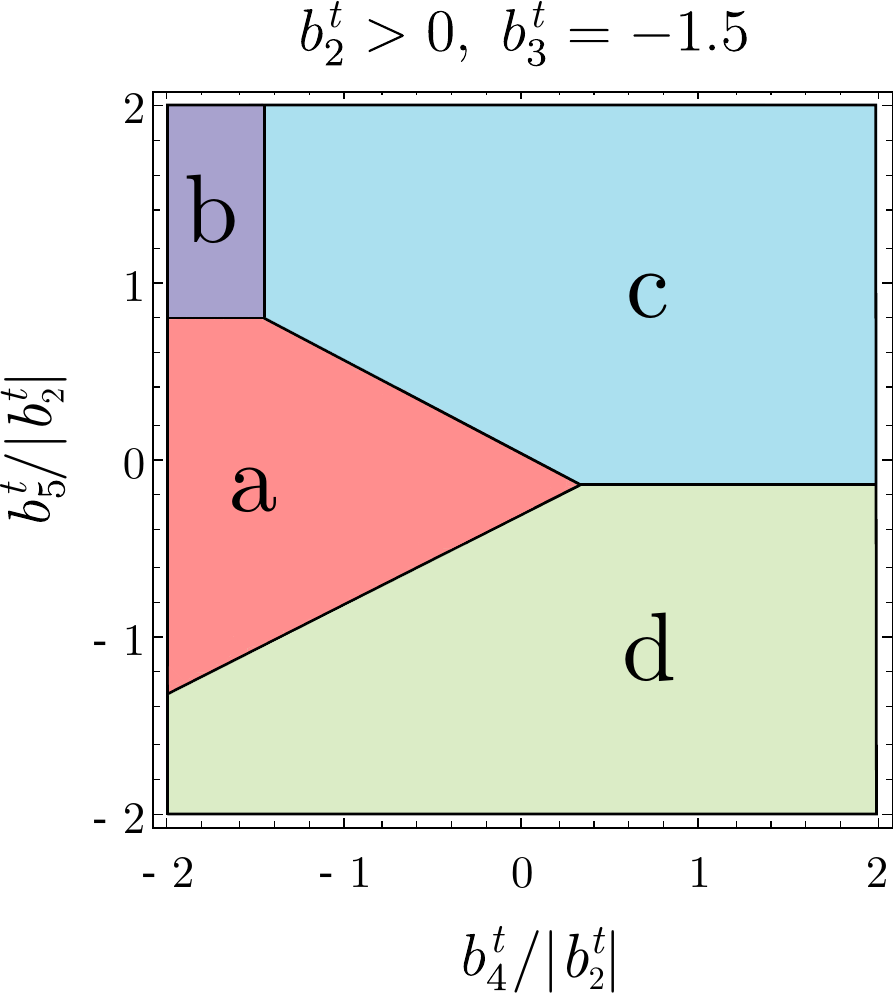}\qquad \qquad
    \includegraphics[width=0.25\linewidth]{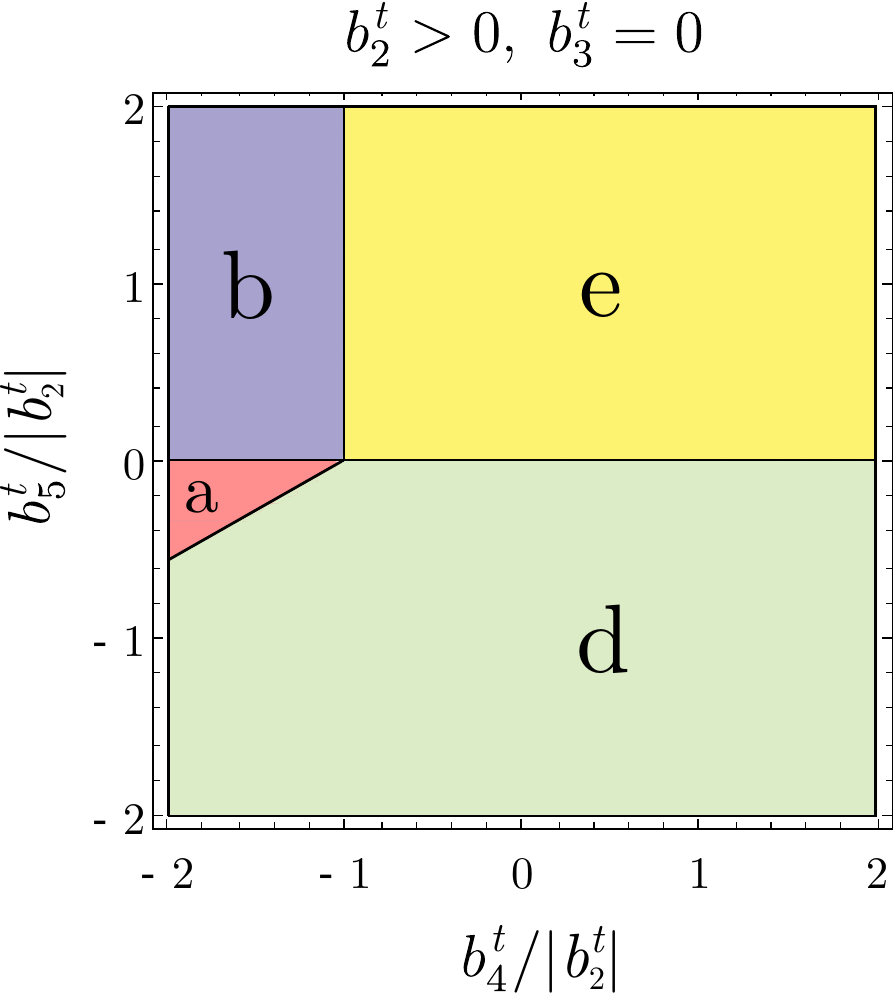}\qquad \qquad
    \includegraphics[width=0.25\linewidth]{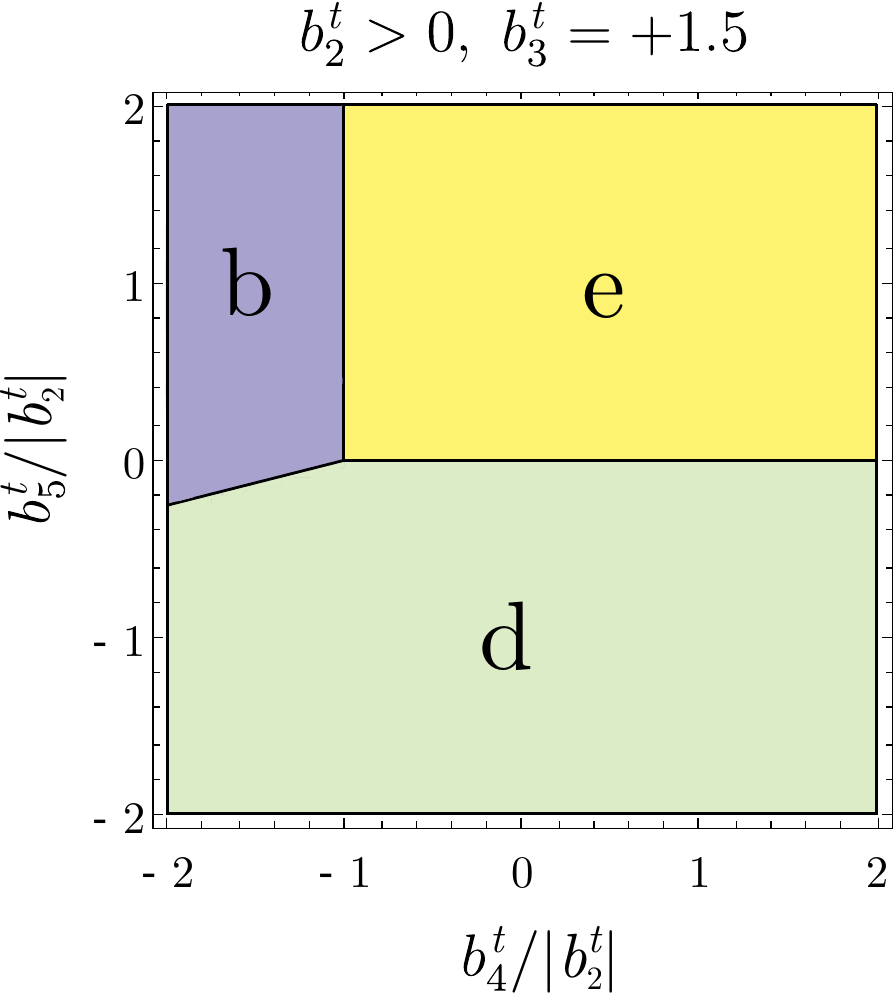}\\
    \vspace{0.4cm}
    \includegraphics[width=0.25\linewidth]{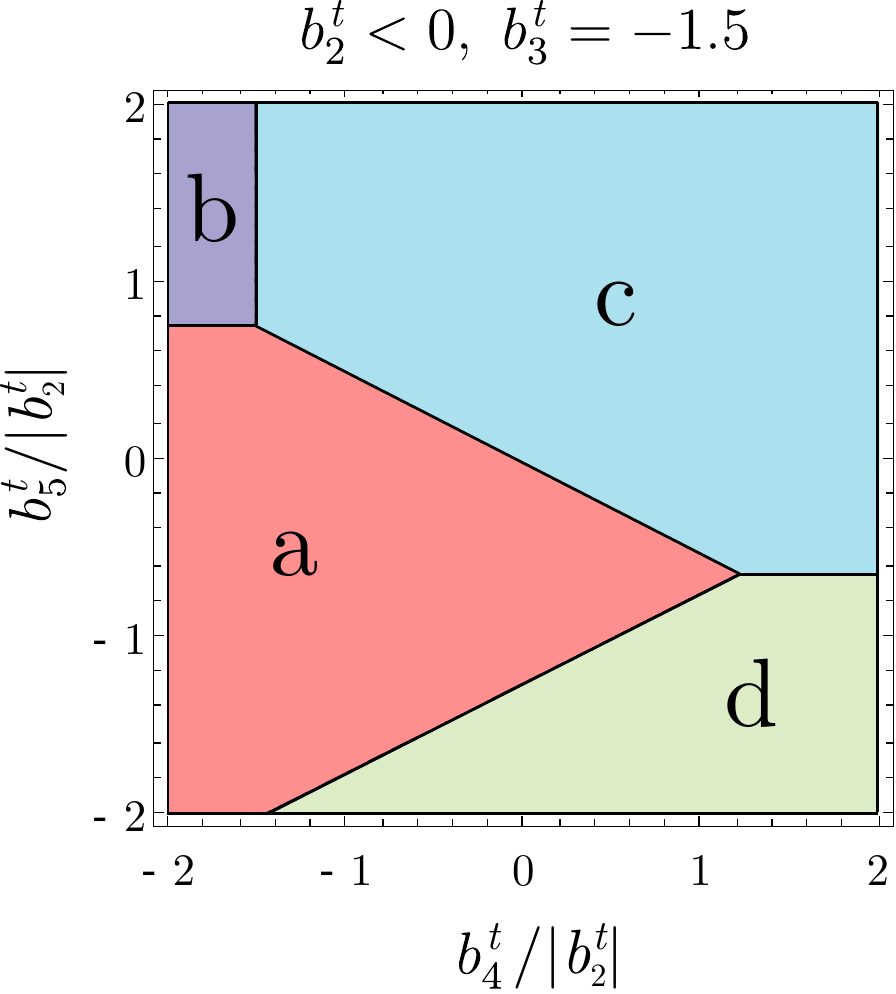}\qquad \qquad
    \includegraphics[width=0.25\linewidth]{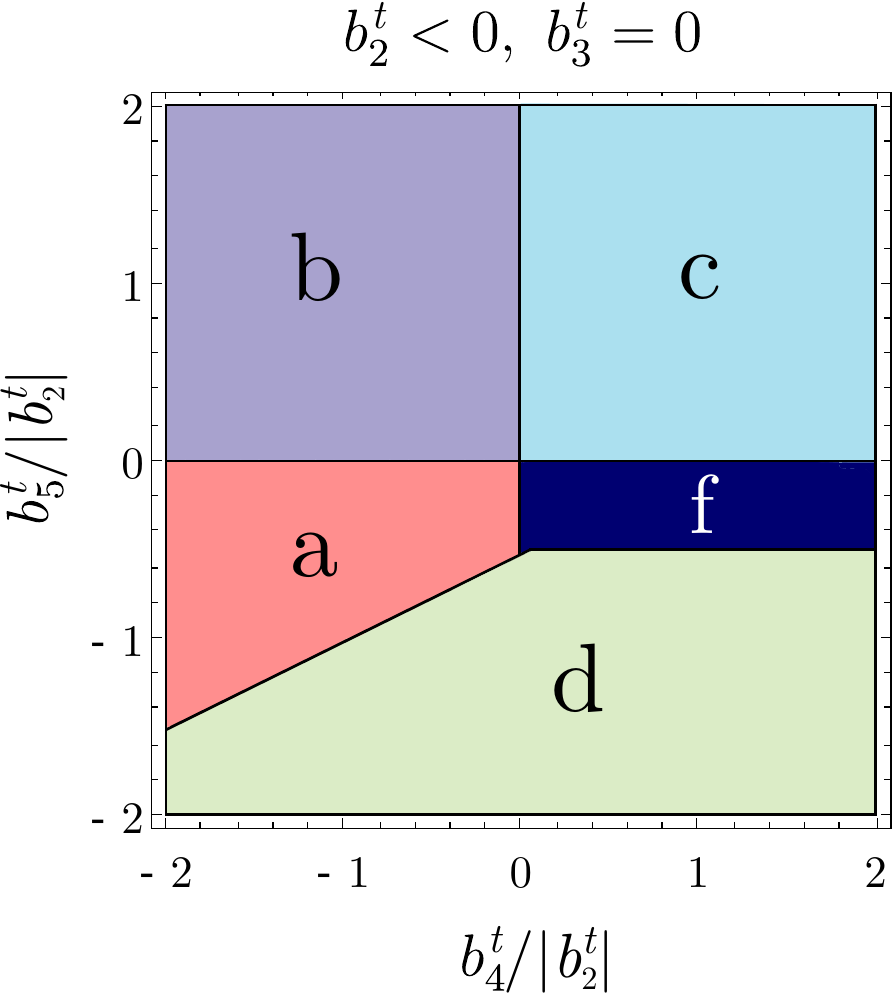}\qquad \qquad
    \includegraphics[width=0.25\linewidth]{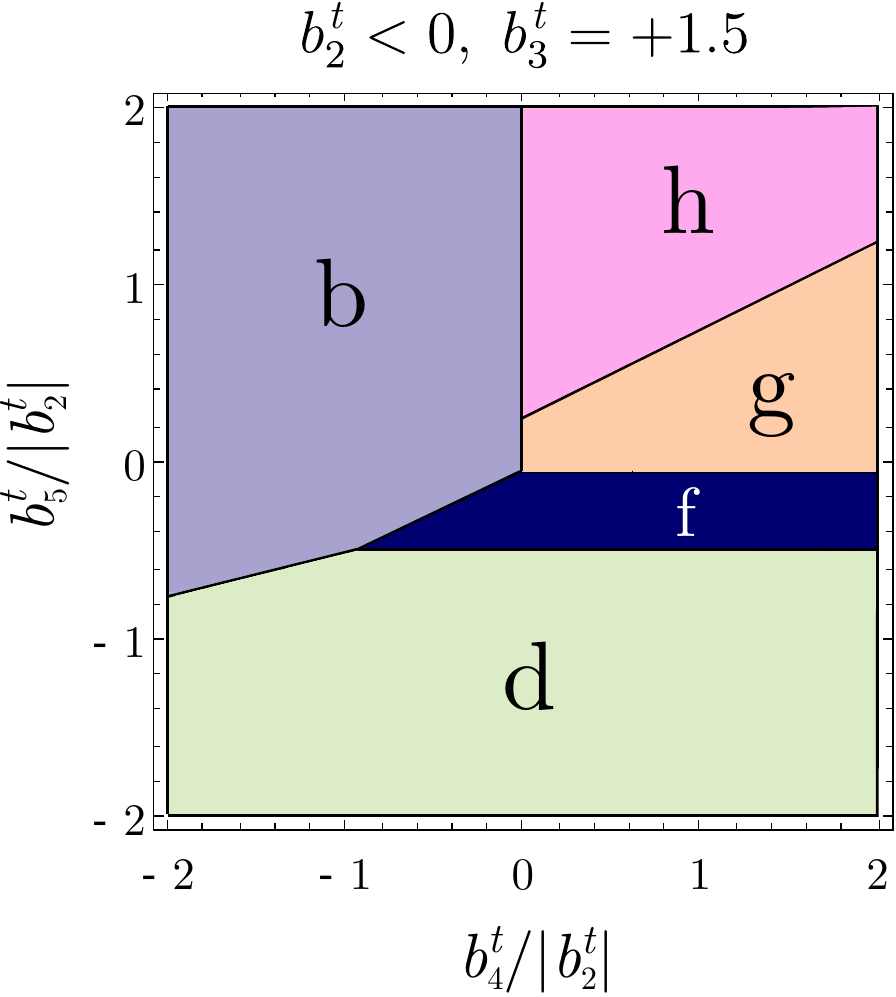}
    \caption{Phase diagram for the free energy in \equref{GLExpansionTripletComplexIR}. The different triplet states labeled (a) to (h) are defined in the main text in \secref{SingeltTripletSepForComplexRep}.}
    \label{fig:my_label}
\end{figure*}

\section{Details for the complex representation}\label{DetailsOfComplexRepresentation}
In this appendix, we present additional details of the different phases transforming under the complex representation $E$ of $C_3$.

As a starting point, it is helpful to chart out a phase diagram describing which of the triplet phases $E^{3_s}(a)$ to $E^{3_s}(h)$ is realized as a function of the quartic terms $b^t_{1,2,3,4,5}$ in \equref{GLExpansionTripletComplexIR}. Upon recognizing that $b_1^t$ does not affect the form of the order parameter (but is assumed to be chosen so as to guarantee the stability of the expansion), we can conveniently display the phases as a function of $b_j^t/|b_2^t|$, $j=3,4,5$, discussing the two possible signs of $b_2^t$ separately. Such a phase diagram is drawn in \figref{fig:my_label}.

As the main text contends, there are no independent terms involving $\sigma_y$ to add to the SU(2)$_+ \times$ SU(2)$_-$-invariant form of the free energy in \equref{MatrixExpansionOfFreeEnergyComplRep}. To see this, we note that it suffices to consider terms involving both $\Delta_+$ and $\Delta_-$ since terms with only $\Delta_+$ (or $\Delta_-$) have already been addressed in \secref{ExactSU2SU2Symmetry}. Among the terms that mix $\Delta_+$ and $\Delta_-$, the following are consistent with time-reversal and $C_3$ symmetry:
\begin{alignat}{1}
    \Delta \mathcal{F}_1 &= \left|\text{tr}\left[\sigma^{}_y \Delta^{}_+ \sigma^{}_y \Delta_-^T \right]\right|^2,\\
    \nonumber\Delta \mathcal{F}_2 &= \text{tr}\left[\Delta^{}_+ \sigma^{}_y \Delta_-^T \Delta_-^* \sigma^{}_y \Delta_+^\dagger \right] + \text{tr}\left[\Delta_-^\dagger \sigma^{}_y \Delta_+^* \Delta_+^T \sigma^{}_y \Delta_-^\pdagger \right],\\
    \nonumber\Delta \mathcal{F}_3 &= \text{tr}\left[\Delta^{}_+ \sigma^{}_y \Delta_-^T \Delta_+^* \sigma^{}_y \Delta_-^\dagger \right] + \text{tr}\left[\Delta_-^\dagger \sigma^{}_y \Delta_+^* \Delta_-^T \sigma^{}_y \Delta_+^\pdagger \right].
\end{alignat}
However, all of these terms can be reformulated as
\begin{widetext}
\begin{align}\begin{split}
    \Delta \mathcal{F}_1 &= \text{tr}\left[\Delta_+^\dagger \Delta_+^\pdagger \right] \text{tr}\left[\Delta_-^\dagger \Delta_-^\pdagger \right] + \left|\text{tr}\left[\Delta_+^\dagger \Delta_-^\pdagger \right] \right|^2  - \left(\text{tr}\left[\Delta_+^\dagger \Delta_+^\pdagger \Delta_-^\dagger \Delta_-^\pdagger \right] + \text{tr}[\Delta_-^\pdagger \Delta_-^\dagger\Delta_+^\pdagger \Delta_+^\dagger] \right) , \end{split}\\
    \Delta \mathcal{F}_2 &= 2\,\text{tr}\left[\Delta_+^\dagger \Delta_+^\pdagger \right] \text{tr}\left[\Delta_-^\dagger \Delta_-^\pdagger \right] - \left(\text{tr}\left[\Delta_+^\dagger \Delta_+^\pdagger \Delta_-^\dagger \Delta_-^\pdagger \right] + \text{tr}[\Delta_-^\pdagger \Delta_-^\dagger\Delta_+^\pdagger \Delta_+^\dagger] \right),\\
    \Delta \mathcal{F}_3 &= 2\left|\text{tr}\left[\Delta_+^\dagger \Delta_-^\pdagger \right] \right|^2 - \left(\text{tr}\left[\Delta_+^\dagger \Delta_+^\pdagger \Delta_-^\dagger \Delta_-^\pdagger \right] + \text{tr}[\Delta_-^\pdagger \Delta_-^\dagger\Delta_+^\pdagger \Delta_+^\dagger] \right),
\end{align}
so they do not constitute independent terms to add to \equref{MatrixExpansionOfFreeEnergyComplRep}.

In concluding this appendix, we present the explicit form of the free-energy (\ref{MatrixExpansionOfFreeEnergyComplRep}) in terms of singlet and triplet components.
Inserting $\Delta_\mu = \sigma_0 \Delta^s_\mu + \vec{\sigma}\cdot \vec{d}_\mu$, $\mu=\pm$ in \equref{MatrixExpansionOfFreeEnergyComplRep} and adding SU(2)$_+$ $\times$ SU(2)$_-$ symmetry breaking only at the level of the quadratic terms, one arrives at
\begin{alignat}{1}
    \nonumber\mathcal{F} &\sim a(T)\sum_{\mu} \left( |\Delta^s_\mu|^2 + \vec{d}_\mu^\dagger\vec{d}^\pdagger_\mu\right) + \delta a\sum_{\mu} \left( |\Delta^s_\mu|^2 - \vec{d}_\mu^\dagger\vec{d}^\pdagger_\mu\right)  
    + \beta_1\left(\sum_\mu |\Delta^s_\mu|^2\right)^2 + \beta_2 |\Delta^s_+|^2|\Delta^s_-|^2 \\
    \nonumber& + \beta_3\left(\sum_{\mu} \vec{d}_\mu^\dagger\vec{d}_\mu\right)^2 + \beta_4 (\vec{d}^\dagger_+\vec{d}^\pdagger_+)(\vec{d}^\dagger_-\vec{d}^\pdagger_-) + \beta_5 |\vec{d}_+^\dagger \vec{d}^\pdagger_-|^2 + \beta_6 |\vec{d}_+^T \vec{d}^\pdagger_-|^2 + \beta_7 \sum_{\mu} |\vec{d}_\mu^T \vec{d}^\pdagger_\mu|^2 \\
    \nonumber& + \beta_8 \sum_\mu |\Delta^s_\mu|^2\vec{d}_\mu^\dagger\vec{d}^\pdagger_\mu + \beta_9 \sum_\mu |\Delta^s_\mu|^2\vec{d}_{\bar{\mu}}^\dagger\vec{d}^\pdagger_{\bar{\mu}} + \beta_{10} \text{Re}\left[\Delta_+^{s*}\Delta_-^{s} \vec{d}_-^\dagger \vec{d}_+^\pdagger \right] \\ 
    &+ \beta_{11} \text{Re}\left[\sum_\mu (\Delta_\mu^{s*})^2 \vec{d}_\mu^T \vec{d}_\mu^{\phantom{T}} \right] + \beta_{12} \text{Re}\left[(\Delta^s_+\Delta^s_-)^* \vec{d}_+^T\vec{d}^{\phantom{T}}_-\right],
\end{alignat}
where $\bar{\mu}=-$ for $\mu=+$ and vice versa. Due to the fewer number of independent parameters in \equref{MatrixExpansionOfFreeEnergyComplRep}, there are many relations between the different coefficients $\beta_1, \dots, \beta_{12}$, namely:
\begin{alignat}{7}
    \nonumber\beta_{1} &= b_1+b_2, \quad
    &&\beta_{2} &&= b_3+b_4+2(b_5-b_2), \quad
    &&\beta_{3} &&= b_1+2b_2, \quad
    &&\beta_{4} &&= b_3+2b_5-4b_2, \\
    \nonumber\beta_{5} &= b_4+2b_5, \quad
    &&\beta_{6} &&= -2b_5, \quad
    &&\beta_{7} &&= -b_2, \quad
    &&\beta_{8} &&= 2(b_1+2b_2), \\
    \beta_{9} &= 2(b_1+b_5)+b_3, \quad
    &&\beta_{10} &&= 2b_4+4b_5, \quad
    &&\beta_{11} &&= 2b_2, \quad
    &&\beta_{12} &&= 4b_5.
\label{DifferentCouplingsBetaB}\end{alignat}    
It is not difficult to observe that the five different purely triplet quartic terms, $\beta_{j=3,4,5,6,7},$ are all independent. Consequently, we can parametrize all twelve $\beta_j$ in terms of the five purely triplet terms and realize all of the triplet states of \secref{SingeltTripletSepForComplexRep}.
\end{widetext}

\end{document}